\documentclass[a4paper,11pt]{article}

\pdfoutput=1 
\usepackage{cite}
\usepackage{jheppubmod}
\usepackage{enumitem}
\usepackage{amsmath}
\usepackage{tensor}
\usepackage{physics}
\usepackage{graphicx}
\usepackage{amsfonts} 
\usepackage{amssymb}
\usepackage[dvipsnames, svgnames, x11names ]{xcolor}
\usepackage{color}
\usepackage{braket}
\graphicspath{ {./1 /} }
\usepackage{subcaption}
\usepackage{graphicx}
\usepackage[T1]{fontenc} 
\usepackage{tikz}
\usepackage{physics}
\usepackage{amsmath}
\usepackage{dsfont}
\newtheorem{theorem}{Theorem}
\usepackage{graphicx}
\usepackage{bbold}
\graphicspath{ {./1 /} }
\colorlet{mygreen}{green!80!black}

\usepackage{tikz}
\usepackage{amsmath} 
\usepackage{mathrsfs} 
\usepackage{xfp} 
\usepackage[outline]{contour} 
\usetikzlibrary{decorations.markings,decorations.pathmorphing}
\usetikzlibrary{angles,quotes} 
\usetikzlibrary{arrows.meta} 
\contourlength{1.4pt}

\tikzset{>=latex} 
\colorlet{myred}{red!80!black}
\colorlet{myblue}{blue!80!black}
\colorlet{mygreen}{green!80!black}
\colorlet{mydarkgreen}{green!30!black}
\colorlet{mydarkred}{red!50!black}
\colorlet{myveryred}{red!30!black}
\colorlet{mydarkblue}{blue!50!black}
\colorlet{mylightblue}{mydarkblue!6}
\colorlet{mypurple}{blue!40!red!80!black}
\colorlet{mydarkpurple}{blue!40!red!50!black}
\colorlet{mylightpurple}{mydarkpurple!80!red!6}
\colorlet{myorange}{orange!40!yellow!95!black}
\tikzstyle{cone}=[mydarkblue,line width=0.2,top color=blue!60!black!30,
                  bottom color=blue!60!black!50!red!30,shading angle=60,fill opacity=0.9]
\tikzstyle{cone back}=[mydarkblue,line width=0.1,dash pattern=on 1pt off 1pt]
\tikzstyle{world line}=[myblue!60,line width=0.4]
\tikzstyle{world line t}=[mypurple!60,line width=0.4]
\tikzstyle{particle}=[mygreen,line width=0.5]
\tikzstyle{photon}=[-{Latex[length=4,width=3]},myorange,line width=0.4,decorate,
                    decoration={snake,amplitude=0.9,segment length=4,post length=3.8}]
\tikzstyle{singularity}=[myred,line width=0.6,decorate,
                         decoration={zigzag,amplitude=2,segment length=6.17}]
\tikzset{declare function={%
  penrose(\x,\c)  = {\fpeval{2/pi*atan( (sqrt((1+tan(\x)^2)^2+4*\c*\c*tan(\x)^2)-1-tan(\x)^2) /(2*\c*tan(\x)^2) )}};%
  penroseu(\x,\t) = {\fpeval{atan(\x+\t)/pi+atan(\x-\t)/pi}};%
  penrosev(\x,\t) = {\fpeval{atan(\x+\t)/pi-atan(\x-\t)/pi}};%
  kruskal(\x,\c)  = {\fpeval{asin( \c*sin(2*\x) )*2/pi}};
}}

\newtheorem{definition}{Definition}[section]
\newtheorem{lemma}[theorem]{Lemma}

\title{\boldmath AdS/CFT, Ultralimits and Baby universes}

\author{Eyoab Bahiru}
\affiliation{Department of Physics, Technion,\\
Haifa 32000, Israel}

\emailAdd{joabdej@gmail.com}

\abstract{We propose boundary theories, CFT$_{p}$ with $p\in \beta\mathbb{N}/\mathbb{N}$, which can be interpreted as an ensemble of theories in the context of AdS/CFT duality, when the boundary spacetime dimension is more than one. These theories are only emergent in the infinite $N$ limit of the boundary CFT and depend on $p$, which corresponds to the different ways one can take a generalized form of the infinite $N$ limit, in the presence of a chaotic or an oscillatory operator. This generalized infinite $N$ limit is called an \emph{ultralimit}, and $p$ is a \emph{free ultrafilter} on $\mathbb{N}$. We propose that the gravitational path integral computes an average of CFT$_{p}$ in each of its sectors; and in the appropriate cases, it produces baby universes and spacetime wormholes. We apply this proposal to the Antonini-Sasieta-Swingle (AS$^{2}$) like states. }

\begin{document}
\maketitle
\flushbottom

\section{Introduction}
\label{sec:intro}

In recent years, it has been suggested that the gravitational path integral `knows more' than what one would naively expect to extract from it. This statement is made primarily due to the contributions of Euclidean wormholes that one can include in the computation of the path integral. This is particularly because Euclidean wormholes seem to reproduce some of the fine grained quantum properties of the quantities that are being computed, which include, famously, the Page curve of black hole evaporation \cite{Penington:2019npb,Almheiri:2019hni,Almheiri:2019psf}. Another related property of Euclidean wormholes is that they introduce Euclidean correlations between expectation values of observables or partition functions from disconnected boundary theories. The boundary interpretation of such correlations is not clear for any finite or infinite value of $N$ (and any other parameters involved) in the examples of the duality where the boundary spacetime dimension is more than one. However, in AdS/CFT the Euclidean wormholes, as is the case for bulk semiclassical spacetime and local quantum fields on it, are expected to emerge only in the infinite $N$ limit. This then raises a question: if Euclidean wormholes are expected to be emergent the infinite $N$ limit, how would they  reproduce effects that are not naively accessible (fine grained properties) or not naively apparent (non factorization of expectation values) in the naive large $N$ expansion\footnote{One may want to address this question by saying Euclidean wormholes are \emph{just} like instanton effects in quantum field theory, that is saddles that provide non perturbative corrections in $1/N$ to observables. However, it is not clear how this would account for the role of spacetime wormholes' contributions in all relevant quantities, notably, the connected correlation functions of expectation values from disconnected holographic CFTs. It is not clear how a non perturbative correction in $1/N$ expansion of a given CFT introduces a correlation with another disconnected independent CFT.}.  In this paper we propose that Euclidean spacetime wormholes (and baby universes) still emerge from the infinite $N$ limit of the boundary, albeit a generalized form of limit called the ultralimit, which agrees with the naively accessible large $N$ data but remains sensitive to some data the standard limit discards. 

Imagine we have a sequence of operators $(O_{N})_{N\in\mathbb{N}}$ with expectation values $\braket{O_{N}}$ chaotic/oscillatory in $N$ ($N$ being for example the rank of the gauge group for the $\mathcal{N}=4$ super Yang Mills theory in four dimensions). We called such sequence, a sequence of chaotic/oscillatory operators. For such sequence of chaotic operators, it is not clear from the bulk perspective, what would capture their large $N$ behavior in the standard AdS/CFT dictionary (check \cite{Kudler-Flam:2026nzz,Liu:2026fnd} for discussion of this question and/or a discussion in general similar in spirit to this paper). The obvious reason for this is that in the regime where AdS/CFT is well understood this sequence of operators are not well defined as the large $N$ limit of the sequence does not exist. However, the ultralimit of any such sequence of expectation values along a particular ultrafilter is always well defined. Roughly speaking, the mechanism this happens is that once one chooses a particular ultrafilter, the ultrafilter unambiguously and consistently chooses a convergent subsequence and assigns that value for the ultralimit\footnote{To be more precise, as we will see, this happens if the chosen ultrafilter is free, otherwise, in the case where it is a principal ultrafilter, the ultralimit will return the corresponding finite $N=N_{0}$ value of the sequence.}. 

Since these concepts of ultrafilters and ultralimits are relatively less familiar in physics, we give heuristic explanations here in the introduction, leaving their more precise discussion to the main body and the appendix. There are several notions of limits used in physics. The most common one is topological limit of a sequence (or net) of objects $(a_{n})_{n\in\mathbb{N}}$ in some topological space $A$. Of course for this limit to be well defined the sequence or the net should satisfy some convergence conditions. The exact convergence condition may depend on the topological space in question; for instance in quantum field theory, we define the limit of a sequence (more precisely net) of operators, $O_{1},O_{2}, ... $ by requiring there matrix elements $\langle\psi|O_{1}|\xi\rangle, \langle\psi|O_{2}|\xi\rangle, ...$ for any two states $\ket{\psi}$ and $\xi$, to converge. The limit of the sequence the operators $O$, reproduces the limiting matrix elements $\langle\psi|O|\xi\rangle$. As we said above, for this limit to exist the sequence has to satisfy some convergence condition and as long as the topological space is closed, convergent sequences will always have a limit. 

The second notion of limits are direct and inverse limits, where on considers a sequence of spaces $(X_{\alpha})_{\alpha \in \mathcal{I}}$ with morphisms $\phi_{\alpha\beta}: X_{\alpha} \rightarrow X_{\beta}$ between them, then one defines the limit of the sequence, a big space where every space in the sequence maps to\footnote{This notion of direct limits (and also inverse limits) arise in category theory and is usually discussed for example in the context of generalized categorical symmetries in physical systems. However, it also seems close to the standard large $N$ limit of AdS/CFT if one restricts focus on the code subspace of $O(1)$ multi trace operators and their corresponding subspaces. The fact that black hole Hilbert spaces do not seems to have a large $N$ limit is because there is no canonical mapping between Hilbert spaces at $N$ and $N+1$ as the dimension of the Hilbert space grows exponentially and most states are new at $N+1$.}. To define the notion of limit we are interested in, ultralimits, we consider a sequence of objects $x_{n}\in X_{n}$ in a sequence of spaces. One does not need a convergence condition or morphisms between the spaces the ultralimit of this sequence always exists. However the limit does not live in any of the spaces in the sequence, $X_{n}$, but in a bigger space called the ultraproduct $\Pi_{n\rightarrow p}X_{n}$, after a choice of a free ultrafilter $p$.  One may consider the scenario where $X_{n}$'s to be the algebra of operators or the Hilbert spaces of the CFTs when the rank of the gauge group is $n$ (which we do in section \ref{disc}). However, at this point and for most of the paper we will be slightly more conservative and consider the sequence of expectation values of chaotic observables. The space these expectation values live in is just the space of real numbers $\mathbb{R}$ or the space of complex numbers $\mathbb{C}$. In this case, $X_{n}$ is $\mathbb{R}$ or $\mathbb{C}$, and if one considers a sequence of bounded expectation values, the ultralimit will also live in $\mathbb{R}$ or $\mathbb{C}$, respectively. The upshot is that, once one chooses an ultrafilter, one gets a notion of a well defined limit even though the sequence does not satisfy a convergence condition.     

A free ultrafilter is just a collection of subsets of the natural numbers $\mathbb{N}$ that we consider `big'. These may be the set of even numbers, prime numbers or multiples $3$. In any case, we consider a collection of such sets that satisfy some axioms. These axioms are quite intuitive conditions for big sets, for instance: no finite subset is considered big; If a set is considered big, then another set larger than this set is also big; the intersection of big sets is also big; and that for any subset $A$, either itself or its complement is big set. Given such set of `big' subsets of $\mathbb{N}$, $p$, one imposes the usual definition of limit on terms indexed by these big sets. For instance, if $(a_{n})_{n}$ is a sequence of alternating $1$ and $-1$ for even $n$ and odd $n$ respectively, then the ultralimit of $a_{n}$ will be $1$ if $p$ contains the set of even numbers or $-1$ if it contains the set of odd numbers. Note that by the axioms of big sets, complementary subsets of $\mathbb{N}$, like the set of even and the set of odd numbers, can not belong to the same free ultrafilter. More generally, we say the ultralimit of $a_{1},a_{2},...$ is $a_{p}$, for the ultrafilter $p$, if and only if for any positive number $\epsilon$, the set of indexes $n$, such that $|a_{n}-a_{p}|<\epsilon$, is a big set in $p$. And due to the axioms we just mentioned, this limit along $p$ is unique and agrees with the usual limit when the usual limit is well defined.         

With this notion ultralimits in hand, we come back to the standard definition of large $N$ limits in AdS/CFT. The large $N$ limit in the bulk corresponds to semiclassical gravitational theory (we will be more precise in the next section) and the dual boundary description arises by looking at the set of appropriately normalized $O(1)$ multitrace operators and considering their expectation values in the vacuum or some other heavy state (created by acting with a heavy ($O(N^{a})$ for $a$ positive number) operator on the vacuum), and taking the large $N$ limit. The infinite $N$ operators, usaully referred to as generalized free fields (GFFs), are operators that reproduce the large $N$ limits of the expectation values of these multitrace operators. They correspond in the bulk to perturbative quantum gravitation fields (if one includes perturbative correction in $1/N$ too) on the classical background corresponding the vacuum or the heavy state that we started with. Schematically we call the collection of these different perturbative quantum gravitational theories, CFT$_{\infty}$. The AdS$_{5}\cross S^{5}$ vacuum state and quantum fields on top of it are part of this `theory', but also eternal black hole spacetimes at different temperatures and quantum fields on top of them are in CFT$_{\infty}$. One can roughly think of this CFT$_{\infty}$ as a direct sum of different super-selection sectors corresponding to the specific classical backgrounds one chooses, with the different representations of the GFFs acting locally in each sector. 

If however one starts with a state created from the vacuum by acting with an oscillating/chaotic operator, so that the expectation values of the $O(1)$ multitrace operators is also oscillating/chaotic, then the standard large $N$ limit can not be taken and such states and operators are not part of CFT$_{\infty}$. However, one can taken the ultralimit along some free ultrafilter, $p$. Then, the expectation values of the $O(1)$ multitrace operators will be well defined. The GNS construction provides a corresponding Hilbert space and von Neumann algebra acting on the Hilbert space. One can also consider perturbative corrections in $1/N$ for these infinite $N$ operators, which again depends on $p$. These perturbative theories are not part of CFT$_{\infty}$, they extend it to a theory we call CFT$_{p}$. Of course CFT$_{p}$ depends on the free ultrafilter chosen, and looking at perturbative sectors with a given background state, this perturbative Hilbert space and algebra of operators depend on $p$. As we will see the space of free ultrafilter is called $\beta\mathbb{N}/\mathbb{N}$, but depending on the starting state we find that there are relatively smaller number of physically distinguishable free ultrafilters. Thus the perturbative Hilbert spaces and algebra of operators labeled by the physically distinguishable free ultrafilters, $p$.

We propose that these distinct perturbative theories are much like the $\alpha$ sectors of Marolf and Maxfield \cite{Marolf:2020xie}, and that the gravitational path integral involving these states computes an average of expectation values over these distinct theories. Thus Euclidean wormholes and baby universes emerge from the ensemble of these theories. In this first paper, we confine ourselves to realizing this proposal in the case of AS$^{2}$ like states \cite{Antonini:2023hdh}.

In the following section, section \ref{sec2}, we review some known topics in the literature and present it as a motivation for the proposal. In section \ref{sec 3}, we discuss ultralimits and introduce more precisely the CFT$_{p}$ boundary theories. In section \ref{sec4} we consider a simple spin chain model that realizes in some way the emergence of baby universes and spacetime wormholes. In the following section, we consider AS$^{2}$ and AS$^{2}$ like states and see how baby universes, Euclidean wormholes emerge. We also provide an argument for the emergence of long wormholes above the Hawking Page temperature in this case. Finally, we comment on the relation of our construction with the different ultraproduct of a sequence of CFT$_{N}$ theories in the context of AdS/CFT. The appendix includes more precise definitions and statements of the theorems used in the main body of the paper, mostly without proofs.

\section{`Traditional' AdS/CFT}\label{sec2}

In the arguably quite successful AdS/CFT duality, at least in the regime of parameters where the bulk theory is better understood, there still remain some pieces that do not seem to fit well. Consider for instance CFT operators whose expectation values oscillate wildly as we take $N$ to infinity, with a variance that does not decay in the limit; precisely because the duality is well understood in the $N \rightarrow \infty$ limit\footnote{Here when we say the $N \rightarrow \infty$ limit, we generally mean the limit where the boundary degrees of freedom become large, even if there is no actual parameter $N$ in the boundary theory.}, it is not completely clear what bulk quantity computes these expectation values, for the simple reason that these operators do not have the `standard' limit as $N$ goes to infinity. Since one of the main goals of AdS/CFT is to understand the bulk theory of quantum gravity at finite $G_{N}\sim \frac{1}{N^{a}}$, for some positive real number $a$, understanding these operators is as important, if not more, as understanding operators with a well defined large $N$ limit. 

\subsection{What integrability teaches us}

To be a bit more precise about these operators with oscillating expectation values, it is better to first understand the sector of the duality where there is very good matching between the two sides\footnote{For most of this subsection we restrict to holographic dualities with boundary dimensions strictly greater than one, and we also will not consider the possible duality of pure gravity in AdS$_{3}$ with an ensemble of CFT's.}. The AdS/CFT duality is a strong -weak coupling duality, which means the strong coupling regime of the bulk theory corresponds to the weak coupling regime of the boundary theory and vice versa. To be more concert, if one considers the bulk theory to be type IIB string theory on AdS$_{5} \cross S^{5}$, the boundary theory would be the maximally symmetric super Yang Mills theory in $4$ dimensions with $SU(N)$ gauge group, defined on the asymptotic conformal boundary of the AdS$_{5}$ spacetime. The relevant parameters for the boundary theory are the 't Hooft coupling $\lambda$ and $N$(as $N^{2}$ counts the total number of degrees of freedom in the gauge theory), while for the bulk theory one has, $\alpha^{'} \sim 1/\sqrt{\lambda}$ which controls the curvature strength in the bulk and the coupling constant for the strings $g_{s} \sim N^{-1} (\alpha^{'})^{-2}$.  Therefore the weak coupling regime of the boundary theory corresponds to a highly curved bulks spacetime; while the regime where the string theory can be understood perturbatively (i.e, when $\alpha^{'} =0$ and around $g_{s}=0$), the boundary gauge theory is strongly coupled. 

There is however an interesting limit where the gauge theory is significantly simplified called the 't Hooft limit. This is the limit where $N$ is taken to go to infinity while $\lambda$ is kept fixed. The simplification arises since all Feynman diagrams in the perturbation theory around $\lambda =0$ that can not be drawn on a plane without lines crossing, are all suppressed in $1/N$, i.e, diagrams that can only be drawn on higher genius 2D surfaces in such a way are suppressed. In this leading order planar limit the growth of the higher order loop diagrams is exponential rather than factorial, and therefore the perturbation theory is so well behaved that it has a finite radius of convergence. Importantly, planar $\mathcal{N}=4$ super Yang Mills theory can be thought of as defined on this $2$D surface and an important tool available for two dimensional models called \emph{integrability} can be applied \cite{Minahan:2010js}. Thus the theory can be mapped to an integrable quantum spin chain and the spectrum (the conformal dimensions) of local operators can be computed, at least algebraic equations of which the spectra are solutions can be given \cite{Sieg:2010jt,Rej:2010ju}. This is done for any local operators fundamental or composite, irrespective of how many constituents are present, using the several techniques of integrability. A crucial point is that this is done for arbitrary values of $\lambda$. Similar work on the same level of generality concerning correction functions and expectation values of extended operators is currently underway.

In the bulk the planar limit corresponds to having free strings ($g_{s}=0$) with arbitrary string tension (or arbitrary $\alpha^{'}$). In this limit, a string propagating in the AdS$_{5}\cross S^{5}$ background is described by a highly non linear sigma model and again using integrability methods exact analytical results can be extracted \cite{Tseytlin:2010jv,McLoughlin:2010jw,Magro:2010jx,Schafer-Nameki:2010qho}. Thus it is now possible to do the matching exercise between the two sides of the duality, and every computation that has been done has found exact matching between these two. 

It is therefore reasonable to assume that observables of the planar $\mathcal{N}=4$ super Yang Mills theory(or the large $N$ limit of holographic CFT's in other dimensions) and their expectation values can be mapped to the bulk string theory(or the relevant quantum theory of gravity) without much subtlety and caveats. A more rigorous proof in this direction is given when the bulk theory is a type IIB string theory on AdS$_{3}\cross S^{3}\cross C_{4}$ where $C_{4}$ is a compact $4$D manifold that is either $T^{4}$, $K3$ or $S^{3}\cross S^{1}$ and the boundary theory is a non linear sigma model with symmetric orbifold target space for the respective $C_{4}$. Thus now $N \sim g_{s}^{-2}$ is the orbifolding parameter while $\alpha^{'}$ has a slightly complicated boundary meaning as a deformation away from the free orbifold point, where the effective tension of the bulk strings vanish. Deforming away from the free point towards $\alpha^{'} \rightarrow 0$, the bulk will be described by a supergravity theory and boundary quantities can be computed by looking at the dominant saddle in the gravitational path integral. The reason for this is that in this limit, roughly speaking, the duality reduces to the following 

\begin{equation}\label{dualitylim}
    Z_{grav}[\phi|_{\partial \text{AdS}} = J] = Z_{M}[J].
\end{equation}

It is clear that the left hand side has to be qualified, and we understand the gravitational path integral as a sum over saddles which are asymptotic to AdS$_{3}\cross S^{3}\cross C_{4}$, with possible semiclassical matter on top of them. The $\phi$'s are sources for the bulk quantum fields that only gravitate perturbatively; and their boundary condition is fixed to $J$. While on the right hand side, $J$ is the source for the dual boundary operators. In addition, $M$ is the $2$D connected manifold the CFT is defined on, which is the conformal boundary of the non compact part of the bulk saddles. As usual correlation functions can be recovered by differentiating with respect to $J$.

This setup in the Euclidean continuation was considered in \cite{Schlenker:2022dyo} and a particular kind of subtlety and caveat alluded to earlier was discussed. In particular the question is, in the computation of the expectation value of a certain boundary observable, what kinds of saddles can contribute? One drawback in the strategy adopted was that, they were only able to consider bulk saddles which look like $\mathcal{M}\cross S^{3}\cross C_{4}$, where $\mathcal{M}$ is asymptotically AdS$_{3}$\footnote{General manifolds with AdS$_{3}\cross S^{3}\cross C_{4}$ asymptotics are really less understood in the literature thus direct consideration of such bulk saddles is not fully available.}. In any case, the authors asked, can manifolds $\mathcal{M}$ with additional boundaries than just $M$ contribute to the observables available in the planar limit? The reason this produces a difficulty for the standard AdS/CFT is that if $\mathcal{M}$ with conformal boundary $M \sqcup M^{'} $ can contribute as a saddle then following \eqref{dualitylim}, it will contribute to a Euclidean correlation between operators `from' $Z_{M}[J]$ and $Z_{M^{'}}[J^{'}]$ after just a differentiation with respect to the boundary conditions for the sources. From the boundary perspective, this correlation is between the expectation values of operators 
sourced by $J$ and $J^{'}$ corresponding to the different theories, which for a given choice of the parameters are each just numbers. The AdS/CFT duality maps a boundary theory with a given choice of the parameters ($N$ and some gauge coupling for instance) to a bulk theory with a corresponding choice its parameters ($g_{s}$ and $\alpha^{'}$). However here, if $\mathcal{M}$ with multiple boundaries is a saddle, a given bulk theory implies a correlation between boundary expectation values.   

It is well known similar situations have arisen in the AdS$_{2}$/CFT$_{1}$ duality (which is also the reason why this particular `subtlety' was considered), and in that case there is a boundary coupling constant that is not exactly mapped to the bulk. Such saddles are understood \cite{Saad:2019lba,Stanford:2019vob} to be implying a correlation between expectation values in a random draw of the Gaussian ensemble of the boundary coupling constants.  

\subsection{Bulk randomness}

\begin{figure}

\tikzset{every picture/.style={line width=0.75pt}} 

\begin{tikzpicture}[x=0.75pt,y=0.75pt,yscale=-1,xscale=1]

\draw   (40,178) .. controls (40,133.82) and (54.1,98) .. (71.5,98) .. controls (88.9,98) and (103,133.82) .. (103,178) .. controls (103,222.18) and (88.9,258) .. (71.5,258) .. controls (54.1,258) and (40,222.18) .. (40,178) -- cycle ;
\draw   (551,175) .. controls (551,130.82) and (565.1,95) .. (582.5,95) .. controls (599.9,95) and (614,130.82) .. (614,175) .. controls (614,219.18) and (599.9,255) .. (582.5,255) .. controls (565.1,255) and (551,219.18) .. (551,175) -- cycle ;
\draw    (71.5,98) .. controls (122,83) and (212,197.5) .. (265,200.5) ;
\draw    (78.5,257) .. controls (134,261) and (210.5,211) .. (263.5,214) ;
\draw    (339,184.5) .. controls (445,131.5) and (529,81) .. (582.5,95) ;
\draw    (340.5,201.5) .. controls (359.5,189.5) and (529,241) .. (582.5,255) ;
\draw   (259,207.25) .. controls (259,203.52) and (261.01,200.5) .. (263.5,200.5) .. controls (265.99,200.5) and (268,203.52) .. (268,207.25) .. controls (268,210.98) and (265.99,214) .. (263.5,214) .. controls (261.01,214) and (259,210.98) .. (259,207.25) -- cycle ;
\draw   (336,193) .. controls (336,188.31) and (338.01,184.5) .. (340.5,184.5) .. controls (342.99,184.5) and (345,188.31) .. (345,193) .. controls (345,197.69) and (342.99,201.5) .. (340.5,201.5) .. controls (338.01,201.5) and (336,197.69) .. (336,193) -- cycle ;
\draw    (263.5,200.5) .. controls (303.5,170.5) and (290,207.5) .. (339,184.5) ;
\draw    (263.5,214) .. controls (303.5,184) and (291.5,224.5) .. (340.5,201.5) ;

\draw (57,160) node [anchor=north west][inner sep=0.75pt]   [align=left] {$M$};
\draw (574,159) node [anchor=north west][inner sep=0.75pt]   [align=left] {$M^{'}$};
\draw (139,149.4) node [anchor=north west][inner sep=0.75pt]    {$\mathcal{M}_{1}$};
\draw (478,149.4) node [anchor=north west][inner sep=0.75pt]    {$\mathcal{M}_{2}$};

\end{tikzpicture}

\caption{Two manifolds, each with single boundary are connected by a single throat forming a manifold with two boundaries.}
    \label{fig 1}
\end{figure}
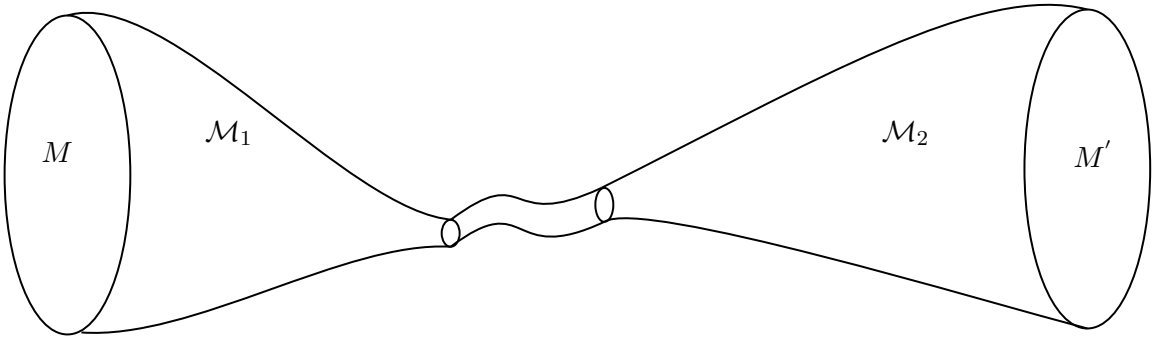

A closely related issue is a possible interpretation of wormholes for the bulk theory first discussed by Coleman \cite{Coleman:1988cy,Giddings:1988wv,Klebanov:1988eh}. Forgetting for the moment wormholes that connect two arbitrary points on the same manifold, let's focus on wormholes that connect bulk manifolds $\mathcal{M}_{1}$ with a single asymptotic boundary $M$ and another bulk manifold $\mathcal{M}_{2}$ with another asymptotic boundary $M^{'}$, fig.\ref{fig 1}. we take $M^{'}$ to be connected and we assume both are asymptotically AdS$_{3}$ spacetimes to be concert\footnote{We neglect the compact spaces since they will not change much of our discussion.}. 

In the infrared limit where the finite size of the wormhole throat can not be resolved, we can assume the wormhole is simply connecting two arbitrary points on $\mathcal{M}_{1}$ and $\mathcal{M}_{2}$. Let's take the direction along the wormhole throat from $\mathcal{M}_{1}$ to $\mathcal{M}_{2}$ as the Euclidean time direction and therefore we can assume the gravitational path integral as states propagating  from $\mathcal{M}_{1}$ to $\mathcal{M}_{2}$ or their conjugate states propagating in the opposite direction. In the limit the throat size is very small, the effect of these propagating states is the same as acting with operators with the same charge and symmetry as the wormhole mouth, weighted by the amplitude for the corresponding state to propagate through the wormhole. In other words the effect of a wormhole is that expectation values in the combined bulk theory for $\mathcal{M}_{1}$ and $\mathcal{M}_{2}$, are computed in the presence of the operator,

\begin{equation}\label{WHaction}
   \sum_{i} c_{i} \int_{\mathcal{M}_{1}} d^{3}x_{1}\int_{\mathcal{M}_{2}} d^{3}x_{2} \;\Phi_{i}(x_{1})\Phi^{\dagger}_{i}(x_{2}).
\end{equation}

\begin{figure}

\tikzset{every picture/.style={line width=0.75pt}} 

\begin{tikzpicture}[x=0.75pt,y=0.75pt,yscale=-1,xscale=1]

\draw   (40,178) .. controls (40,133.82) and (54.1,98) .. (71.5,98) .. controls (88.9,98) and (103,133.82) .. (103,178) .. controls (103,222.18) and (88.9,258) .. (71.5,258) .. controls (54.1,258) and (40,222.18) .. (40,178) -- cycle ;
\draw   (551,175) .. controls (551,130.82) and (565.1,95) .. (582.5,95) .. controls (599.9,95) and (614,130.82) .. (614,175) .. controls (614,219.18) and (599.9,255) .. (582.5,255) .. controls (565.1,255) and (551,219.18) .. (551,175) -- cycle ;
\draw    (71.5,98) .. controls (122,83) and (212,197.5) .. (265,200.5) ;
\draw    (78.5,257) .. controls (134,261) and (210.5,211) .. (263.5,214) ;
\draw    (339,184.5) .. controls (445,131.5) and (529,81) .. (582.5,95) ;
\draw    (340.5,201.5) .. controls (359.5,189.5) and (529,241) .. (582.5,255) ;
\draw   (259,207.25) .. controls (259,203.52) and (261.01,200.5) .. (263.5,200.5) .. controls (265.99,200.5) and (268,203.52) .. (268,207.25) .. controls (268,210.98) and (265.99,214) .. (263.5,214) .. controls (261.01,214) and (259,210.98) .. (259,207.25) -- cycle ;
\draw   (336,193) .. controls (336,188.31) and (338.01,184.5) .. (340.5,184.5) .. controls (342.99,184.5) and (345,188.31) .. (345,193) .. controls (345,197.69) and (342.99,201.5) .. (340.5,201.5) .. controls (338.01,201.5) and (336,197.69) .. (336,193) -- cycle ;
\draw    (263.5,200.5) .. controls (303.5,170.5) and (290,207.5) .. (339,184.5) ;
\draw    (263.5,214) .. controls (303.5,184) and (291.5,224.5) .. (340.5,201.5) ;
\draw    (134,122.5) .. controls (174,92.5) and (409,63.5) .. (509,119.5) ;
\draw    (141,128.5) .. controls (181,98.5) and (395.96,81.47) .. (499.96,125.47) ;
\draw    (181,160.5) .. controls (234,112.5) and (346,107.5) .. (457,145.5) ;
\draw    (191.02,169.12) .. controls (254.02,123.12) and (331.01,122.48) .. (444.01,154.48) ;
\draw    (197.58,219.17) .. controls (270.58,246.17) and (345.37,241.95) .. (471.37,201.95) ;
\draw    (184.41,227.87) .. controls (257.41,254.87) and (336.62,255.09) .. (462.62,215.09) ;
\draw   (134.56,121.76) .. controls (135.06,121.1) and (137.11,121.8) .. (139.14,123.33) .. controls (141.16,124.87) and (142.4,126.64) .. (141.9,127.3) .. controls (141.4,127.96) and (139.36,127.26) .. (137.33,125.73) .. controls (135.3,124.19) and (134.06,122.42) .. (134.56,121.76) -- cycle ;
\draw   (180.5,161.17) .. controls (181.59,159.72) and (184.83,160.33) .. (187.74,162.52) .. controls (190.64,164.72) and (192.11,167.67) .. (191.02,169.12) .. controls (189.92,170.57) and (186.68,169.96) .. (183.77,167.77) .. controls (180.87,165.57) and (179.4,162.62) .. (180.5,161.17) -- cycle ;
\draw   (444.01,154.48) .. controls (442.9,152.81) and (444.96,149.51) .. (448.6,147.11) .. controls (452.24,144.7) and (456.08,144.11) .. (457.18,145.78) .. controls (458.29,147.45) and (456.23,150.75) .. (452.59,153.15) .. controls (448.95,155.55) and (445.11,156.15) .. (444.01,154.48) -- cycle ;
\draw   (499.96,125.47) .. controls (498.87,123.81) and (500.01,121.14) .. (502.5,119.49) .. controls (505,117.84) and (507.91,117.85) .. (509,119.5) .. controls (510.09,121.15) and (508.95,123.83) .. (506.46,125.48) .. controls (503.96,127.13) and (501.05,127.12) .. (499.96,125.47) -- cycle ;
\draw   (184.41,227.87) .. controls (183.3,226.2) and (185.36,222.9) .. (189,220.5) .. controls (192.64,218.1) and (196.48,217.5) .. (197.58,219.17) .. controls (198.69,220.84) and (196.63,224.14) .. (192.99,226.54) .. controls (189.35,228.95) and (185.51,229.54) .. (184.41,227.87) -- cycle ;
\draw   (462.62,215.09) .. controls (460.95,213.98) and (461.56,210.14) .. (463.98,206.51) .. controls (466.4,202.89) and (469.71,200.84) .. (471.37,201.95) .. controls (473.04,203.06) and (472.43,206.9) .. (470.01,210.53) .. controls (467.59,214.16) and (464.28,216.2) .. (462.62,215.09) -- cycle ;
\draw  [color={rgb, 255:red, 255; green, 255; blue, 255 }  ,draw opacity=1 ][line width=3] [line join = round][line cap = round] (190,158.5) .. controls (189.34,158.5) and (186.15,157.35) .. (187,156.5) .. controls (187.8,155.7) and (195.04,158.5) .. (197,158.5) ;
\draw  [color={rgb, 255:red, 255; green, 255; blue, 255 }  ,draw opacity=1 ][line width=3] [line join = round][line cap = round] (194,160.5) .. controls (194,161.97) and (193,162.03) .. (193,163.5) ;
\draw  [color={rgb, 255:red, 255; green, 255; blue, 255 }  ,draw opacity=1 ][line width=3] [line join = round][line cap = round] (423,141.5) .. controls (394.9,141.5) and (422.65,143.91) .. (429,145.5) .. controls (430.97,145.99) and (433.26,147.54) .. (435,146.5) .. controls (435.86,145.99) and (435.32,144.45) .. (435,143.5) .. controls (434.89,143.18) and (434.3,143.65) .. (434,143.5) .. controls (429.73,141.36) and (424.95,140.47) .. (421,138.5) .. controls (420.4,138.2) and (419.67,138.5) .. (419,138.5) ;
\draw  [color={rgb, 255:red, 255; green, 255; blue, 255 }  ,draw opacity=1 ][line width=3] [line join = round][line cap = round] (431,139.5) .. controls (429,139.5) and (427,139.5) .. (425,139.5) ;
\draw  [color={rgb, 255:red, 255; green, 255; blue, 255 }  ,draw opacity=1 ][line width=3] [line join = round][line cap = round] (490,114.5) .. controls (485.32,114.5) and (479.38,112.35) .. (476,111.5) .. controls (474.67,111.17) and (470.77,111.11) .. (472,110.5) .. controls (474.18,109.41) and (480.9,112.95) .. (482,113.5) .. controls (484.17,114.59) and (491.43,115.5) .. (489,115.5) .. controls (485.65,115.5) and (481.37,116.87) .. (479,114.5) .. controls (478.15,113.65) and (474.8,112.57) .. (476,112.5) .. controls (481.66,112.17) and (487.34,112.17) .. (493,112.5) .. controls (493.74,112.54) and (495.75,113.5) .. (495,113.5) .. controls (492.1,113.5) and (489.12,111.5) .. (487,111.5) ;
\draw  [color={rgb, 255:red, 255; green, 255; blue, 255 }  ,draw opacity=1 ][line width=3] [line join = round][line cap = round] (442,215.5) .. controls (440.11,215.5) and (432,218.5) .. (432,218.5) .. controls (432,218.5) and (440.61,213.84) .. (443,213.5) .. controls (444.98,213.22) and (447.03,213.83) .. (449,213.5) .. controls (449.74,213.38) and (451.75,212.5) .. (451,212.5) .. controls (446.89,212.5) and (437.03,216.5) .. (446,216.5) ;
\draw  [color={rgb, 255:red, 255; green, 255; blue, 255 }  ,draw opacity=1 ][line width=3] [line join = round][line cap = round] (203,229.5) .. controls (203,227.54) and (202.62,225.99) .. (202,223.5) .. controls (201.82,222.78) and (201,220.75) .. (201,221.5) .. controls (201,224.5) and (201.33,227.52) .. (201,230.5) .. controls (200.85,231.87) and (195.47,231.91) .. (195,230.5) .. controls (193.57,226.21) and (210.6,224.5) .. (206,224.5) ;

\draw (57,160) node [anchor=north west][inner sep=0.75pt]   [align=left] {M};
\draw (574,159) node [anchor=north west][inner sep=0.75pt]   [align=left] {M'};
\draw (139,149.4) node [anchor=north west][inner sep=0.75pt]    {$\mathcal{M}_{1}$};
\draw (478,149.4) node [anchor=north west][inner sep=0.75pt]    {$\mathcal{M}_{2}$};

\end{tikzpicture}

    \caption{The same manifolds are now connected by multiple wormholes.}
    \label{fig 2}
\end{figure}
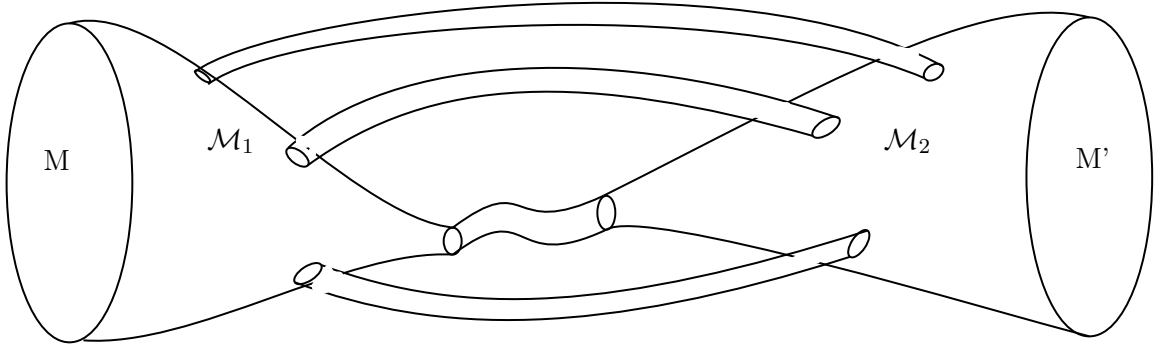

As one considers multiple wormholes, the combined effect exponentiates, and the full effect of such wormholes with multiple throats, fig. \ref{fig 2}, would be to just introduce the above nonlocal coupling to the combined action of the two disconnected bulk theories.

However, one can rewrite the exponential of \eqref{WHaction} as,

\begin{equation}
    \int \prod_{i,j}dp_{i}\;dq_{j}\;e^{-\frac{1}{2} c^{-1}_{i} p_{i}q_{i}} \; e^{\int_{\mathcal{M}_{1}}d^{3}x_{1}\; p_{i}\Phi_{i}(x_{1})}e^{\int_{\mathcal{M}_{2}}d^{3}x_{2}\;q_{i}\Phi^{\dagger}_{i}(x_{2})}
\end{equation}

where the repeated indices in the exponential are summed over. Therefore instead of a bilocal interaction between the theory defined on $\mathcal{M}_{1}$ and on $\mathcal{M}_{2}$, one can consider modifying the couplings for the operators $\Phi_{i}(x_{1})$ and $\Phi^{\dagger}_{i}(x_{2})$ and consider the correlated average (sometimes called the Hubbard–Stratonovich representation) in these couplings of the two theories. Since the couplings $p_{i}$ and $q_{i}$ are constant in spacetime, no local operator can change them and they correspond to different super-selection sectors as they take different values. Thus the effect of the wormholes can be summarized by the connected correlation between expectation values for the two theories at these random couplings,

\begin{equation}\label{bulkensambe}
    \int \prod_{i,j}dp_{i}dq_{j}\; \mu(p_{i},q_{i})\; \langle \Phi_{1}\rangle_{p}\langle \Phi_{2}\rangle_{q},
\end{equation}

where $\mu$ is the measure now including the effect coming from the wormholes connecting points on each of the individual universes, and $\Phi_{1}$ is some operator on $\mathcal{M}_{1}$ while $\Phi_{2}$ is an operator on $\mathcal{M}_{2}$. We will use this bulk intuition that at the large $N$ limit (semiclassically) we expect a wormhole which can be interpreted as the existence of such correlations between expectation values; which we relate to the manner we take the large $N$ limit in the boundary, i.e, ultralimits, and an average in this procedure. In particular ultralimits will provide the boundary analog of \eqref{bulkensambe}. The above discussion can also be generalized to more than two boundaries.   

As discussed in the previous subsection, to recover the boundary version of \eqref{bulkensambe}, one has to note that there are certain operators which integrability suggests are not participants in correlations involving wormholes with multiple boundaries. Thus an important question is how the varied ultralimits correspond to the correlated average of the $p$ and $q$ shifted theories, which we discuss following the next subsection. 

We would like to also mention Euclidean wormhole solutions indeed exist in string theory. Particularly, in compactifications of string theory that lead to non linear sigmal models coupled to gravity with a moduli space that has a metric with an indefinite signature. Such effective field theories typically arise in axion-dilaton systems, where the timelike direction on the moduli space corresponds to the wrong sign picked up by the kinetic term of the axions after Euclidean continuation. A more careful discussion of this and other constructions can be found in the references \cite{Giddings:1987cg,Arkani-Hamed:2007cpn,Maldacena:2004rf,Bergman:2007ss,Bergshoeff:2004pg,Loges:2023ypl,Hertog:2017owm,Astesiano:2023iql}.

\subsection{The boundary story}

Coming to the boundary discussion in the higher dimensional case, where resolutions applied to AdS$_{2}$/CFT$_{1}$ are not available, the existence of such a saddle is `a piece that does not fit'. However \cite{Schlenker:2022dyo} showed that for the planar limit (and $\alpha^{'}\rightarrow 0$) such saddles do not contribute to the computations of the observables. A typical expectation value that we would like to compute would be the thermal expectation value of some local operator $\mathcal{O}$,

\begin{equation}\label{expect}
    \text{Tr}(e^{-\beta H}\mathcal{O})= \sum_{n}e^{-\beta E_{n}} \bra{n}\mathcal{O}\ket{n}
\end{equation}

where $H$ is the Hamiltonian and $E_{n}$ are its eigenvalues for its eigenvectors $\ket{n}$. The state operator correspondence implies that an operator $O_{i}$ on $\mathbf{R}^{d}$ can be mapped to a state $\Psi_{O_{i}}$ on $\mathbf{R}\cross S^{d-1}$ whose energy is given by\footnote{We have assumed the sphere has radius one.},

\begin{equation}
    E_{i} = \Delta_{i} - \frac{c}{12}
\end{equation}

where $c$ is the central charge of the theory and $c=6N$, while $\Delta_{i}$ is the conformal dimension of the operator $O_{i}$ and we set the sphere radius to one. If $O$ is an operator which receives contributions only from states with $E_{i}<0$, i.e, $\Delta_{i} < \frac{c}{12}$, which is most definitely satisfied for operators with a planar limit, then the expectation value \eqref{expect} blows up in the limit $\beta \rightarrow \infty$. Since the operators $O_{i}$'s may also have a non zero spin, $J_{i}$, in $\mathbf{R}^{d}$, one can slightly generalize \eqref{expect} to $\text{Tr}(e^{-\beta H + i\alpha R} \;O)$ where $R$ generates rotation around $S^{d-1}$. Then the left hand side will have and extra $e^{i\alpha J_{i}}$ term in the sum. We can further generalize \eqref{expect} by considering expectation values like 

\begin{align}\label{genaexpe}
    \text{Tr}(e^{-\beta_{1} H + i\alpha_{1} R} \;&O_{1} e^{-\beta_{2} H + i\alpha_{2} R} \;O_{2} \cdots  e^{-\beta_{r} H + i\alpha_{r} R} \;O_{r}) = \\ & \sum_{n,m_{1},\cdots,m_{r-1} } e^{-\beta_{1}E_{n}+i\alpha_{1} J_{n}}\bra{n}O_{1}\ket{m_{1}}\cdots e^{-\beta_{r}E_{m_{r-1}}+i\alpha_{r} J_{m_{r-1}}}\bra{m_{r-1}}O_{r}\ket{n}
\end{align}

Using the OPE expansion of the theory, the above expectation value can be written in terms of the OPE coefficients after writing $\bra{m_{i}}O_{j}\ket{m_{k}}$ as $c_{ijk}$, which correspond to a path integral on a 'pair of pants' where each holes corresponds to $i,j$ and $k$. For instance, \eqref{expect} is given by a torus with a hole, which corresponds to $O$, and the sum over $n$ creates a torus from a cylinder with length $\beta$. Therefore, a general expectation value like \eqref{genaexpe} we might want to compute is given by a partition function on some genus $ r$ Riemann surface with possible punctures on it, which is equivalent to turning on sources and differentiating the partition function on the genius $r$ surface with respect to the sources before setting them to zero; and each lengths across the holes given by $\beta_{i}$ and a twist of $\alpha_{i}$. 

Again if the above expectation value does not receive contribution from states (operators) with $\Delta_{i} < \frac{c}{12}$, in particular for states that survive the planar limit, it will blow up as any of the $\beta_{i} \rightarrow \infty $ for $i= 1,\cdots,r$. If we imagine $M_{r}$ to be a Riemann surface with genius $r$, and \eqref{genaexpe} is computed by differentiating $Z_{M_{r}}[J_{1},\cdots,J_{r}]$, which in the bulk is given by $Z_{grav}$ with boundary conditions for bulk sources set to $[J_{1},\cdots,J_{r}]$. These sources modify the bulk saddle only perturbatively and we may as well consider the case with no sources turned on. More generally, bulk matter fields gives correction to the bulk saddle only perturbatively and therefore we can consider the case with no matter fields at all, for our purposes.   

Thus, what we would like to find out is if a bulk manifold of type $\mathcal{M}\cross S^{3}\cross C_{4}$, where a conformal boundary of $\mathcal{M}$ is $M_{r}$, can have additional boundaries like $M^{'}$ and at the same time its partition function blows up when any of the $\beta_{i}$'s are taken to infinity. The bulk manifold should be a saddle and therefore it satisfies the Einstein equations. 

Using the equations of motion, with no matter fields, the action reduces to 

\begin{equation}
    I = \frac{1}{4\pi Gl^{2}} \int_{\mathcal{M}}d^{3}x\;\sqrt{g} -\frac{1}{8\pi G}\int_{\partial\mathcal{M}}d^{2}x\; \sqrt{h}K
\end{equation}
where $h$ is the induced metric and $K$ is the extrinsic curvature of $\partial\mathcal{M}$. Here we see one of the qualifications for \eqref{dualitylim} that we mentioned earlier, since the above action is infinite for hyperbolic 3 manifolds with boundary. Thus one has to adopt a renormalization scheme and add counterterms which cancel the boundary term and renormalize the infinite volume coming from the first term above \cite{Schlenker:2022dyo,Krasnov:2006jb}. One gets $$e^{-I_{R}}=e^{-V_{R}(\mathcal{M})/4\pi G}.$$ where $V_{R}(\mathcal{M})$ is the renormalized volume of $\mathcal{M}$ which can be negative and depends of the choice of $\beta_{i}$'s.

On \cite{Schlenker:2022dyo}, the authors proved that the manifold $\mathcal{M}$ whose renormalized volume goes to $-\infty$ when any of the $\beta_{i}$'s go to $\infty$, is the one the `fills' all of the cylinders specified by the length $\beta_{i}$, in other words, if you cut across any of these cylinders, each is a boundary to a disk in $\mathcal{M}$, usually called a Schottky manifold. In particular, $\mathcal{M}$ can not be a manifold with additional boundary other than $M_{r}$. In particular, assuming off shell bulk configuration do not contribute, a correlation between expectation values like \eqref{genaexpe} and similar expectation values on different boundary manifolds is not predicted by the bulk theory\footnote{Here, as we mentioned before we are neglecting general 10 dimensional manifolds which are asymptotically AdS$_{3}\cross S^{3}\cross C_{4}$.}, for operators that survive in the planar limit. This computation in fact implies any operator which does not receive contributions from states with $E_{i}> c/12$, that is, according to the analysis of BTZ \cite{Banados:1992wn} of states with AdS$_{3}$ asymptotics, black hole states, such correlations are not present.  

To prove the lesson we were drawing from integrability, one needs to show that either on shell or off shell geometries, which can possibly contribute to correlations between expectation values, with disconnected boundaries will not contribute for observables that survive the planar limit, for any $\alpha^{'}$. However, we take the above computation as supporting evidence and assume that observables that persist in the planar limit and have well defined large $N$ limit do not involve any subtlety for the traditional AdS/CFT correspondence.

\section{Oscillating/chaotic observables}\label{sec 3}

Consider for instance we want to compute expectation values like \eqref{expect} in the boundary gauge theory on two distinct tori $M$ and $M^{'}$, with the length of the thermal circles given by $\beta $ and $\beta^{'}$. If we take $\beta$ and $\beta^{'}$ to be very large, we have 

\begin{equation}\label{boundarycorre}
    \text{Tr}(e^{-\beta H}\mathcal{O})\text{Tr}(e^{-\beta^{'} H}\mathcal{O}^{'})= \bra{0}\mathcal{O}\ket{0}\bra{0}\mathcal{O}^{'}\ket{0}+ O(e^{-(\beta+\beta^{'})E_{1}})
\end{equation}
assuming we have non degenerate vacuum and $E_{1}$ is the energy of the first excited state. However, if there is a bulk saddle connecting $M$ and $M^{'}$, \eqref{bulkensambe} implies that we have,

\begin{equation}
     \int \prod_{i,j}dp_{i}dq_{j}\; \mu(p_{i},q_{i})\; \bra{0} \mathcal{O}\ket{0}_{p}\bra{0} \mathcal{O}^{'}\ket{0}_{q} + O(e^{-\beta E_{1,p}-\beta^{'}E_{1,q}})
\end{equation}

It is however important to note a couple of points. One is that the above expression is expected to match with \eqref{boundarycorre} in the large $N$ limit, i.e, at $N=\infty$ or perturbation theory around $1/N = 0$. The second point is the implication of the discussion in the previous section, which is that wormhole saddles contribute for only for operators not accessible in the planar limit, in other words operators whose expectation values and couplings depend on $N$, may oscillate and not have well defined large $N$ limit. Therefore whatever boundary effect reproduces the above bulk result, it should concern these operators and their large $N$ limit. In particular, precisely speaking, \eqref{boundarycorre} is not what is troubling us, what's `troubling' is the large $N$ limit of \eqref{boundarycorre} for oscillating operators. We claim that taking the large $N$ limit along a particular ultrafilter(more precisely non principal ultrafilters), which is always well defined, corresponds to what are usually called $\alpha$ sectors in most of the literature, a' la Coleman and later Marolf and Maxfield; and the proposal is that the above bulk result, which is equivalent to having a wormhole geometry, is reproduced by averaging over the physically distinguished free ultrafilters. For each ultrafilter, one can also define the perturbation theory in 
$1/N$.

\subsection{Ultrafilters}

As is foreshadowed above the aim is to discuss the appropriate notion of limit for the oscillating observables one is interested in. This notion in particular has to be uniform for a given boundary theory whichever oscillating observable we want to consider. In other words it can not be just arbitrarily choosing the limit point of a convergent subsequence\footnote{ More generally, a limit point of a convergent subnet.}. Thus one needs a notion of `big' sets so that even if some sequence $a_{N}$, which is for us the expectation value of some bounded operator that receives contribution from states with $N$ dependent energy and couplings, does not converge when $N \in \mathbb{N}$ (set of natural numbers), it converges in some yet-to-be-defined `big' set in a consistent way. This notion is provided by what are called free (non-principal) ultrafilters. A more detailed discussion of topics raised here can be found in Appendix \ref{ultra}.

Here we introduce ultrafilters\footnote{Notions of filters and ultrafilters were introduced by Tarski and later by Cartan as a way to generalize limits. After the discovery of Łoś's theorem and ultraproducts though, they have become quite essential in model theory and other branches of mathematics.}. The first point would be such `big' set $A \subset \mathbb{N}$ should not be an empty set. In addition, if we call $A$ a `big' set and there is some other set $B$ where $A \subset B$, then $B$ should also be a `big' set. Finally we need such a set to generalize the notion of a neighborhood too, therefore the set of `big' sets should be closed under intersection. That is we consider $p$, a set of subsets of $\mathbb{N}$ where $\emptyset \notin p$; if $A, C \in p$ and $A \subset B \subset \mathbb{N}$ then $B \in p$ and $A \cap C \in p$. Such a $p$ is called a \emph{filter}. However, we add one more important property which is to require that $p$ is the maximal or the largest of such sets. Then $p$ is said to be an \emph{ultrafilter}. 

The set of neighborhoods around some point on the real line for instance is an example of a filter. An empty set is not a neighborhood and an intersection of neighborhoods around a given point is also a neighborhood around the point, in addition a set strictly containing some neighborhood around a point is also a neighborhood\footnote{Here we can take a neighborhood to be not just an open set around a point but a set containing an open set around a point.}. Another well known example is cofinite filter (Fr\'echet filter), which is the set of subsets, of $\mathbb{N}$ for example, whose compliment is finite. One can easily check that this set satisfies all the axioms of a filter.   

However the above two examples are not ultrafilters. This follows from the fact that imposing the condition that a filter is maximal necessarily implies that it should contain any subset of $\mathbb{N}$ or its compliment. 
In other words, if $p$ is an ultrafilter then either $A \subset \mathbb{N}$ or $\mathbb{N}\backslash A$ is in $p$ \footnote{To show this we first consider a filter $p$ on $\mathbb{N}$ such that for any subset $A \subset \mathbb{N}$, either $A\in p$ or $\mathbb{N}\backslash A \in p$. Then $p$ is an ultrafilter. The reason is the following. Let's take another filter $q$ on $\mathbb{N}$. If $p \subseteq q$ and $p \neq q$ then there is some $B \subset \mathbb{N}$ which is an element of $q\backslash p$. But then because of the defining property of $p$, $\mathbb{N}\backslash B \in p$. 
Since $p$ is a subset of $q$, we have $\mathbb{N}\backslash B,\;B \in q$. However, $(\mathbb{N}\backslash B) \cap B = \emptyset$, which is a contradiction and $q$ fails to be a filter.  

One can similarly show that if $p$ is an ultrafilter, then for any $A \subset \mathbb{N}$, either $A\in p$ or $\mathbb{N}\backslash A \in p$. To show this assume for some $\emptyset \neq A \subset \mathbb{N}$, both $A$ and $\mathbb{N}\backslash A$ are not in $ p$. Now consider another set $r$ which we define by adding $A$ and any $B \subset \mathbb{N}$ that includes $A$ to $p$. In addition, we also add to $p$ the set

$$\{X\subset \mathbb{N}|\; A \cap C \subseteq X, \text{ for any } C\in p\}$$

The goal would be to show that $r$ is in fact a filter and since $p$ is a subset of $r$, $p$ is not a maximal filter and so not an ultrafilter. By definition $p$ is a filter, and for any set we added to $p$, we have also added above any set that includes it. In addition one can also see that for any two subsets of $\mathbb{N}$, their intersection is also included (This point is easy to check and left for the reader). Finally, $r$ should not include $\emptyset$ to be considered a filter. Note that all the subsets that we have added to $p$ in $r$ are either super sets of $A$ or include $A \cap C$. Thus it is enough to show that $A \cap C\neq \emptyset$ for any $C\in p$. But it is easy to see that one can not have $A \cap C= \emptyset$, since this would imply that $C \subset \mathbb{N}\backslash A$. But since $p$ is a filter any subset of $\mathbb{N}$ that includes $C\in p$ is also in $p$, which contradicts the assumption that $\mathbb{N}\backslash A \notin p$. Thus $A \cap C\neq \emptyset$. Therefore we see that $r$ is a filter and $p$ is its subset, thus $p$ is not maximal and can not be an ultrfilter.}. We can now easily see why the cofinite filter is not an ultrafilter. There are subsets of $\mathbb{N}$, for example the set of even numbers, which are not in the filter and neither are their complements. One can similarly argue for the filter of neighborhoods of a point on $\mathbb{R}$.   

In summary an ultrafilter is a filter, that satisfies the above three axioms, that contains either $A$ or $\mathbb{N}\backslash A$, for any $A\subset \mathbb{N}$. A simple example of an ultrafilter is the set of all subsets of $\mathbb{N}$ that contain $x \in \mathbb{N}$. If $x\in \mathbb{N}$
$$e(x) = \{A\subset\mathbb{N}|\; x\in A\}$$
is an ultrafilter which is generated by $x$.

$e(x)$ can be seen to satisfy all the axioms of a filter and any subset of $\mathbb{N}$ either contains $x$ or not. Therefore, for any $A\subset \mathbb{N}$, either $A$ or $\mathbb{N}\backslash A$ is in $e(x)$. One can see that the intersection of all the elements of $e(x)$ is just $\{x\}$. Such an ultrafilter is called a principal ultrafilter and for each element of $\mathbb{N}$, there is a corresponding principal ultrafilter. We generically denote a principal ultrafilter as $e$ if we do not have to specify which element it is associated with. 

One can see that $e(x)$ is an ultrafilter that contains a finite subset of $\mathbb{N}$, in particular it contains $\{x\}$. Thus one finds that principal ultrafilters fall short for our purposes, which was to define a notion of `big' sets. It is in fact easy to show that an ultrafilter which contains any finite subset of $\mathbb{N}$ is a principal ultrafilter. 

If for instance an ultrafilter, $p$, contains a given finite subset of $\mathbb{N}$, $\{x_{1},x_{2}, \cdots, x_{k}\}$, then it should not contain the set $\mathbb{N}\backslash \{x_{i}\}$ for at least one $i$, where $i$ takes values from $1,\cdots,k$. Otherwise, one can have 
$$\bigcap_{i=1, \cdots,k}\mathbb{N}\backslash \{x_{i}\} = \mathbb{N}\backslash \{x_{1},x_{2}, \cdots, x_{k}\}$$
as an element of $p$. This is contradictory since we assumed that $\{x_{1},x_{2}, \cdots, x_{k}\} \in p$ and its intersection with the above set is $\emptyset$.

Therefore we have, if $p$ contains $\{x_{1},x_{2}, \cdots, x_{k}\}$ then it should not contain $\mathbb{N}\backslash \{x_{i}\}$ for some $i$. But since $p$ is an ultrafilter, it should then contain $\{x_{i}\}$. On the other hand, since the intersection of any member of $p$ with $\{x_{i}\}$ can not be $\emptyset$, all members of $p$ contain $x_{i}$, in other words $p=e(x_{i})$. 

Any ultrafilter which is not a principal ultrafilter, is called non principal or free ultrafilter. Equivalently, a filter that only contains infinite subsets of $\mathbb{N}$, and contains either $A$ or $\mathbb{N}\backslash A$ for any infinite set $A\subseteq \mathbb{N}$, is called a free ultrafilter. For instance a free ultrafilter contains either the set of even numbers or the set of odd numbers. 

The construction of free ultrafilters is more or less straight forward. The first point is that any free ultrafilter does not contain finite subsets, thus all cofinite subsets of $\mathbb{N}$ are elements. Then one chooses to include in the ultrafilter between the set of even and odd numbers. If one chooses to include even number for example, any set that contains the set of even numbers will also be added. Then one chooses between the sets $0,1$ or $2$ mod $3$; and add all supersets of the choice. One continues making choices like this and finally constructs a free ultrafilter\footnote{One may notice that this constructions assumes the axiom of choice, which may bother some mathematicians, but we will not be concerned with it here. The existence of free ultrafilters does not depend on the axiom of choice, only their explicit construction does.}.  

\subsection{Ultralimits}

As we mentioned before the goal is to generalize the notion of limit for chaotic expectation value or partition function, which we denote by $\braket{O}_{N}$, that does not converge as we take the large $N$ limit. The way we do this is to rather define the limit as the sequence gets close to some particular values not as the index runs overs the natural numbers but over `big' subsets of the natural numbers. The `big' sets are provided by free ultrafilters as discussed above. Free ultrafilters (or ultrafilters generally) are essential particularly because for any bounded sequence, such a limit is unique for a given ultrafilter. Thus the limit is better defined for a given free ultrafilter, not a particular infinite subset of the natural numbers. 

To be a bit more precise, we define the limit along a particular ultrafilter $p$ as,
\begin{equation}\label{ultralim}
    \lim_{N\xrightarrow[p]{}\infty} \braket{O}_{N} = \braket{O}_{p}
\end{equation}
if and only if for any neighborhood $\mathcal{V}$ of $\braket{O}_{p}$, we have $\{N \in \mathbb{N}|\; \braket{O}_{N} \in \mathcal{V}\} \in p$. The neighborhood $\mathcal{V}$ is in the appropriate topological space for the expectation values or partition function, which is usually either the complex plane or the real line. 

This limit is unique. One can show this by first assuming the above limit converges to two distinct values $\braket{O}_{p}$ and $\braket{O}_{p}^{'}$. Then for any neighborhood $\mathcal{V}$ of $\braket{O}_{p}$ and $\mathcal{V}^{'}$ of $\braket{O}_{p}^{'}$), both sets $$\{N \in \mathbb{N}|\; \braket{O}_{N} \in \mathcal{V}\}\; \text{ and }\; \{N^{'} \in \mathbb{N}|\; \braket{O}_{N^{'}} \in \mathcal{V}^{'}\}$$ are in $p$. 

However since $\braket{O}_{p}$ and $\braket{O}_{p}^{'}$ are distinct in the complex plane( or the real line) one can choose  $\mathcal{V}$ and  $\mathcal{V}^{'}$ to be disjoint open sets\footnote{The actual condition that allows us to do this is that the space the sequence takes values in has to be Hausdorff.}. But since $p$ is a filter, the above sets has to have a non empty intersection. That is, there is at least one $N_{0}\in \mathbb{N}$ in the intersection of the above two sets and $\braket{O}_{N_{0}}$ belongs to both $ \mathcal{V}$ and $ \mathcal{V}^{'}$, which is a contradiction. Therefore $\braket{O}_{p}$ and $\braket{O}_{p}^{'}$ can not be distinct and the limit is unique.

Until now we have not restricted the ultrafilters to be free. Indeed if one takes the limit along a principal ultrafilter $e(N_{0})$, one always has 
\begin{equation}
    \lim_{N\xrightarrow[e(N_{0})]{}\infty} \braket{O}_{N} = \braket{O}_{N_{0}}.
\end{equation}
Heuristically this is clear, since $N_{0}$ is always present for any member of $e(N_{0})$, however small one makes the neighborhood $\mathcal{V}$, $\braket{O}_{N_{0}}$ is always its element.

Following this, for our purposes, we can identify $e(N_{0})$ with $N_{0}$ itself and the free ultrafilters as all the points at infinity\footnote{This is also customary practice in the mathematical literature.}, and we rewrite ultralimit along $p$, \eqref{ultralim}, as
\begin{equation}\label{limbeta}
   \lim_{N\xrightarrow[]{}p} \braket{O}_{N} =\braket{O}_{p}.
\end{equation}
with $e(N_{0})\simeq N_{0}$.

The above notation will be clear if we look at the space of ultrafilters on $\mathbb{N}$,
\begin{equation}
    \beta\mathbb{N} = \{p\;|\; p \text{ is an ultrafilter on }\mathbb{N}\}
\end{equation}
where again the principal ultrafilters are identified with the natural numbers they are associated with, the free ultrafilters  correspond to the extreme large $N$ points. Then \eqref{limbeta} can be understood as a limit in the space $\beta\mathbb{N}$ (more on the properties of $\beta\mathbb{N}$ can be found in the appendix \ref{ultra}). 

The space $\beta\mathbb{N}$ is extremely huge with a cardinality much bigger than the continuum (as what one would suspect following the construction we provided for free ultrafilters). However, assuming for example the expectation values take values in $\mathbb{R}$, for a bounded sequence of expectation values we have, $\braket{O}_{N} \in [-M,M]$ for some $M \in \mathbb{R}$. Therefore we have $\braket{O}_{p} \in [-M,M]$ and most $p$'s are redundant. Thus in the computation involving a particular observable, only the physically distinguishable free ultrafilters can participate. 

The principal ultrafilters will just return $\braket{O}_{N_{0}}$ if $N_{0}$ is the natural number that generate the principal ultrafilter. For any possible operator in the CFT of which one might want to compute the limit, one will always get the observable at $N=N_{0}$. In addition, if we take $\beta\mathbb{N}$ to label a space of boundary theories, where for instance the principal ultrafilter $e(N_{0})$ corresponds to the CFT where the parameter of the boundary field theory $N$ is finite and equal to $N_{0}$ then, we can take the free ultrafilters to correspond to all the possible $N=\infty$ boundary gauge theories. To put it in other words, the space $\beta\mathbb{N}$ is best understood as the maximal compactification of the discrete space $\mathbb{N}$ (see appendix \ref{ultra}). For each natural number, we have for example the corresponding $\mathcal{N}=4 $ super Yang Mills theory in four dimensions with the rank of the gauge group given by that natural number. We look at this space as the space of CFT's label by the rank of the gauge group. The maximal compactification of this space is the space of boundary theories labeled by $\beta\mathbb{N}$ where the free ultrafilters indicate all the possible boundary theories at the strict large $N$ limit.  \emph{We claim that these different $N=\infty$ theories (at least some sectors of these theories), CFT$_{p}$, are the ones captured by wormhole contributions in the bulk}\footnote{Here we have used to term CFT$_{p}$ slightly liberally to call these theories at the strict large $N$ limit. The only reason at the moment we call them CFT$_{p}$ is to show that they arise from the ultralimit of a CFT.}.  

\subsection{CFT$_{p}$, the boundary theories labeled by $p$}

In the standard AdS/CFT duality the large $N$ limit of the boundary, which is expected to be dual to the semiclassical bulk theory, is defined by considering single trace operators in the boundary theory (say $\mathcal{N}=4$ super Yang Mills in four dimensions to be concert). With the appropriate normalization of the fields, single trace operators like Tr$F^{2}$ do not have explicit factors of $N$ and are good bases for the operators in the large $N$ limit.

The standard large $N$ limit is then described by considering bounded functions of these single trace operators, we call $O$, with a well defined expectation values in the limit,

\begin{equation}\label{normlim}
    \lim_{N\rightarrow\infty} \braket{O}_{N} = \braket{O},
\end{equation}

where the $\braket{}_{N}$ in the vacuum state of the super Yang Mills theory when the rank of the gauge group is $N-1$. These set of operators are generalized free fields, form a type III$_{1}$ von Neumann algebra and are exactly the planar operators that we discussed. As we said, there is a huge amount of evidence these are dual to the usual semiclassical bulk fields on the anti de Sitter spacetime background.  

Consider however a bounded function of the single trace operators, we call $O_{N}$, whose vacuum expectation value is oscillating in $N$. At the moment we are not taking it to be a heavy operator with $O(N)$ or $O(N^{2})$ energy, but a bounded operator. These kinds of $O(1)$ oscillatory operators are highly non generic as far as we know \footnote{An example is the exact $\mathbb{S}^{3}$ canonical partition function in ABJM theory \cite{Drukker:2011zy}. However the periodicity in $N$ is essentially inherited for the discreteness of the Chern-Simons level, as one transforms from the grand potential to $Z_{k}(N)$, $k$ being the level.}. If we act on the vacuum with this operator and consider the expectation values of $O$, the standard limit will in general not exist since $O_{N}$ is oscillatory.

\begin{equation}
    \lim_{N\rightarrow\infty}  \braket{O}_{N,\text{mod}} = \lim_{N\rightarrow\infty} \braket{O_{N}^{\dagger}\;O\;O_{N}}_{N} = \text{indefinite}
\end{equation}

where the $\braket{}_{N,\text{mod}}$ in the vacuum state acted upon by $O_{N}$. However the ultralimit along some free ultrafilter, $p$, is well defined.

\begin{equation}\label{ultralimit}
    \lim_{N\rightarrow p}  \braket{O}_{N,\text{mod}} = \lim_{N\rightarrow p} \braket{O_{N}^{\dagger}\;O\;O_{N}}_{N} = \braket{O}_{p,\text{mod}}
\end{equation}

In addition, the expectation value of $O_{N}$ computes the overlap of the vacuum and the new state we have constructed. As we will discuss in detail in the following sections, the generalized free fields mentioned earlier still correspond to semiclassical bulk fields in this case too. However, one has to take the `ensemble' average of $\braket{O}_{p,\text{mod}}$ over the physically distinguished ultrafilters to recover the bulk dual. The resulting bulk dual naturally involves baby universe and spacetime wormholes. 

Coming back to the single $p$ case, the equation \eqref{ultralimit} can be understood as enlarging the theory to include operators like $O_{N}$ into the strict large $N$ limit, which apriori only included generalized free fields. However, the generalized free fields acting on the vacuum is not the full story of the standard large $N$ AdS/CFT. In particular there are states that are semiclassically different from the vacuum, i.e, distinct bulk backgrounds involving black holes, stars, galaxies and so on that are still asymptotically AdS (tensored with some more compact manifolds). Such states are usually created by the action of heavy operators, say $O(N^{2})$, on the vacuum state. Some examples of such semiclassical states can be found for example in \cite{Bahiru:2023zlc}\footnote{The authors of the paper used such states to define gauge invariant operators in the CFT that create localized perturbation in the bulk. The states are relevant because they break most of, if not all, symmetries of the theory.}.

Again here, the action of the generalized free fields introduced above on such states correspond to bulk fluctuations on the corresponding semiclassical backgrounds, say a black hole or a star. The expectation values of the generalized free fields on these backgrounds are constructed so that they have a well defined large $N$ limit. That is we preform the necessary normalization to the fields $O$ so that their expectation values in the heavy state, we call $\ket{\psi}$ is well defined. That is we define the normalized version of $O$, $\tilde{O}$ so that 
\begin{equation}
    \lim_{N\rightarrow\infty} \bra{\psi_{_{N}}}\tilde{O}\ket{\psi_{N}} = \braket{\tilde{O}}
\end{equation}
we can take for instance $\tilde{O}= O - \bra{\psi_{N}}O\ket{\psi_{N}}$ for the case of a black hole. Now $\tilde{O}$'s correspond to bulk fluctuations. We can thus say that in addition of the generalized free fields, $O$, the standard large $N$ AdS/CFT also involves appropriately normalized versions of heavy operators like $\mathcal{O}^{\dagger}O\mathcal{O}$, where $\mathcal{O}$ is the operator used to create the heavy state from the vacuum. 

Now consider acting on the vacuum with some heavy oscillating/chaotic operator, $\mathcal{O}_{N}$, that changes the energy of the state by an $O(N^{2})$ energy, like what we did above. Then the standard large $N$ of generalized free fields is not well defined in this modified state but the ultralimit will always be well defined.  

\begin{equation}
     \lim_{N\rightarrow\infty} \bra{\psi_{_{N,\text{mod}}}}\tilde{O}\ket{\psi_{N,\text{mod}}}  = \lim_{N\rightarrow\infty} \langle\mathcal{O}_{N}^{\dagger}\;\tilde{O}\;\mathcal{O}_{N}\rangle \; \text{ indefinite}.
\end{equation}
but,
\begin{equation}
     \lim_{N\rightarrow p} \bra{\psi_{_{N,\text{mod}}}}\tilde{O}\ket{\psi_{N,\text{mod}}}  = \lim_{N\rightarrow p} \langle\mathcal{O}_{N}^{\dagger}\;\tilde{O}\;\mathcal{O}_{N}\rangle = \braket{\tilde{O}}_{p,\text{mod}}.
\end{equation}

Therefore the CFT$_{p}$ theory now includes oscillating/chaotic heavy operators like $\mathcal{O}_{N}^{\dagger}\;\tilde{O}\;\mathcal{O}_{N}$ since they have a well defined expectation values. Roughly speaking the perturbative and `background' states in CFT$_{p}$ are created by the action of oscillating/chaotic operators like $O_{N}$ or $\mathcal{O}_{N}$ on the vacuum, or on another heavy operators like $\mathcal{O}$ that are acting on the vacuum. The normalized generalized free field operators correspond to bulk quantum fields on those backgrounds. A more precise version of this description of CFT$_{p}$ would be to just consider a specific sector (with a specific background, say the vacuum) of the theory and perform the GNS construction from the set of expectation values of the generalized free fields in this limit. The set of expectation values defines a state on the $*$ algebra of the generalized free fields. One can thus define the Hilbert space of the sector then the von Neumann algebra of operators that act on the Hilbert space. The ultraliimit of expectation values the GFF's depend in general on the choice of free ultrafilter $p$ in the presence of oscillating operators, thus the theory at the strict large $N$ limit, conveniently named CFT$_{p}$, depend on $p$. The proposal is that when the gravitational path integral computes an expectation value it compute the average of the expectation values of the operator in these CFT$_{p}$ with the appropriate weight. In the following sections we see this in example. Wormholes and baby universes are natural in this proposal, though they do not arise always.

To summarize, schematically the standard large $N$ limit of the boundary CFT is defined by the expectation values of GFF's and GFF's in the background of heavy operators

\begin{equation}
    CFT_{\infty} \equiv \{\{O\}, \{\mathcal{O}^{\dagger}\tilde{O}\mathcal{O}\}\}
\end{equation}

while CFT$_{p}$ is defined by the ultralimit along free ultrafilter $p$ of the following set of operators

\begin{equation}
    CFT_{p} \equiv \{\{O\}, \{O_{N}\}, \{\mathcal{O}^{\dagger}\tilde{O}\mathcal{O}\}, \{\mathcal{O}^{\dagger}_{N}\;\tilde{O}\;\mathcal{O}_{N}\}\}
\end{equation}

Of course for this discussion to be consistent, ultralimits along different ultrfilters $p$'s should all be indistinguishable (particularly the free ultrafilters) for observables that do not receive contributions from wormholes. In the last section, we discussed which observables do not receive contributions from wormholes. These are expectation values of operators that survive the planar limit and not receive contributions from black hole states. These are operators with a well defined large $N$ limit, in the `traditional' sense. It is easy to see that if a sequence of expectation values has a well defined limit, then all the free ultralimits (ultralimits along a free ultrafilter) of the sequence agree with the original limit.      

This follows from the fact that any free ultrafilter contains the cofinite ultrafilter. A convergence of a sequence expectation values in the traditional sense, \eqref{normlim},
implies that for any neighborhood $\mathcal{U}$ of $\braket{O}$ there is a natural number $L$ such that 
\begin{equation}
    \{\braket{O}_{N} \in \mathcal{U}, \text{ for } N\in \mathbb{N} \text{ and }N>L\}.
\end{equation}
But this implies that for any neighborhood $\mathcal{U}$ of $\braket{O}$, the set
\begin{equation}\label{cofin}
    \{N\in \mathbb{N}| \; \braket{O}_{N} \in \mathcal{U}\}
\end{equation}
is a cofinite subset of natural numbers. However, since any free ultrafilter does not contain finite subsets of natural numbers, all free ultrafilters contain all cofinite subsets of $\mathbb{N}$, that is \eqref{cofin} is an element of any free ultrafilter. Therefore, \eqref{normlim} implies that 
\begin{equation}
    \lim_{N\rightarrow p}\braket{O}_{N} = \braket{O}
\end{equation}
for any free ultrafilter $p$. 

The simplest example one can think of to illustrate ultrafilters is the alternating sequence $a_{N}=(-1)^{N}$. This is a periodic sequence that is $1$ for even $N$ and $-1$ for odd $n$. Thus this sequence has only two `physically distinguishable' free ultrafilters $p_{e}$ and $p_{o}$. $p_{e}$ denotes any free ultrafilter that contains the set of even numbers while $p_{o}$ denotes any free ultrafilter that contains the set of odd numbers. Since the set of even and odd numbers are complements of each other in the set of natural numbers therefore no free ultrafilter can contain both. Thus one has,
\begin{equation}
    \lim_{N\rightarrow p_{e}} a_{N} = 1, \;\text{ and }\;  \lim_{N\rightarrow p_{o}} a_{N} = -1
\end{equation}

A more complicated example would be a sequence like $a_{N}=\alpha N$ (mod 1), where $\alpha$ is an irrational number. Since $\alpha$ is an irrational number, according to theorem due to Weyl the sequence $a_{N}$ is equidistributed on a circle. This means for any interval $[a,b] \subset [0,1)$ on a circle, the number of indices with $N<M$, which we call $P_{M}$, where $a_{N}\in [a,b]$ is proportional to $M$. More precisely, if $a_{N}$ is equidistributed, then
\begin{equation}
    \lim_{M\rightarrow\infty}\frac{P_{M}(a_{N}\in [a,b])}{M} = b-a.
\end{equation}

Therefore if we choose some point $x \in [0,1)$, then for any real number $\epsilon>0$ the set 
\begin{equation}
    A_{\epsilon} = \{N\in \mathbb{N}|\; |a_{N}-x|<\epsilon\}
\end{equation}
is infinite. One can consider the set of all $A_{\epsilon}$ with $\epsilon>0$ and check that all their finite intersections are also infinite. Thus one can construct free ultrafilters starting from the set 
\begin{equation}
    A = \{A_{\epsilon} |\;  \epsilon>0 \}
\end{equation}
and adding for instance the cofinite subset of $\mathbb{N}$ and any other infinite subset whose intersection with each $A_{\epsilon}$ is infinite until one reaches the maximal one which is a free ultrafilter $p_{x}$\footnote{This is in fact a theorem, which states that any subset of the power set of the natural numbers, that satisfies the finite intersection property (which means any finite intersection of sets inside the set of sets is non empty) can be extended to form an ultrafilter.}. In particular, for any $x\in [0,1)$, there is a free ultrafilter $p_{x}$ such that 
\begin{equation}
    \lim_{N\rightarrow p_{x}}a_{N} = x
\end{equation}

For the same reason of the equidistribution property of $a_{N}$ above, one can also check that there is a free ultrafilter $p_{y}$ such that,
\begin{equation}
    \lim_{N\rightarrow p_{y}} \text{cos}(N) = y,
\end{equation}
for any $y\in [-1,1]$.

A slightly easier example that stands between the alternating sequence and $a_{N}=\alpha N$ (mod 1) would be the sequence $c_{N}=\lfloor\sqrt{2}N\rfloor$ (mod $4$). This sequence takes values in $\{1,2,3\}$ and since $\sqrt{2}$ is irrational it is not a periodic sequence. However, there will be three physically distinguishable free ultrafilters $p_{k}$ where $k=1,2$ or $3$, and we will have $\lim_{N\rightarrow p_{k}}c_{N}=k$. 

\subsection{Perturbative and non-perturbative corrections}

Similar arguments as in the previous section will also show that the $O(1/N)$ corrections in the standard limit match order by order with perturbative corrections from the ultralimits.

If one has asymptotic expansion for $\braket{O}_{N}$,
\begin{equation}
    \braket{O}+\frac{1}{N}\braket{O}^{(1)} + \frac{1}{N^{2}}\braket{O}^{(2)}+ \cdots,
\end{equation}
then the coefficient of the $k^{th}$ order, i.e, the coefficient for $1/N^{k}$ is given by
\begin{equation}\label{norpert}
   \braket{O}^{(k)} = \lim_{N\rightarrow\infty}N^{k}\bigg(\braket{O}_{N}- \sum_{n=0}^{k-1}\frac{1}{N^{n}}\braket{O}^{(n)}\bigg)
\end{equation}
where $\braket{O}^{(0)}\equiv\braket{O}$ and $k>0$. 

Since the large $N$ limit of $N^{k}\bigg(\braket{O}_{N}- \sum_{n=0}^{k-1}\frac{1}{N^{n}}\braket{O}^{(n)}\bigg)$ is well defined, as we said before, the limit along any free ultrafilter matches with the standard limit.
\begin{equation}
   \braket{O}^{(k)} = \lim_{N\rightarrow p}N^{k}\bigg(\braket{O}_{N}- \sum_{n=0}^{k-1}\frac{1}{N^{n}}\braket{O}^{(n)}\bigg)
\end{equation}
for any free ultrafilter $p$. 

Therefore perturbative expansion around $N=\infty$ of operators that do not receive wormhole contribution for the CFT$_{p}$ (for $p$ a free ultrafilter) is also the same as the perturbative expansion around the standard $N=\infty$ theory which is usually a generalized free field theory. However, bounded sequence of expectation values of chaotic operators do not have large $N$ limit in the generalized free field theory but they have (sometimes distinct) large $N$ limits in the CFT$_{p}$'s and thus can have distinct perturbative expansion around the strict large $N$ expectation value. Thus we define the $1/N$ expansion around $\braket{O}_{p}$ as
\begin{equation}\label{ppertexpa}
     \braket{O}_{p}+\frac{1}{N}\braket{O}^{(1)}_{p} + \frac{1}{N^{2}}\braket{O}^{(2)}_{p}+ \cdots
\end{equation}
where 
\begin{equation}\label{pdeppert}
   \braket{O}^{(k)}_{p} = \lim_{N\rightarrow p}N^{k}\bigg(\braket{O}_{N}- \sum_{n=0}^{k-1}\frac{1}{N^{n}}\braket{O}^{(n)}_{p}\bigg)
\end{equation}
and $\braket{O}^{(0)}_{p}\equiv\braket{O}_{p}$.

We have to note that for consistency, the symbol $1/N$ is an infinitesimal number along an ultrafilter $p$. To clarify what we mean by this statement we recall that when we do perturbation theory $1/N$ is not actually a real number, rather it is a number smaller than any positive real number but greater than zero. This is a re-phrasal of the statement that the perturbative series we come across in most quantum field theories are not formal power series with finite radius of convergence but asymptotic series that are well defined when the coupling is taken to zero (we will come back to this point in the following subsection). The symbol $1/N$ is actually an element of an extension of the real numbers. To introduce this extension, we use methods similar to what one uses to complete rational numbers into real numbers. One can represent real numbers as a Cauchy sequence of rational numbers with some equivalence condition. A Cauchy sequence of rational numbers is a sequence of rational numbers $(a_{1},a_{2},\cdots)$ such that for any rational number $\epsilon>0$, there is $M$ such that for all $n,m>M$, $|a_{n}-a_{m}|<\epsilon$. For instance, $(3,\;3.1,\;3.14,\;3.141,\cdots)$ is a Cauchy sequence. On the other hand the equivalence condition is that if two sequences $(a_{n})$ and $(b_{m})$ are such that for any rational number $\epsilon>0$, there is $M$ such that for all $n>M$, $|a_{n}-b_{n}|<\epsilon$, then we say $(a_{n}) \sim (b_{m})$. Thus $\pi= [(3,\;3.1,\;3.14,\;3.141,\cdots)]$, where $[.]$ denotes an equivalence class. Constant sequences, more precisely their equivalence classes, are identified with rational numbers themselves.     

Similarly we extend the real numbers to include the couplings we use in perturbation theory by considering a sequence of real numbers with some equivalence relations. The equivalence class of constant sequences will correspond to the real numbers themselves. The equivalence condition now however will depend of a choice of a free ultrafilter $p$. In other words, $(a_{1},a_{2},\cdots)$ is in the same $p-$equivalence class as $(b_{1},b_{2},\cdots)$ if and only if they agree on one of the `big' subsets of $\mathbb{N}$ defined by $p$, that is the set of indices $\{n \in \mathbb{N}|a_{n}=b_{n}\} \in p$.

For example, the number $5$ is given by the equivalence class of the constant sequence $(5,5,\cdots)$. If the free ultrafilter $p$ includes the set of even numbers, then the sequence $(x, 5,y,5,z,5, \cdots)$ is also in the same $p-$equivalence class. In this context, the perturbative coupling $1/N$ is defined as the $p-$equivalence class of $(1,1/2,1/3,\cdots)$. Not just the equivalence condition but also ordering relation between these new numbers depend on the filter. That is $[a_{n}]>[b_{n}]$ if and only if $\{n \in \mathbb{N}|a_{n}>b_{n}\} \in p$. In particular, the infinitesimal number $1/N= [1/n]$ is greater than zero and smaller than any real number, since $1/n>0$ for any natural number $n$, and $[1/n]$ is smaller than any real number $x$ since eventually, for $n>M$ where $M$ is some big enough natural number, we have $1/n<x$. In the same sense, $1/N^{2}:= [1/n^{2}]<[1/n]$ and more generally,
\begin{equation}
    1/N^{k} := [1/n^{k}]< [1/n^{k-1}], \text{ for real number }k
\end{equation}
It is precisely this notion that we used to write \eqref{ppertexpa} and the infinitesimal number $1/N$ is defined with respect to the specific free ultrafilter used in the limit.

One on the other hand can define the inverse of the infinitesimal number $[1/n]$, as $H:=[n]$ which is an infinite number larger than any real number. We will make use of such numbers later. These numbers are usually called non-standard numbers while the constant sequence equivalence classes, which are just real numbers, are called standard numbers \cite{keisler2012elementary}\footnote{A non standard number $[a_{n}]$ that is less than some real number is called limited number. This corresponds to a bounded sequence of real numbers. For such numbers one can separate the standard part of the number from its non standard part by taking the ultralimit of the sequence. The standard part of $[a_{n}]$ is given by, $$\text{st}(a_{n})=\lim_{n\rightarrow p}a_{n}.$$ In other words the non standard number $[a_{n}]$ can be written as $[a=\text{st}(a_{n})] + \eta$ where $\eta$ is an infinitesimal number.}.   

\subsection*{Non perturbative corrections}

We won't have much to say on non perturbative correction to the $p$ dependent expansion of the expectation value of an operator \eqref{ppertexpa} until the end of this subsection, except that under some assumptions about the perturbative expansion the usual resurgence story can be applied, which we will review for completeness in the present context.

The observable $\braket{O}_{N}$ is unique, however, the perturbative expansion \eqref{ppertexpa} in general depends on the choice of free ultrafilter. In addition, we expect from general considerations that the perturbative expansion will not correspond to a single finite observable, that the expansion has asymptotic nature. Thus the perturbative expansion can not be the whole story. The first step is that one has to include non perturbative corrections to the perturbative expansion. This is true even for operators with a well defined planar limit and thus no ambiguity due to the choice of a free ultrafilter. This ambiguity of the perturbative expansion (more precisely the ambiguity in trying to reproduce $\braket{O}_{N}$ from it) is less obvious and an important aspect of resurgence theory. However, even after we apply resurgence we do not get the finite $N$ object, but only part of the information in the function, $\braket{O}_{N}$. As we have said before this is so since the ultralimit chooses a particular subsequence-like sector of the sequence. We first discuss this issue then discuss the second type of ambiguity later. 

We focus on operators with a trans-series expansion that we commonly encounter in quantum field theories and quantum gravity. We consider both operators with a well defined large $N$ limit in the standard sense and chaotic operators, along some free ultrafilter $p$. For planar limit operators, this expansion is the same for any other free ultrafilter and with the standard planar expansion, while for chaotic operators it is in general different for different free ultrafilters.  Let's consider, in addition to the perturbative expansion around the large $N$ value, including a non perturbative correction like,
\begin{equation}\label{expa}
    \braket{O}_{p}+\frac{1}{N}\braket{O}^{(1)}_{p} + \frac{1}{N^{2}}\braket{O}^{(2)}_{p}+ \cdots + \braket{O}^{exp}_{p}\;e^{-aN}+ \cdots,
\end{equation}
where the dots at the end represent corrections even smaller than $e^{-aN}:=e^{-a[n]}$. We of course expect a full perturbative expansion around the instanton saddle, $A\; e^{-aN}(\sum_{k}a_{k}/N^{k})$, which is included in the ellipsis at the end. The term $\braket{O}^{exp}_{p}$ is the product of the overall factor $A$ with the one instanton amplitude $a_{0}$.

A naive proposal to extract $\braket{O}^{exp}_{p}$ from the sequence $\braket{O}_{N}$, following \eqref{pdeppert}, would be
\begin{equation}\label{naive}
    \braket{O}^{exp}_{p} \overset{?}{=}  \lim_{N\rightarrow p}e^{aN}\bigg(\braket{O}_{N}- \sum_{n=0}^{\infty}\frac{1}{N^{n}}\braket{O}^{(n)}_{p}\bigg).
\end{equation}
Of course the issue with this proposal is that the series $\sum_{n=0}^{\infty}\frac{1}{N^{n}}\braket{O}^{(n)}_{p}$ is usually not well defined for any finite value of $N$ in quantum field theory (and quantum gravity) therefore the above limit as written does not make sense. The reason for this is that the coefficients $\braket{O}^{(n)}_{p}$'s themselves depend on $n$ strongly. In QFT in general this is because the number of Feynman diagrams grow with increasing the perturbative order. Similar arguments can be made from the bulk and boundary perspectives in our case too \cite{Hawking:1978jz,Dorigoni:2024dhy}. Thus we assume the series to be an asymptotic series in general rather than a convergent one.

One rather uses a new convergent series derived from this asymptotic series by applying a Borel transform. For instance, if the growth of the coefficients for large $n$ is given by $\braket{O}^{(n)}_{p} \sim n!a^{-n}$\footnote{This follows arguments by Dyson and Lipatov \cite{Dyson:1952tj,Lipatov:1976ny}, which we assume also apply here with at most mild modifications\cite{Drukker:2011zy,Hatsuda:2015gca,Grassi:2014vwa}. A more general condition on the perturbative expansion is that for large $n$,$$\braket{O}^{(n)}_{p} \leq \alpha\; (n!)^{m}a^{-n}, $$ for some constants $\alpha,m$ and $a$. Consequentially the Borel transform will be, $$B[O](t)= \sum_{n=0}^{\infty}\frac{t^{n}}{\Gamma(1+n/k)}\braket{O}^{(n)}_{p}.$$} for some positive real number $a$,
\begin{equation}
    B[O](t)= \sum_{n=0}^{\infty}\frac{t^{n}}{n!}\braket{O}^{(n)}_{p}
\end{equation}
is the Borel transform and is a well defined series within its radius of convergence\footnote{A more precise statement of its property is that $B[O](t)$ is a simple resurgent function, which means that it is an analytic function that is \emph{endlessly countable}. That is, roughly, it has only isolated singularities along a path that starts from the origin and goes to infinity, in any direction in the complex $t$ plane, so that they can be avoided by deforming the same path. In addition, these singularities are \emph{simple}, i.e, in a small enough neighborhood of each singularity one has $$B[O](t)= \frac{\alpha}{2\pi i (t-a)}+\frac{1}{2\pi i} B^{1}[O](t-a)\text{ log}(t-a) + \text{non singular function in }(t-a). $$ The function $ B^{1}[O](t-a)$ is an analytic function close to $t-a=0$ and can be computed by $B[O](t)- \mathcal{M}B[O](t)$, where $\mathcal{M}$ is the monodromy operator, going around a point by a $2\pi$ angle in the complex plane.}.

The procedure to recover $\braket{O}^{exp}_{p}$ (or any additional perturbative correction on top of this saddle) that is suggested by resurgence theory \cite{Dorigoni:2014hea,Aniceto:2018bis}, even though it has similar idea, is much more subtle than \eqref{naive}. One uses the above convergent series to define what is called the Borel resummation of the asymptotic series $\sum_{n=0}^{\infty}\frac{1}{N^{n}}\braket{O}^{(n)}_{p}$ as,
\begin{equation}
    \mathcal{O}_{p}(N)= \int_{0}^{\infty}dt\;e^{-t}B[O](t/N) =N \int_{0}^{\infty}dt\;e^{-Nt}B[O](t),
\end{equation}
where we renamed the integration parameter in the second equality. Then one aims reproduces how much $\mathcal{O}_{p}(N)$ fails to reproduce $\braket{O}_{N}$, if it were the case that $\braket{O}_{N}$ has a trans-series expansion like \eqref{expa}. Again, since $\braket{O}_{N}$ is unique, its difference with $\mathcal{O}_{p}(N)$ will depend on $p$. 

It is easy to see that the failure of $\mathcal{O}(N)$ is at a non perturbative order. One can Taylor expand $\mathcal{O}(N)$ around $1/N=0$, and since $B[O](t/N)$ inside the integral has no singularity in this limit, one finds that it has the same perturbative expansion as $\braket{O}_{N}$. This failure is also directly related to the singularities one encounter when doing the integral over $t$, which for the Borel transform above, is encountered at $t=a$. Absent these singularities (and some addition growth conditions), the Borel resummation exactly recovers $\braket{O}_{N}$ and can be considered as its unique analytic continuation\footnote{A theorem by the Watson-Nevalinna-Sokal implies that if $\braket{O}_{N}$ is analytic in $|\text{arg }(1/N)|< \pi/2 + \epsilon$ and that $$|\braket{O}_{N}-\sum_{n}^{k-1}\braket{O}^{(n)}/N^{n}|\leq c^{k}k!N^{-k},$$ then $\braket{O}_{N}$ is the unique function with that asymptotic series and $\mathcal{O}(N) = \braket{O}_{N}$ on their common domain of definition.}. Indeed applying inverse Laplace transform to the expansion \eqref{expa} reproduces a singularity at $t=a$ with coefficient $\braket{O}^{exp}$. 

Therefore as one does the integral over $t$ for $\mathcal{O}_{p}(N)$, one has to avoid this branch cut singularity at $t=a$ by deforming the contour away from the real line. For instance since $B[O](t)$ is analytic in the complex plane with positive real axis, one can go either above or below the branch cut,
\begin{equation}
     \mathcal{O}_{p}(N)_{\pm}= N \int_{0}^{\infty\pm i\epsilon}dt\;e^{-Nt}B[O](t)
\end{equation}
Therefore, because of the singularities of the Borel transform, this analytic continuation is not going to be unique. A unique continuation is only possible for a trans-series that includes non perturbative corrections, as the ambiguity between the different continuations is non perturbative. For operators with a  well defined large $N$ limit in the standard sense, all perturbative expansions associated with different free ultrafilters agree. Therefore this is the only source of ambiguity for the resummation. This ambiguity can be resolved however if one considers a trans-series rather than just a perturbative asymptotic series. We follow the same steps for chaotic operators, however now the trans-series itself will be $p$ dependent while the ambiguity due to the singularity in the Borel plane ($t$ plane) is resolved. 

The difference between the two resummations, called lateral Borel resummations, is given by 
\begin{equation}
     \mathcal{O}_{p}(N)_{+}-\mathcal{O}_{p}(N)_{-}= N \int_{a}^{\infty}dt\;e^{-Nt}\text{ disc}_{a} B[O](t).
\end{equation}
Expanding around $a$ as $t=a + s^{'}$ and absorbing $N$ into the integration parameter, one has
\begin{equation}\label{nonperp}
     \mathcal{O}_{p}(N)_{+}-\mathcal{O}_{p}(N)_{-}= e^{-aN}\int_{0}^{\infty}ds\;e^{-s}\text{ disc}_{a} B[O](a+s/N) = e^{-aN}\int_{0}^{\infty}ds\;e^{-s} B^{1}[O](a+s/N)
\end{equation}
where disc$_{a} B[O](a+s/N) = B[O](a+s/N + i\epsilon) - B[O](a+s/N - i\epsilon)$ and $s=N\;s^{'}$. 

The fact that $B[O](t)$ is a simple resurgent function, implies that disc$_{a} B[O](s)$ is another Borel transform and \eqref{nonperp} is just a new Borel resummation with the prefactor $e^{-aN}$. In fact the consistency condition on the trans-series, in particular the Bridge equation, implies that the new Borel resummation has a perturbative expansion that is the same as the perturbative expansion on top of the one instanton saddle we mentioned previously. As one may imagine, the new Borel resummation will also have singularities and similar discontinuities which reproduce perturbation theory on top of other instanton backgrounds. In any case, for our proposes the above discontinuity reproduces $\braket{O}^{exp}_{p}$\footnote{Practically an important bottle neck in this procedure is that one doesn't usually have $B[O](t)$ in a closed form since one has access to only few orders in the vacuum perturbation theory. A good discussion of the issues involved and several approximation methods can be found for example in \cite{caliceti2007useful}.}.  

Imagine for a moment that one has recovered the full transeries $\mathcal{O}(N)_{p}$. This will not match for positive integer $N$ with the chaotic function $\braket{O}_{N}$, since the choice of the free ultrafilter $p$ removes some of the information in $\braket{O}_{N}$. However, one might then hope to get the full $\braket{O}_{N}$ from all transseries $\mathcal{O}(N)_{p}$, $p\in\beta\mathbb{N}/\mathbb{N}$. Given this, one can recover for example all the possible values $\braket{O}_{N}$ takes for positive integer $N$. However, there will still be some more information needed to reproduce the exact sequence, for instance the relationship between the different physically distinguishable free ultrfilters one has. 

A simple example would be the alternating sequence $a_{n}= (-1)^{n}$. There are no perturbative or non perturbative corrections. One has $\lim_{n\rightarrow p}a_{n}=1$ or $-1$, depending on $p$ includes the set of even numbers or not. However knowing only $a_{p1}=1$ and $a_{p2}=-1$ will not enable us to recover the exact sequence but if we also know that $p1$ is a free ultrafilter containing the set of even numbers, while $p2$ \emph{does not} contain the set of even number, it is possible to reconstruct the sequence\footnote{We suspect that this is an interesting mathematical problem that needs further exploring.}. Therefore we see that one has to go beyond just the transeries expansion to recover the finite $N$ object \cite{Liu:2026fnd}. 

In summary we get a trans-series expansion for each free ultrafilter and these expansions for chaotic observables distinguish the different CFT$_{p}$ theories. The wormhole contribution for chaotic observables is reproduced by a sort of `averaging' over the physically distinguishable free ultrafilters at each order in the trans-series. This `averaging' is expected to give a trans-series expansion one would get, in the bulk, on the background of a wormhole connecting the two boundary theories. In the following section, we see how this happens in several examples.

\section{Spin chain toy models}\label{sec4}

To simulate the kind of behavior that we encounter in the case of the large $N$ limit of AdS/CFT, we consider a simple example of a chain of $N$ qubits all in the spin-Z up state, $\xi_{0}=\bigl( \begin{smallmatrix} 1 \\ 0 \end{smallmatrix} \bigr)$, and an oscillating operator acting on them. We denote the Pauli operators $\vec{\sigma}_{n}= (\sigma_{x},\sigma_{y},\sigma_{z})_{n}$. The operator we consider is 
\begin{equation}
    O_{N}=\prod_{n=1}^{N}\vec{\sigma}_{n}.\vec{a}_{N}, \text{ where } a_{N} = \frac{1}{2}\biggl( \begin{smallmatrix} 1 + (-1)^{N}\\0 \\ 1-(-1)^{N} \end{smallmatrix} \biggr)
\end{equation}
that is $a_{N}= \biggl( \begin{smallmatrix} 1\\0 \\ 0 \end{smallmatrix} \biggr)$ if $N$ is even and $a_{N}= \biggl( \begin{smallmatrix} 0\\0 \\ 1 \end{smallmatrix} \biggr)$ if $N$ is odd. Then we consider the state of the $N$ qubits,
\begin{equation}
   \xi_{N} = O_{N}\otimes_{n=1}^{N}\xi_{0,n} = \otimes_{n=1}^{N} \vec{\sigma}_{n}.\vec{a}\; \xi_{0,n}.
\end{equation}
As a result of the oscillating operator $O_{N}$, the state $\xi_{N}$ does not asymptote to a particular state as $N$ goes to infinity. This state can be thought of as the ground state of an oscillating Hamiltonian $H_{N}= (-1)^{N}\sum_{n=1}^{N}\sigma_{z}^{(n)}$, all spin Z up when $N$ is even and all spin Z down when $N$ is odd. 

If one is interested in the von Neumann algebra of operators acting on an infinite spin chain one proceeds by first constructing an algebra of sums of finite products of the Pauli operators $\vec{b}_{n}.\vec{\sigma}_{n}$ acting on finite number of sites. That is,
\begin{equation}
    \mathcal{A}_{0}= \{\mathbb{I},\vec{b}_{1}.\vec{\sigma}_{n_{1}}\vec{b}_{2}.\vec{\sigma}_{n_{2}}\cdots \vec{b}_{r}.\vec{\sigma}_{n_{r}}\}
\end{equation}
for any arbitrary but finite $r$. Then one considers the expectation values of these operators on some chosen product state of the chain of qubits as defining a state on the algebra\footnote{Strictly speaking, one first defines the norm closure of $\mathcal{A}_{0}$ to form a C$^{*}$ algebra. The state is then defined as a map from the C$^{*}$ algebra to complex numbers, which can be chosen to be an expectation value of the operators in some product state. One then gets a von Neumann algerba as the GNS representation the C$^{*}$ algebra with respect to the chosen state.}. However, for the above Hamiltonian, since there is no well defined standard limit for the state, the expectation values of these operators is not well defined. Another way to see this would be since the above operators receive contributions from an oscillating operator like $O_{N}$, when seen as operators acting on the spin-Z up ground state $\xi_{0}=\otimes_{n}\xi_{0,n}$, the expectation values of the operators do not have a well defined standard limit. 

Thus if $O\in \mathcal{A}_{0}$ then the expectation value,
\begin{equation}
    \lim_{N\rightarrow\infty}\bra{\xi_{N}}O\ket{\xi_{N}} \text{ does not exist} 
\end{equation}
But as we have been discussing, the limit along an ultrafilter always exists, in particular if $p$ is a free ultrafilter then it either includes the set of even numbers or the set of odd numbers and if we let $\mathcal{O}=O_{N}^{\dagger}O\;O_{N}$ then we have,
\begin{equation}\label{oscilim}
    \lim_{N\rightarrow p}\bra{\xi_{N}}O\ket{\xi_{N}} = \lim_{N\rightarrow p}\bra{\xi_{0}}\mathcal{O}\ket{\xi_{0}} =\omega_{p} (O)
\end{equation}
Thus it is clear that eventhough there are $\beta\mathbb{N}\backslash\mathbb{N}$ worth of free ultrafilters parameterizing the large $N$ theories, there are only two physically distinguishable free ultrafilters, namely a free ultrafilter containing the set of even numbers $p_{e}$ and one containing the set of odd numbers $p_{o}$, and the thermodynamic limit of the above ground state is actually described by theories labeled by these ultrafilters. Thus for each $p$ there is a state $\omega_{p}$ and we can apply the GNS construction to get a von Neumann algebra. That is the state vector $\ket\Omega_{p}$ is such that
\begin{equation}
    \bra{\Omega_{p}}O\ket{\Omega_{p}} = \omega_{p}(O)
\end{equation}
and any state is defined by acting with operators in $\mathcal{A}_{0}$ on $\ket{\Omega_{p}}$, $\ket{\Psi_{p}}= O_{1}\ket{\Omega_{p}}$ and $\ket{\Phi_{p}}= O_{2}\ket{\Omega_{p}}$ for $O_{1,2}\in \mathcal{A}_{0}$ and 
\begin{equation}
    \langle\Psi_{p}|\Phi_{p}\rangle = \omega_{p}(O_{1}^{\dagger}O_{2}) = \lim_{N\rightarrow p}\bra{\xi_{N}}O^{\dagger}_{1}O_{2}\ket{\xi_{N}}
\end{equation}
\footnote{Here we have not been too careful with our notation. More precisely, any operator $O$ in C$^{*}$ algebra constructed from $\mathcal{A}_{0}$ has a representation $\pi_{p}(O)$ such that $$\bra{\Omega_{p}}\pi_{p}(O)\ket{\Omega_{p}} = \omega_{p}(O).$$ The Hilbert space is then constructed by the completion of $\{\pi_{p}(O)\ket{\Omega_{p}}\}$. The GNS representation is the triplet $(\mathcal{H}_{p},\pi_{p},\ket{\Omega_{p}})$.  }Completing the dense set of states defined above with respect to this inner product, will give us a Hilbert space $\mathcal{H}_{p}$ and completing the algebra $\mathcal{A}_{0}$ with respect to the Hilbert space $\mathcal{H}_{p}$ will give a von Neumann algebra, $\mathcal{A}_{p}$, associated with each physically distinguishable free ultrafilter. The states $\ket{\Omega_{p_{e}}}$ and $\ket{\Omega_{p_{e}}}$ are large $N$ limits of all spin up and all spin down states respectively and are orthonormal to each other. Even more, they correspond to two different superselection sectors since no local operator (in other words operators that do not depend on $N$ as are operator in $\mathcal{A}_{0}$) can change one to the other. Therefore as alluded to before, these theories analogous to the $\alpha$ sectors of Marolf and Maxfield \cite{Marolf:2020xie}, which are the individual members of the ensemble. 

\subsection{Baby universes}

The averaging that we would like to perform is thus going to be on these two sets of theories. In particular the averaged value of any operator $O \in \mathcal{A}_{0}$ is given by
\begin{equation}\label{qubaver}
    \overline{\langle O\rangle} = \alpha\; \omega_{p_{e}}(O)+(1-\alpha)\omega_{p_{o}}(O)
\end{equation}
where $0\leq\alpha\leq 1$. If there was a bulk dual to system in question, $\alpha$ will be read from the bulk computation. These particular bulk computations would be computations involving wormholes in a single universe. More of this will be discussed in the next section.

Since \eqref{qubaver} is a convex combination of two states, it actually defines another state, $\omega_{PI}$, on the algebra $\mathcal{A}_{0}$. With respect to this state one once again can build a Hilbert space $\mathcal{H}_{PI}$ and a von Neumann algebra of operators $\mathcal{A}_{PI}$ in the same way we described above. This theory is analogous to the semi classical theory defined by the bulk path integral. It can be checked that this representation of $\mathcal{A}_{0}$ is in fact a reducible representation and that it has a commutant. This commutant is exactly what we expect in AdS/CFT when the standard large $N$ limits do not seem to be well defined, for instance in the AS$^{2}$ or a single sided black hole cases \cite{Kudler-Flam:2025cki,Schlenker:2022dyo} or even the evaporating black hole. For the AS$^{2}$ case there is a baby universe created by the heavy operator creating a shock wave and for the single sided black hole case (to be more concrete for example a big black hole in AdS) and the evaporating black hole, there is the interior of the black hole. Following the nomenclature in those situations we say that the commutant corresponds to the algebra describing baby universe entangled with this averaged system.

To see that $\mathcal{A}_{PI}$ indeed has a non trivial commutant acting on $\mathcal{H}_{PI}$, we can proceed as follows. Since $\alpha \in [0,1]$, for any $O \in \mathcal{A}_{0}$ we have
\begin{equation}
     \omega_{p_{e}}(O) \leq \omega_{PI}(O),
\end{equation}
then by the Cauchy-Schwarz inequality we have
\begin{equation}
     |\omega_{p_{e}}(O_{1}^{\dagger}O_{2})|^{2} \leq \omega_{p_{e}}(O_{1}^{\dagger}O_{1})\;\omega_{p_{e}}(O_{2}^{\dagger}O_{2}) \leq\omega_{PI}(O_{1}^{\dagger}O_{1})\;\omega_{PI}(O_{2}^{\dagger}O_{2}). 
\end{equation}
If $\ket{\Omega_{PI}}$ is the vacuum state in $\mathcal{H}_{PI}$, then the above equation implies,
\begin{equation}
     |\omega_{p_{e}}(O_{1}^{\dagger}O_{2})|^{2}\leq \langle O_{1}\Omega_{PI}|O_{1}\Omega_{PI}\rangle \langle O_{2}\Omega_{PI}|O_{2}\Omega_{PI}\rangle.
\end{equation}
Thus we can define a sesquilinear, bounded map as $(.,.): \mathcal{H}_{PI}\cross \mathcal{H}_{PI} \rightarrow \mathbb{C}$ by $O_{1}\ket{\Omega_{PI}} \cross O_{2}\ket{\Omega_{PI}} \mapsto \omega_{p_{e}}(O_{1}^{\dagger}O_{2})$. Since states like $O\ket{\Omega_{PI}}$ are dense in $\mathcal{H}_{PI}$, this map is densly defined. Then following the Riesz representation theorem for such maps, one can represent the map in terms of the standard inner product on $\mathcal{H}_{PI}$ using a unique linear and bounded operator $T$ acting on $\mathcal{H}_{PI}$,
\begin{equation}
    \omega_{p_{e}}(O_{1}^{\dagger}O_{2})=  \langle O_{1}\Omega_{PI}|T|O_{2}\Omega_{PI}\rangle
\end{equation}
We claim that $T$ is an operator in the commutant of $\mathcal{A}_{PI}$ and it is non trivial as long as $\alpha\neq 1$, where $\omega_{p_{e}}$ is not proportional to $\omega_{PI}$. We show this for the dense subspace of $\mathcal{H}_{PI}$ which then can be extended to the full Hilbert space. Thus for any two states $\ket{O_{1}\Omega_{PI}}$ and $\ket{O_{2}\Omega_{PI}}$ and $O_{3} \in \mathcal{A}_{PI}$ we have,
\begin{equation}
    \langle O_{1}\Omega_{PI}|O_{3}\;T|O_{2}\Omega_{PI}\rangle = \omega_{p_{e}}((O_{3}^{\dagger}O_{1})^{\dagger}O_{2}) = \omega_{p_{e}}(O_{1}^{\dagger}O_{3}O_{2})= \langle O_{1}\Omega_{PI}| T\;O_{3}|O_{2}\Omega_{PI}\rangle. 
\end{equation}
Thus $T\in \mathcal{A}_{BU}:= \mathcal{A}_{PI}^{'} $. Together with similar arguments for $\omega_{p_{o}}$, we find that as long as $\alpha\neq1,0$, $\mathcal{A}_{PI}$ has a non trivial commutant, $\mathcal{A}_{BU}$. We see that the baby universe algebra of operators emerge in the large $N$ ultralimit of a theory with chaotic operators. The free ultrafilters can be considered as points at infinity of the natural numbers and in this sense baby universes are configurations that exist in the large $N$ and not finite $N$. 

This simple example we have chosen however has a drawback which is that the asymptotic states along the ultralimits are disjoint and a convex combination of the two states will give a state whose representation is just a direct sum of the representation of each of the states \cite{bratteli2012operator}. Thus the resulting commutant will just be the same as the center of $\mathcal{A}_{PI}$, whose content is just `labeling' the physically distinguished ultrfilters (in the case where the commutants of the algebras along the individual ultrafilters is trivial). In a more physical examples that we will consider later, the states will not be disjoint and $\mathcal{A}_{PI}^{'}$ has more content than just the center. 

\subsection{Wormholes}

Another perspective at \eqref{qubaver}, as opposed to a definition of a state on $\mathcal{A}_{0}$, is that the overline is the third quantization average over the different superselection sectors which we compute by inserting a complete basis states in this third quantized theory \cite{Marolf:2020xie}. That is,
\begin{equation}
   \overline{\langle O\rangle} = \langle\Omega_{PI}|O|\Omega_{PI}\rangle = |\langle\Omega_{PI}|\Omega_{p_{e}}\rangle|^{2}\langle\Omega_{p_{e}}|O|\Omega_{p_{e}}\rangle+|\langle\Omega_{PI}|\Omega_{p_{o}}\rangle|^{2}\langle\Omega_{p_{o}}|O|\Omega_{p_{o}}\rangle
\end{equation}
Thus we read that $\sqrt{\alpha}= |\langle\Omega_{PI}|\Omega_{p_{e}}\rangle|$ and $\sqrt{1-\alpha}= |\langle\Omega_{PI}|\Omega_{p_{o}}\rangle|$.

In the cases where the would-be bulk theory implies a correlation between expectation values by a Euclidean wormhole saddle, according to our prescription, the boundary computation proceeds as follows,
\begin{equation}\label{worm}
    \overline{\langle O_{1}\rangle \langle O_{2}\rangle} = \alpha\; \omega_{p_{e}}(O_{1})\omega_{p_{e}}(O_{2}) + (1-\alpha)\;\omega_{p_{o}}(O_{1})\omega_{p_{o}}(O_{2}).
\end{equation}

This would follow again by inserting the orthonormal basis state of the third quantized theory for each boundary as,
\begin{align}
     \overline{\langle O_{1}\rangle \langle O_{2}\rangle} &=|\langle\Omega_{PI}|\Omega_{p_{e}}\rangle|^{2}\langle\Omega_{p_{e}}|O_{1}|\Omega_{p_{e}}\rangle \langle\Omega_{p_{e}}|O_{2}|\Omega_{p_{e}}\rangle \\&+|\langle\Omega_{PI}|\Omega_{p_{o}}\rangle|^{2}\langle\Omega_{p_{o}}|O_{1}|\Omega_{p_{o}}\rangle \langle\Omega_{p_{o}}|O_{2}|\Omega_{p_{o}}\rangle
\end{align}
Note that $O_{1,2}$ does not depend on $N$ and only acts a finite number of qubits. As we will see these will correspond to the semiclassical operators that we are interested in more realistic cases. But note that $O_{1}$ is a perturbative operator (in this spin chain scenario, a local operator) that is defined on an oscillatory state so the full expectation value is still oscillatory. As mentioned in \eqref{oscilim}, the full operator $\mathcal{O}$ is still $N$ dependent and oscillatory. The same will be true for the chaotic operators that we discuss in AdS/CFT, and as long as the additional operators we act on top these operators are not chaotic and $N$ dependent too, the averaging argument carries through.  

If the operators in the two distinct `boundary theories', $O_{1}$ and $O_{2}$, do not receive contributions from chaotic states or operators, then the expectation values in the state $\omega_{p_{e}}$ will be the same as $\omega_{p_{o}}$, since all free ultralimits agree with the standard large $N$ limit and we will have just the disconnected correlation function. That is,
\begin{equation}
     \overline{\langle O_{1}\rangle \langle O_{2}\rangle} = \omega_{N}(O_{1})\omega_{N}(O_{2})
\end{equation}
when $O_{1}$ and $O_{2}$ are what we called planar limit operators and do not receive contributions from chaotic operators. By $\omega_{N}(.)$ we mean the standard large $N$ limit of the expectation values of the operators. We see that \eqref{worm} is indeed how the free ultralimits can produce wormholes. We can also generalize it as
\begin{equation}
    \overline{\langle O_{1}\rangle\cdots \langle O_{n}\rangle} = \alpha\; \omega_{p_{e}}(O_{1})\cdots\omega_{p_{e}}(O_{2}) + (1-\alpha)\;\omega_{p_{o}}(O_{1})\cdots\omega_{p_{o}}(O_{n}).
\end{equation}

One can get a possible generalization of the above example of a chain of qubits that oscillates as the size of the chain increases, by choosing $\vec{a}_{N}$ to be a unit vector picked from a spherical multivariant Gaussian distribution rather than being an oscillating unit vector. Since we have fixed $\vec{a}_{N}$ to be a unit vector, the distribution has the same measure as the uniform Haar measure on $\mathbb{S}^{2}$. 

We can write $\vec{a}_{N}=(\text{sin}\theta\text{ cos}\phi, \text{sin}\theta\text{ sin}\phi, \text{cos}\theta)^{t}$ with $\theta \in [0,\pi)$ and $\phi\in [0,2\pi)$, and the state of the qubit chain for a given $N$ is 
\begin{equation}
    \xi_{N}= \bigl( \begin{smallmatrix} \text{cos }\theta \\ \text{sin }\theta \;e^{i\phi} \end{smallmatrix} \bigr)^{\otimes N}, 
\end{equation}
and $\theta$ and $\phi$ are chosen from the distribution for each $N$. Thus for two distinct doublets $(\theta_{i},\phi_{i})$, where $i=1,2$, the overlap between states vanish in the large $N$ limit. Therefore they define two distinct infinite chains and the number of physically distinguishable free ultrailters are at most parametrized by $[0,\pi)\cross[0,2\pi)$, in other words a point on $\mathbb{S}^{2}$. Now we can proceed with the same discussion as above on what the would be bulk theory baby universes and wormhole correspond to, except that now we have a continuum p-sectors rather than discrete. 
\begin{equation}
    \overline{\langle O_{1}\rangle\cdots \langle O_{n}\rangle}= \int d\mu_{p}\;\omega_{p}(O_{1})\cdots\omega_{p}(O_{n})
\end{equation}
for $O_{1},\cdots O_{n} \in \mathcal{A}_{0}$.

It is also good to mention that the probability density function for a standard, independent 3 dimensional Gaussian unit vector is equivalent to the normalized Haar measure,
\begin{equation}
    d\mu=\frac{1}{4\pi} \text{sin}\theta\; d\theta \;d\phi.
\end{equation}
Thus we can see that the average with respect to this choice $\theta$ and $\phi$'s, of each of the components of the $\xi_{N}$ is zero. On the other hand the state is normalized to one.
\begin{equation}
    \langle\xi_{N}|\xi_{N}\rangle = 1
\end{equation}
Thus the components must have large fluctuations with significant variance as we change $N$ and take the standard large $N$ limit. This point is a good segue into the next section.   

\section{The AS$^{2}$ states}

\subsection{Below the Hawking Page temperature}

We consider states like the AS$^{2}$ states, which are bulk configuration first proposed by Antonini, Sasieta, and Swingle \cite{Antonini:2023hdh}. Their bottom up construction follows from considering two thermally entangled states, where each of these states are entangling two copies of thermal gas of particles in AdS. These thermally entangled states are the bulk duals of the thermofield double states of the boundary theory below the Hawking -Page temperature, say $\beta_{1}$ and $\beta_{2}$. The AS$^{2}$ states are then constructed by cutting out finite regions of one the two entangled AdS spaces for each states, and gluing them along an infinitely thin matter shell. The resulting state will be two thermal gas of particles in AdS at different temperatures of $\beta_{1}$ and $\beta_{2}$, each (presumably) entangled with a closed baby universe. The Euclidean preparation of this state involves the usual bullk Euclidean preparation of thermofield states, except that now there is a heavy operator insertion in the Euclidean cylinder, creating the shock wave along which the finite AdS regions are glued; and the slices where the thermal gas state at temperature $\beta_{1}$ is at a distance $\beta_{1}$ in Euclidean time from the operator insertion while the other thermal gas state in second AdS is prepared on a slice $\beta_{2}$ away from the operator insertion. Therefore, at the boundary, this state is expected to be dual to partially entangled thermal states (PETS) \cite{Goel:2018ubv}. 

Thus, following our spin chain toy model, we start with the following thermally entangled state in the boundary with a well defined standard large $N$ limit,
\begin{equation}\label{psi0}
    \ket{\Psi_{0}}= \frac{1}{\sqrt{Z}}\sum_{i} e^{-\beta_{1} H_{N}/2}\; \ket{i}_{1}\ket{i}_{2}
\end{equation}

The states $\ket{i}$ are perturbative states on top of the vacuum of the boudnary CFT's, defined on $\mathbb{R} \cross \mathbb{S}^{d-1}$, or some other semiclassical state with a bulk dual that is an asymptotically AdS semiclassical configuration which is not a black hole. The CFt's are taken to be CPT conjugates of each other. Such classical states can be constructed in the usual method of Euclidean path integration in the boundary, where to construct non-vacuum classical backgrounds one inserts the appropriate operator to the past of the Cauchy slice on which the states are prepared \cite{Horowitz:1998ha,Bahiru:2023zlc,Belin:2018fxe}. The perturbative states are then constructed by acting with $O(1)$ multi trace operators on this background. The state $\ket{\Psi_{0}}$ is a thermally entangled state between two of such semiclassical spacetimes\footnote{We have assumed that the classical backgrounds are such that a stable thermal state can be defined.}. $H_{N}$ is the Hamiltonian of the CFT and for its eigenstate $\ket{i}$, $E_{i}= E_{cl}+\delta E_{i}$, where $E_{cl}$ corresponds to the energy of the background spacetime. While we have $Z=\sum_{i}e^{-\beta_{1}(E_{cl}+\delta E_{i})}$. Such states are expected to have a well defined standard large $N$ limit for large enough $\beta_{1}$. Above the Hawking Page temperature however, the state $\ket{\Psi_{0}}$ will not have a large $N$ limit and the proper way to define the large $N$ theory is look at the expectation values of simple operators (subtracted $O(1)$ multi trace operators) and go through the GNS construction. But for our purposes, we take $\beta_{1}>\beta_{HP}$, where $\beta_{HP}$ is inverse Hawking-Page temperature.  

The algebra of simple operators that observers in each of the two CFT's are going to measure are given by the bounded functions of the subtracted single trace operators,
\begin{equation}
    O=\tilde{O}-\langle \tilde{O}\rangle_{0}
\end{equation}
where $\tilde{O}$ is any single trace operator in either CFT, and $\langle \tilde{O}\rangle_{0}$ is the expectation value of the single trace operator in the ground state of the perturbative states, $\ket{i}$. The reason for this subtraction is because the expectation value of the simple operator may grow with $N$ because of the insertion of the operator preparing the no trivial semiclassical state. Constructing an algebra of operators from the bounded functions of such $O(1)$ operators for each CFT, we get the full algebra,
\begin{equation}\label{fullalge}
    \mathcal{A}_{0}=\mathcal{A}_{0}^{1}\cup \mathcal{A}_{0}^{2}
\end{equation}
where $\mathcal{A}_{0}^{1}$ is the algebra of bounded functions of subtracted single trace operators for CFT$_{1}$, and $\mathcal{A}_{0}^{2}$ is the same for CFT$_{2}$. 
Note that this algebra is not yet a von Neumann algebra.

At this point we consider an oscillatory or chaotic (as a function of $N$) operator, $\mathcal{O}_{N}= e^{-\beta_{2}H_{N}/2}\;O_{N}$, and act with it on the state $\ket{\Psi_{0}}$. That is after we normalize the state with $\mathcal{N}_{N}$ we have,
\begin{equation}
    \ket{\Psi_{N}}=\mathcal{N}_{N}\mathcal{O}_{N} \ket{\Psi_{0}} = \frac{1}{\sqrt{Z_{N}}}\sum_{i} e^{-\beta_{2}H_{N}/2}\;O_{N}e^{-\beta_{1} H_{N}/2}\; \ket{i}_{1}\ket{i}_{2},
\end{equation}
while $\mathcal{N}_{N}=\frac{\sqrt{Z}}{\sqrt{Z_{N}}}$ and $\langle\Psi_{N}|\Psi_{N}\rangle = 1$. This state is indeed PETS when the ground state in each universe is the CFT vacuum, and we can also see that, 
\begin{align}
  \ket{\Psi_{N}} &=\frac{1}{\sqrt{Z_{N}}}\sum_{i} \big(\sum_{m}\ket{m}_{2}\bra{m}\big)\; e^{-\beta_{2}H_{N}/2}\;O_{N}e^{-\beta_{1} H_{N}/2}\; \big(\sum_{j}\ket{j}_{1}\bra{j}\big) \;\ket{i}_{1}\ket{i}_{2}\\
   &=\frac{1}{\sqrt{Z_{N}}}\sum_{i,m} \bra{m}e^{-\beta_{2}H_{N}/2}\;O_{N}e^{-\beta_{1} H_{N}/2}\ket{i}\; \ket{i}_{1}\ket{m}_{2}\\
   &=\frac{1}{\sqrt{Z_{N}}}\sum_{i,m} e^{-\beta_{2}E_{m}/2\; -\beta_{1} E_{i}/2}\bra{m}\;O_{N}\ket{i}\; \ket{i}_{1}\ket{m}_{2}
\end{align}
The state $\ket{\Psi_{N}}$ does not have the standard large $N$ limit because of our choice of the operator $\mathcal{O}_{N}$. In particular, the matrix elements/the wave function coefficients  $$Z^{-1/2}_{N}\bra{m}e^{-\beta_{2}H_{N}/2}\;O_{N}e^{-\beta_{1} H_{N}/2}\ket{i}$$ are oscillatory or chaotic functions of $N$. Note that $$Z_{N}=\sum_{i,m} \bra{m}e^{-\beta_{2}H_{N}/2}\;O_{N}e^{-\beta_{1} H_{N}/2}\ket{i}\big(\bra{m}e^{-\beta_{2}H_{N}/2}\;O_{N}e^{-\beta_{1} H_{N}/2}\ket{i}\big)^{\dagger}.$$ As a result, for any simple operator $O\in \mathcal{A}_{0}$, we have 
\begin{equation}
    \lim_{N\rightarrow\infty}\bra{\Psi_{N}}O\ket{\Psi_{N}}=\lim_{N\rightarrow\infty} \mathcal{N}^{2}_{N} \bra{\Psi_{0}}\mathcal{O}_{N}^{\dagger}\;O\;\mathcal{O}_{N} \ket{\Psi_{0}}\; \text{ does not exist}
\end{equation}
On the other hand, the ultralimit of the expectation value along any free ultrafilter $p$ does exist,
\begin{equation}
    \lim_{N\rightarrow p}\bra{\Psi_{N}}O\ket{\Psi_{N}}=\omega_{p}(O).
\end{equation}
Taking the completion of the algebra $\mathcal{A}_{0}$ with respect to $\omega_{p}(.)$ will give us the algebra of operators for each $p$-sector, $\mathcal{A}_{p}$. 

The GNS vacuum is the state vector that reproduces the above expectation value for any operator in $\mathcal{A}_{p}$, 
\begin{equation}
    \bra{\Omega_{p}}O\ket{\Omega_{p}} = \omega_{p}(O)
\end{equation}

In addition if for a moment we take the operator $O_{N}$ to be a bounded operator, we have 
\begin{equation}
   | \bra{m}e^{-\beta_{2}H_{N}/2}\;O_{N}e^{-\beta_{1} H_{N}/2}\ket{i}|\leq M_{im}
\end{equation}
for some real number $M_{im}$ for any $N$. Therefore, as we started by saying that the states $\ket{i},\ket{m}$ have a well defined standard large $N$ limit, we can write $\ket{\Omega_{p}}$ up to a total phase as,
\begin{equation}\label{psip}
    \ket{\Omega_{p}}=\frac{1}{\sqrt{Z_{p}}}\sum_{i,m} \tilde{c}_{im}^{p} \; \ket{i}_{1}\ket{m}_{2} =\sum_{i,m} c_{im}^{p} \; \ket{i}_{1}\ket{m}_{2}
\end{equation}
where $|\tilde{c}^{p}_{im}|\leq M_{im}$ and $Z_{p}= \sum |\tilde{c}^{p}_{im}|^{2} \leq \sum |M_{im}|^{2}=M$. We have also taken the standard large $N$ limit for $\ket{i}$ and $\ket{m}$. Whether or not the norm of $\tilde{c}^{p}_{im}$ explores all the values in $[0,M_{im}]$ depends on how chaotic/oscillatory our chosen operator $O_{N}$ is. In the case where $O_{N}$ is not a bounded operator in $N$, we have $|\tilde{c}^{p}_{im}|< \infty$, however, we are still going to have $|c^{p}_{im}|\leq 1$.

If on the other and $O_{N}$ is a heavy operator, i.e, $O(N^{a})$ for positive $a$, then we can write 
\begin{equation}
   | \bra{m}e^{-\beta_{2}H_{N}/2}\;O_{N}e^{-\beta_{1} H_{N}/2}\ket{i}|= f(N)\frac{ | \bra{m}e^{-\beta_{2}H_{N}/2}\;O_{N}e^{-\beta_{1} H_{N}/2}\ket{i}|}{f(N)}
\end{equation}
so that $f(N)$ is $O(N^{a})$ and thus $\frac{ | \bra{m}e^{-\beta_{2}H_{N}/2}\;O_{N}e^{-\beta_{1} H_{N}/2}\ket{i}|}{f(N)}$ is order one. We can similarly do the this for the partition function, and since we are below the Hawking Page transition, its $N$ dependence will only be captured by the overall factor $f(N)^{2}$. Thus along any ultrafilter $p$, we will arrive at the last equality of equation \eqref{psip} with $|c^{p}_{im}|\leq 1$.

Now we can figure out what the physically distinguishable free ultrafilters are, which is given as parameterizing the orthonormal set of the coefficients $\{\{c^{p}_{im}\},\{c^{q}_{im}\},\cdots\} $ in the standard $\ell^{2}$ norm. That is, for any two sets of coefficients $\{c^{q}_{im}\}$, $\{c^{p}_{im}\}$,
\begin{equation}
    \sum_{im} (c^{p}_{im})^{*}c^{q}_{im} = \delta_{p,q}.
\end{equation}
Of course, depending on the choice of $O_{N}$ there can even be only two physically distinguished free ultrafilters or there can be an infinite of them.  

A simple case one might consider is the situation where the matrix elements $\langle m|O_{N}|i\rangle $ (or its normalized version as above, if $O_{N}$ is an $O(N^{a})$ operator) does not converge as we take the standard large $N$ limit, but it has two accumulation points. For instance, for even $N$  the limit point of the sequence is $\delta_{mi}\text{ cos}^{2}(m\pi/2)$ and for odd $N$, the subsequence has a limit point $\delta_{mi}\text{ sin}^{2}(m\pi/2)$. Thus we have for any $O(1)$ operator $O\in \mathcal{A}_{0}$,
\begin{align}
    \lim_{N\rightarrow p_{e}}\bra{\Psi_{N}}O\ket{\Psi_{N}} &= \frac{1}{Z_{p_{e}}}\sum_{m,n} e^{-(\beta_{2}\; +\beta_{1}) (E_{m}+ E_{n})/2}\text{ cos}^{2}(m\pi/2)\text{ cos}^{2}(n\pi/2)\; \bra{n}_{1}\bra{n}_{2} O\ket{m}_{1}\ket{m}_{2}\\
    &=\omega_{p_{e}}(O) 
\end{align}
where $p_{e}$ is any free ultrafilter that contains the set of even numbers. On the other hand,
\begin{align}
    \lim_{N\rightarrow p_{o}}\bra{\Psi_{N}}O\ket{\Psi_{N}} &= \frac{1}{Z_{p_{o}}}\sum_{m,n} e^{-(\beta_{2}\; +\beta_{1}) (E_{m}+ E_{n})/2}\text{ sin}^{2}(m\pi/2)\text{ sin}^{2}(n\pi/2)\; \bra{n}_{1}\bra{n}_{2} O\ket{m}_{1}\ket{m}_{2}\\
    &=\omega_{p_{o}}(O) 
\end{align}
where now $p_{o}$ is any free ultrafilter that includes the set of odd numbers. In particular one can complete $\mathcal{A}_{0}$ in to a von Neumann algebra with respect to each of the two `types' of the physical free ultrafilters, and get $\mathcal{A}_{p}$. Of course $Z_{p_{e/o}}$ are the partition function with the overall $N$ dependent factor is extracted for the case of a heavy operator $O_{N}$.

Thus the GNS vacua can be written as 
\begin{equation}
    \ket{\Omega_{p_{e}}} = \frac{1}{\sqrt{Z_{p_{e}}}}\sum_{m} e^{-(\beta_{2}\; +\beta_{1}) E_{m}/2}\text{ cos}^{2}(m\pi/2)\; \ket{m}_{1}\ket{m}_{2} 
\end{equation}
up to a total phase in front of $\ket{\Omega_{p_{e}}}$. Similarly, we have (again up to a total phase) 
\begin{equation}
    \ket{\Omega_{p_{o}}} = \frac{1}{\sqrt{Z_{p_{o}}}}\sum_{m} e^{-(\beta_{2}\; +\beta_{1}) E_{m}/2}\text{ sin}^{2}(m\pi/2)\; \ket{m}_{1}\ket{m}_{2} .
\end{equation}

Then the Hilbert space for the each $p-$sector is given by
\begin{equation}
    \mathcal{H}_{p_{e}/p_{o}} = \overline{\{\mathcal{A}_{p_{e}/p_{o}}\ket{\Omega_{p_{e}/p_{o}}}\}}.
\end{equation}
The over line above is to indicate that we are taking the closure of the space with respect to the inner product provided by $\omega_{p_{e}/p_{o}}(.)$.

However the gravitational path integral computes an average value for the expectation value of simple operators in $\mathcal{A}_{0}$. For $O\in \mathcal{A}_{0}$,
\begin{equation}\label{PIstate}
    \overline{\braket{O}} = \alpha \;\omega_{p_{e}}(O)+ (1-\alpha)\;\omega_{p_{o}}(O)
\end{equation}
Following the same type of arguments used in the previous section of the spin chain toy model, we see that the above expression defines a state on $\mathcal{A}_{0}$, with respect to which one can take the closure of algebra to get the von Neumann algebra $\mathcal{A}_{PI}$. According to our proposal this is indeed the algebra that corresponds to the algebra of operators that act on the two copies of asymptotically AdS spacetimes that are present in the AS$^{2}$ construction. If $\alpha \in (0,1)$, as proved in the previous section, we can be sure that there is a non trivial commutant to algebra of operators acting in the asymptotically AdS regions, $\mathcal{A}_{PI}$. This commutant corresponds to the algebra of operators acting in bulk region not causally connected to the boundary. The reason for this is that $\mathcal{A}_{PI}$ can be used to reconstruct operators in this bulk region by the HKLL reconstruction. Thus the operators in the bulk regions causally connected to the boundary are equivalent(dual) to $\mathcal{A}_{PI}$. Therefore the commutant, which we call $\mathcal{A}_{BU} = \mathcal{A}_{PI}^{'}$ correspond to the von Neumann algebra of operators acting on a region causally disconnected from the boundary. Note that ,
\begin{equation}
    \mathcal{A}_{PI} = \mathcal{A}_{0}^{1} \vee \mathcal{A}_{0}^{2}
\end{equation}
where $\mathcal{A}_{0}^{1} \vee \mathcal{A}_{0}^{2} := \overline{\mathcal{A}_{0}^{1} \cup \mathcal{A}_{0}^{2}}$, and the overline here is closure w.r.t. the state defined by \eqref{PIstate}. 

We want to show that $\mathcal{A}_{PI}$ is a type I von Neumann algebra. Since if $\mathcal{A}_{PI}$ is a type I von Neumann algebra, then $\mathcal{A}_{BU}$ must also be a type I von Neumann algebra. It follows then, $\mathcal{A}_{BU}$ acts in a spacetime region that is disconnected from where $\mathcal{A}_{PI}$ acts on (in other words we can specify states in this region and where $\mathcal{A}_{PI}$ acts on, independently without any entanglement between the states). Since we have said above that this region is also not connected to the boundary, we conclude that $\mathcal{A}_{BU}$ acts on a disconnected universe with no boundary, i.e, a baby universe. We will see later the case where $\mathcal{A}_{BU}$ acts is not a disconnected universe even though it is not connected with the boundary. 

To show that $\mathcal{A}_{PI}$ is type I we look at the state that we started the discussion with, $\ket{\Psi_{0}}$. \footnote{The steps we show here are also in view of what we need for the case where $\beta_{1,2}$ are smaller that $\beta_{HP}$.}. As we mentioned before, for $\beta_{1}>\beta_{HP}$, this state is expected to describe an entangled state of a thermal gas of particles living in two disconnected universes. The entanglement entropy  between the two spacetimes (and the partition function $Z$) is $O(1)$, and there is no spacetime connection between them. For this reason, the state vector itself has a well defined standard large $N$ limit, and the closure of the algebras of operators $\mathcal{A}_{0}^{1} \cup \mathcal{A}_{0}^{2}$ is a type I von Neumann algerba \cite{Maldacena:2001kr,Witten:2021jzq}, so are the closures of each of the individual factors in the union. This implies that there is a minimal projection $\Pi_{i}^{(1)}$ in $\overline{\mathcal{A}_{0}^{1}}$ and $\Pi_{j}^{(2)}$ in $\overline{\mathcal{A}_{0}^{2}}$, and thus a minimal projection $\Pi_{i}^{(1)}\Pi_{j}^{(2)}$ in $\mathcal{A}_{0}^{1} \vee \mathcal{A}_{0}^{2}$, where the closure is taken with respect to $\ket{\Psi_{0}}$. As usual, the expectation values of these projection operators are just the large $N$ limits of operators in the finite $N$ entangled theory, in particular the projection operators of the finite $N$ theory. 

Coming back to the state $\ket{\Psi_{N}}$, we can consider these finite $N$ operators, for instance the one that projects onto an energy eigenstate labeled by an even index $m$, then we can take the free ultralimit along $p_{e}$ of this operator (by which we mean that we define the the operator `at' $p_{e}$ by stating its expectation values as the ultralimits of the expectation values of these finite $N$ operators along $p_{e}$). This will give us a minimal projector (up to a prefactor) in the algebra $\mathcal{A}_{p_{e}}$. The same can be said about $\mathcal{A}_{p_{o}}$, where now we take the finite $N$ projector onto an energy eigenstates labeled by an odd $m$ (for $m$ even, the finite $N$ projection operator is zero in the ultralimit). Since both $\mathcal{A}_{p_{e}}$ and $\mathcal{A}_{p_{o}}$ contain minimal projection operators, they are type I von Neumann algebras. However, note that $\mathcal{A}_{p_{e}/p_{o}}$ are von Neumann algebras constructed with respect to $\omega_{p_{e}/p_{o}}(.)$. Therefore, since the state with respect to which $\mathcal{A}_{PI}$ is constructed, is a convex combination of $\omega_{p_{e}}$ and $\omega_{p_{o}}$ (which is also what we do in the case where more than two physically distinguished $p$'s are present); and since both $\mathcal{A}_{p_{e}}$ and $\mathcal{A}_{p_{o}}$ are type I algebras, the algebra $\mathcal{A}_{PI}$ is also a type I von Neumann algebra. Thus we conclude the proof.

Given that the states $\omega_{p_{e/o}}$ in the equation \eqref{PIstate} are not disjoint it can be thought of defining a density matrix of rank 2. In general, the rank of this density matrix is some number $d$ which is related to the number of physically distinguished free ultrafilters leading to pair wise non disjoint states in the ultralimit. And given the rank of the density matrix associated with $\omega_{PI}$ is $d$, the von Neumann algerba of the baby universe $\mathcal{A}_{BU}$ will be a type I$_{d}$ von Neumann algebra. Note that $d$ can be infinite if the rank of the density matrix is infinite.        

It follows from the above construction that baby universes are emergent strictly in the large $N$ limit. They are results of the fact that the ultralimits of the boundary CFT along different free ultrafilters do not always agree. Then semiclassically accessible operators will have a commutant, which corresponds to semiclassical operators that act in the baby universe. 

\subsection{Above the Hawking Page temperature}

In the large $N$ limit, the thermofield double state, \eqref{psi0}, goes through a first order phase transition as we increase the temperature beyond the Hawking Page temperature. The state will become an entangled state between the boundary theories in the deconfined phase, with a huge amount of entanglement. The entanglement entropy for $\mathcal{N}=4$ super Yang Mills in four dimensions for example will be $O(N^{2})$ while the partition function goes like $e^{N^{2}f}$ where $f \sim O(1)$. 

As a result of this exponential dependence in $N^{2}$ of the partition function, the state vector \eqref{psi0} will not have a standard large $N$ limit. Unlike the case where the indefiniteness of the limit comes from an oscillatory/chaotic operator, this indefiniteness of the limit is apparently not cured by taking the ultralimit, since the ultralimit of exponentially (or even any powers of $N$) growing quantity still diverges, in particular it is an unlimited number. The same goes for the entanglement entropy. The resolution to this issue with defining a state in the large $N$ limit has played a crucial role in the recent works to reemphasis the role of von Neumann algebras in quantum gravity \cite{Leutheusser:2021frk,Leutheusser:2021qhd,Witten:2021unn}. 

Along with the growth of the partition function, the density of the energy eigen states in some $O(1)$ band $\Delta E$ also diverges. As a result, some operators, in particular the subtracted single trace operators do have a well defined expectation values in the large $N$ limit, even though the state vector does not. This collection of expectation values defines a state on the algebra of subtracted single trace operators at $N=\infty$, by the GNS construction. The resulting von Neumann algebras in each of the CFT's will be type III$_{1}$ \cite{Leutheusser:2021frk,Leutheusser:2021qhd,Furuya:2023fei} which are expected to be dual to the two exterior spacetime regions of the eternal black hole in the bulk, and matches with the general expectation about the type of a von Neumann algebra associated with a subregion in QFT.  


For the state defined as the large $N$ limit of the thermofield double state, $\omega_{0}$, if $|\Omega_{0}\rangle $ is its state vector representation in the GNS Hilbert space, the associated modular operator, $\Delta_{0}$, characteristically acts on a dense set of states as follows. If $a \in \bar{\mathcal{A}_{0}^{1}}$ (the closure is with respect to $\omega_{0}$),
\begin{equation}
    \Delta_{0}^{1/2}a\ket{\Omega_{0}} = (a^{\dagger})^{'}\ket{\Omega_{0}}
\end{equation}
where $a^{'}\in \bar{\mathcal{A}_{0}^{2}}= \bar{\mathcal{A}_{0}^{1}}^{'}$ is the modular conjugation of $a$. This modular operator is large $N$ limit of the modular operator of the finite $N$ theory. To proceed further and generalize our discussion for the AS$^{2}$ state, let's discuss a slightly different version of this story (but equivalent to the GNS construction) where we use a representation of thermofield double state with a Hilbert space, and thus the state vectors it contains, that has a well defined (standard) large $N$ limit. We discussed above is the case where the Hilbert space does not have a standard large $N$ limit.   

For any given finite value of $N$, we consider operators, $\tilde\kappa$, in our holographic CFT whose square is a trace class operator. The trace here is just the standard quantum mechanical trace over basis states of the CFT in discussion. We associate with such an operator a state vector normalized to one. That is,

\begin{equation}
    \langle{\kappa}\ket{\kappa} = \text{tr} (\kappa^{\dagger}\kappa) = \frac{\text{tr} (\tilde{\kappa}^{\dagger}\tilde\kappa)}{\sqrt{\text{tr} (\tilde{\kappa}^{\dagger}\tilde\kappa)}\sqrt{\text{tr} (\tilde{\kappa}^{\dagger}\tilde\kappa)}} = 1
\end{equation}

We then build the pre Hilbert space by acting on this state with other trace class operators in our theory, which form an $*-$algebra $\mathcal{B}$\footnote{ A $*-$algebra is an algebra of operators, i.e, a set where a product of operators and a multiplication of operators by a complex number are also in the set, with additional property that the `dagger' of an operator is also an operator in the same set. The presence of an identity is not required for it to be a $*-$ algebra and here identity is not an element since it has infinite trace, therefore this alegrba is called non unital $*-$ algebra.}. However, there is two distinct ways that the operators can act on our state and produce a new normalizable state. That is, if $O$ is a trace class operator in our CFT, it can act on $\ket{\kappa}$ from the left or from the right, $\ket{O\kappa}$ or $\ket{\kappa O}$ respectively, and give two distinct states as long as $O$ and $k$ do not commute. (These two ways for the action of our operator $O$ correspond to the operators from two CFT's that are entangled in \eqref{psi0}). We thus call acting from the left first representation of $O$, $\pi_{1}(O)$, and acting from the right, the second representation of $O$, $\pi_{2}(O)$. 
\begin{equation}
    \pi_{1}(O)\ket{\kappa}= \ket{O\kappa}\; \text{ and }\pi_{2}(O)\ket{\kappa}= \ket{\kappa O^{\dagger}}.
\end{equation}

Note that generally we take the two CFT's in \eqref{psi0} to be CPT conjugates of each other  to get the desired Hartle Hawking state on the eternal black hole background. It is easy to see that the two representation commute with each other and,
\begin{equation}
    \bra{\kappa}\pi_{1}(O)\ket{\kappa} = \overline{\bra{\kappa}\pi_{2}(O)\ket{\kappa}}.
\end{equation}
One can also easily see that these representations are faithful representation of the algebra of trace class operators in the CFT that preserve all its algebraic structures, i.e, representation of a product of operators is the product of the representation of each of the operators, multiplication of a representation of an operator by a complex number is the representation of the multiplication of the operator by the complex number, and the `dagger' of a representation is the representation of a `dagger' of an operator. 

If we define an anti unitary operator $J$ as
\begin{equation}
    J\ket{\rho} = \ket{\rho^{*}}
\end{equation}
it is then easy to see that 
\begin{equation}
    J \pi_{1}(\mathcal{B})J = \pi_{2}(\mathcal{B}),
\end{equation}
assuming that $\kappa$ is Hermitian. Thus $J$ plays the role of modular conjugation as $\pi_{1}$ and $\pi_{2}$ are commutants. 

Now one can take the operator $K= \pi_{1}(\text{log }\kappa)-\pi_{2}(\text{log }\kappa)$. We can see that this operator has a well defined expectation value, even in the large $N$ limit, which is zero. One can recover  now the action of a modular operator,
\begin{equation}
    e^{K}\pi_{1}(O)\ket{\kappa} = \pi_{1}(\kappa)\pi_{2}(\kappa^{-1})\ket{O\kappa}= \ket{\kappa O}= \pi_{2}(O)^{\dagger}\ket{\kappa}.
\end{equation}
That is we have the modular operator $\Delta_{\kappa}=  e^{K}$\footnote{It can be seen that $J\Delta^{1/2}_{k} \pi_{1}(O)\ket{\kappa}=\pi_{1}(O^{\dagger})\ket{\kappa}$, which gives the Tomita operator.}. The state vector $\ket{\kappa}$ and the Hilbert space that we build by acting with operators as described above have a well defined standard large $N$ limit.

In fact if we set $\kappa= e^{-\beta_{1}H_{N}/2}/Z^{1/2}$, the above story describes the thermofield double state \eqref{psi0} exactly and the large $N$ limit of the finite $N$ Hilbert space found there \emph{is} the GNS construction we described at the beginning of this sub section as a way to get a well defined large $N$ limit. In short, first we described GNS construction at infinite $N$ starting from the usual story of thermofield double state, then we described a GNS equivalent construction at finite $N$, then took the large $N$ limit to arrive at the same GNS construction at infinite $N$.  The modular operator in this case will be 
\begin{equation}
    \Delta_{0} = \pi_{1}(e^{-\beta_{1}H_{N}/2})\pi_{2}(e^{\beta_{1}H_{N}/2}).
\end{equation}

Coming back to the state $\ket{\Psi_{N}}$, we set
\begin{equation}
   \kappa_{N}= \frac{e^{-\beta_{2}H_{N}/2}\;O_{N}e^{-\beta_{1} H_{N}/2}}{Z_{N}^{1/2}}
\end{equation}
and then the above exposition describes the AS$^{2}$ like states for any finite $N$. To find out the corresponding modular operator we notice that,
\begin{equation}
    K= \pi_{1}\big(\text{log }(e^{-\beta_{2}H_{N}/2}\;O_{N}e^{-\beta_{1} H_{N}/2})\big)- \pi_{2}\big(\text{log }(e^{-\beta_{2}H_{N}/2}\;O_{N}e^{-\beta_{1} H_{N}/2})\big)
\end{equation}
where the $Z_{N}^{1/2}$ cancels out, and $\Delta_{N} = e^{K}$, here we are assuming $O_{N}$ is Hermitian.

For non oscillating or non chaotic operator, as we said above, we can simply adopt the above discussion of the finite $N$ Hilbert space and modular operator, then take the large $N$ limit to get the large $N$ GNS Hilbert space as we do for the thermofield double state. However, Since $O_{N}$ is an oscillatory or chaotic function of $N$, we will not have a well defined large $N$ limit. What we can do however is to look at expectation values and take the ultralimit,
\begin{align}
   \lim_{N\rightarrow p} \langle\kappa_{N}|\pi_{1}(O)|\kappa_{N}\rangle =& \lim_{N\rightarrow p} \frac{1}{Z_{N}}\text{tr }(e^{-\beta_{1}H_{N}/2}\;O_{N}e^{-\beta_{2} H_{N}/2} Oe^{-\beta_{2}H_{N}/2}\;O_{N}e^{-\beta_{1} H_{N}/2})\\ =& \lim_{N\rightarrow p} \frac{1}{Z_{N}}\sum_{i,m,n}e^{-\beta_{1}E_{i}}\bra{i}\;O_{N}\ket{m}\bra{m}e^{-\beta_{2} H_{N}/2} Oe^{-\beta_{2}H_{N}/2}\ket{n}\bra{n}O_{N}\ket{i}
\end{align}

As we said, had $\bra{n}O_{N}\ket{i}$ been a convergent sequence of expectation values, the above expectation values will have an ultralimit that's the same as the standard large $N$ limit. On the other hand, if $\bra{n}O_{N}\ket{i}$ (or its normlaized form if $O_{N}$ is taken to be a heavy operator) is a oscillatory or chaotic sequence (in $N$) of expectation values, it will only have a well defined ultralimit value, say $M_{ni}^{p}$ along any arbitrary free ultrafilter $p$. That is, the difference of $\bra{n}O_{N}\ket{i}$ (or normalized $\bra{n}O_{N}\ket{i}$) and $M^{p}_{ni}$ vanishes in the large $N$ limit along some infinite subset of the natural numbers in $p$. In other words, as far as the large $N$ ultralimit along $p$ is concerned we may as well replace $\bra{n}O_{N}\ket{i}$ or its normalized form in the relevant case by $M^{p}_{ni}$. However, the same issues we mentioned for the thermofield double state above the Hawking Page temperature arise here too. Thus, $Z_{N}$ and the density of the states $\ket{i}$ still diverges in the ultralimit. However, for $O(1)$ operators, the limit exists \cite{Witten:2021unn,Witten:2021jzq}, roughly as a result of the cancellation of these two effects. The same discussion follows for $\langle\kappa_{N}|\pi_{2}(O)|\kappa_{N}\rangle$ with the obvious changes on the action of the operator and in the following we only discuss $\pi_{1}$. In the particular example that we were considering earlier, where the sequence of expectation values has two accumulation points, we will have,
\begin{equation}
    \lim_{N\rightarrow p_{e/o}} \langle\kappa_{N}|\pi_{1}(O)|\kappa_{N}\rangle = \omega_{p_{e/o}}(O).
\end{equation}

We get the von Neumann algebra of operators for the first boundary, $\mathcal{A}_{p_{e/o}}^{1}$, along an ultrafilter $p_{e/o}$ by taking the completion of the set of operators $\{\pi_{1}(O)\}$, where $O$'s are operators in the holographic CFT with which we started, with respect to $\omega_{p_{e/o}}$.

Now we want to show that $\Delta_{N}$ is not an inner automorphism, and thus the algebras $\mathcal{A}_{p_{e/o}}^{1}$ are type III von Neumann algebras. To see this, we look at the operator $\pi_{1}\big(\text{log }(e^{-\beta_{2}H_{N}/2}\;O_{N}e^{-\beta_{1} H_{N}/2})\big)$. This operator must not be in $\mathcal{A}_{p_{e/o}}^{1}$ in the large $N$ limit. This can be checked by computing its expectation value in the large $N$ limit (rather ultralimit) and seeing if it is a finite number or a limited number. The expectation value at some finite $N$ is given by,

\begin{align}
   \langle\kappa_{N}|\pi_{1}\big(L_{N}\big)|\kappa_{N}\rangle\\
   &=\frac{1}{Z_{N}} \sum_{i,m,n}e^{-\beta_{1}E_{i}-\beta_{2}(E_{m}+E_{n})/2}\bra{i}\;O_{N}\ket{m}\bra{m}\text{log }L_{N}\ket{n}\bra{n}O_{N}\ket{i}\\
   & = \frac{1}{\text{tr}(L_{N}^{\dagger}L_{N})} \text{tr}\big(L_{N}^{\dagger}(\text{log }L_{N})L_{N}\big)
\end{align}

To directly show that the above quantity diverges is slightly complicated and one may need more conditions on the sequence of operators $O_{N}$. For us, to simplify the computation, we assume that not only 
\begin{equation}\label{weakO}
    \lim_{N\rightarrow p_{e}}\bra{i}O_{N}\ket{j} = \delta_{ij} \text{ cos}^{2}(j\pi/2)
\end{equation}
but also that we can replace $e^{-\beta_{2}H_{N}/2}\;O_{N}e^{-\beta_{1} H_{N}/2}$ by $C\;\Pi_{even}e^{-(\beta_{1}+\beta_{2})H_{N}/2}$ inside the log and trace as we take the ultralimit along $p_{e}$, for some constant $C$. For this we need the trace norm,
\begin{equation}
  \lim_{N\rightarrow p_{e}}  ||e^{-\beta_{2}H_{N}/2}\;O_{N}e^{-\beta_{1} H_{N}/2} - C\;\Pi_{even}e^{-(\beta_{1}+\beta_{2})H_{N}/2}||_{1} = 0
\end{equation}
In this case, we have 
\begin{equation}
    \lim_{N\rightarrow p_{e}} \text{tr}(L_{N}^{\dagger}L_{N})= C\;\lim_{N\rightarrow p_{e}}\text{tr}_{even}(e^{-(\beta_{1}+\beta_{2})H_{N}})
\end{equation}
However, since we labeled the energy levels even and odd by such that at any value of $N$ half are even and half are odd, the density of energy eigen states still diverges even if we consider only the even energy eigen states. In addition, 
\begin{equation}
     \lim_{N\rightarrow p_{e}}\frac{1}{\text{tr}(L_{N}^{\dagger}L_{N})} \text{tr}\big(L_{N}^{\dagger}(\text{log }L_{N})L_{N}\big) =  \lim_{N\rightarrow p_{e}}\frac{\text{tr}_{even}(e^{-(\beta_{1}+\beta_{2})H_{N}}H_{N})+O(1)}{\text{tr}_{even}(e^{-(\beta_{1}+\beta_{2})H_{N}})}.
\end{equation}
The Hamiltonian is $O(N^{a})$, for some positive real number $a$, and it still diverges in the ultralimit even if we only consider energy eigen states labeled even. For a heavy operator $O_{N}$, \eqref{weakO} will have an $N$ dependent factor while $C$ will now be an $O(N^{a})$ coefficient which will to the $O(N^{a})$ divergence of the modular Hamiltonian. Thus the final conclusion of the discussion will still be the same.

Therefore we find that since the Hamiltonian of the thermofield double state has divergent expectation value above temperatures above Hawking Page temperature, operator $\pi_{1}(L_{N})$ has a divergent expectation value that is it is not an element of the von Neumann algebra $\mathcal{A}_{p_{e}}^{1}$ and the modular operator $\Delta_{N}$ is not inner automorphism in the ultralimit. 

The argument proceeds just like above for $\mathcal{A}_{p_{e}}^{2}$, which is the completion of the set $\{\pi_{2}(O)\}$ and for the ultrfilter $p_{o}$. Thus we conclude that the algebras $\mathcal{A}_{p_{e/o}}^{1/2}$ are type III von Neumann algebras. We proceed as above and define the averaged or path integral state as,
\begin{equation}
    \omega_{PI}(O) = \alpha\; \omega_{p_{e}}(O)+(1-\alpha)\; \omega_{p_{o}}(O)
\end{equation}
where $O$ is an operator in the holographic CFT we started with. To be more explicit in writing the equation above, $\pi_{1}(O)$ corresponds to an operator in boundary theory 1 acting on the state \eqref{psi0}; and the expectation value of this operator in the path integral/averaged state is  
\begin{equation}
    \bra{\Omega_{PI}}\pi_{1}(O)\ket{\Omega_{PI}} = \alpha \lim_{N\rightarrow p_{e}}\bra{\kappa_{N}}\pi_{1}(O)\ket{\kappa_{N}} + (1-\alpha) \lim_{N\rightarrow p_{o}}\bra{\kappa_{N}}\pi_{1}(O)\ket{\kappa_{N}}
\end{equation}
and similarly for $\pi_{2}(O)$. The state vector $\ket{\Omega_{PI}}$ is defined such that,
$$\bra{\Omega_{PI}}\pi_{1}(O)\ket{\Omega_{PI}}= \overline{\bra{\Omega_{PI}}\pi_{2}(O)\ket{\Omega_{PI}}}= \omega_{PI}(O)$$
We take the completion of the observables $\pi_{1}(\mathcal{B}) \cup \pi_{2}(\mathcal{B})$ with respect to $\omega_{PI}$ to get a von Neumann algebra $\mathcal{A}_{PI}$ and the associated Hilbert space $\mathcal{H}_{PI}$. We can also write $\mathcal{A}_{PI} = \mathcal{A}_{PI}^{1} \vee \mathcal{A}_{PI}^{2}$, where $\mathcal{A}_{PI}^{1/2}$ are von Neumann algebras where $\pi_{1/2}(\mathcal{B})$ are completed with respect to $\omega_{PI}$ individually. 

As we have shown above, $\mathcal{A}_{p_{e/o}}^{1}$ and $\mathcal{A}_{p_{e/o}}^{2}$ are all type III von Neumann algebras, and thus $\mathcal{A}_{PI}^{1}$ and $\mathcal{A}_{PI}^{2}$ are both type III, since the state $\omega_{PI}$ is a convex combination of the states that correspond to each physically distinguished ultrafilters (the same is true even when the state $\omega_{PI}$ is an integral of states corresponding to physically distinguished ultrafilters with positive measure). According to our proposal, these are supposed to be duals to the algebra of operators in the bulk region that are causally connected to the first boundary theory and the second boundary theory, respectively. Which is to say the regions exterior to the horizon. Thus, we generally expect them to be a type III$_{1}$ von Neumann algebras at leading order (although it is possible that perturbative corrections change the type of the algebra). 

Following the same arguments we gave in the previous subsection, we can see that $\mathcal{A}_{PI}$ has a non trivial commutant, which we call $\mathcal{A}_{In} = \mathcal{A}_{PI}^{'}$. On the other hand both $\mathcal{A}_{PI}^{1}$ and $\mathcal{A}_{PI}^{2}$ are type III and have commutants that are again type III. That is, we have $\mathcal{A}_{PI}^{2},\mathcal{A}_{In} \subset(\mathcal{A}_{PI}^{1})^{'} $ and $\mathcal{A}_{PI}^{1},\mathcal{A}_{In} \subset(\mathcal{A}_{PI}^{2})^{'} $.

If Haag duality is satisfied in the bulk, we find that $(\mathcal{A}_{PI}^{1})^{'} $ is a type III$_{1}$ algebra associated with the region complementary to the subregion casually connected to boundary 1. We note that since $\mathcal{A}_{In} = \mathcal{A}_{PI}^{'}$, we have $(\mathcal{A}_{PI}^{1})^{'} = \mathcal{A}_{In} \vee \mathcal{A}_{PI}^{2}$; and since $\mathcal{A}_{In}$ is non trivial it can not be $\mathbb{C}I$ neither will it just be $\mathcal{Z}(\mathcal{A}_{PI}^{1})$  or $\mathcal{Z}(\mathcal{A}_{PI}^{2})$ if the resulting states along the ultrfilters are no disjoint, rather it is a big algebra associated to the bulk degrees of freedom that are not casually connected to either boundary 1 or boundary 2, which nonetheless live in a bulk region complementary to the two exterior regions. We are tempted to say that these are operators that live in the interior region of a long wormhole, even though our argument above is not a definitive proof that $\mathcal{A}_{In}$ is associated to such a region inside the long wormhole, since technically it can even be a type I von Neumann algebra. 

A different heuristic argument for a type III$_{1}$ nature of $\mathcal{A}_{In}$ is present for the case that $O_{N}$ is a heavy operator. The reason is that one expects $\omega_{p_{e/o}}$ behave much like a boundary state with a long wormhole dual in the bulk. Note that naively the projection onto the even(odd) labeled states at finite $N$ and taking the limit will not change general behavior of the general behavior of the state, since for example the partition function the density of the energy eigenstates in some band still grow with $N$ to some positive power. For the usual long wormhole algebra of exterior operators, the reason for the algebraic union of the left exterior and right exterior operators to be a type III$_{1}$ can be understood as it being entangled with the operators creating the shock waves sustaining the long wormhole. In the case where this is studied in detail \cite{Chandrasekaran:2022eqq}, this is because the some fixed frame the left operators at some fixed $O(1)$ time in the boundary and left operators at some $O(t>> t_{s})$ time, where $t_{s}$ is scrambling time, in the boundary form a free product algebra. The same goes for the right side. It is this full free product algebra that is entangled with the free product algebra on the right. In particular the algebraic union of the standard subtracted single trace operators inserted at $O(1)$ time from the left and the right CFT is not $\mathcal{B(H)}$ and do not have a trivial commutant. The fact the necessitated the use of free product algebra is the vanishing of out of time order correlators in the particular limit that is relevant for long wormholes. This property of the OTOC will not change if we consider adding, at finite $N$ a projection operators onto energy eigen states labeled `even'. Thus for the case of \cite{Chandrasekaran:2022eqq}, including the projection operator will not change the relevant behavior of the state. Thus we expect similar thing to happen for the long wormhole constructed by gluing the exterior regions of two eternal wormholes along a high energy shock wave. If it is the case that, just like the long wormhole geometry, the algebraic union of $\mathcal{A}_{p_{e/o}}^{1}$ and $\mathcal{A}_{p_{e/o}}^{2}$ is type III$_{1}$, then $(\mathcal{A}_{PI}^{1})^{'}$ will also have to be type III, and a general expectation in QFT, due to Buchholz–D'Antoni–Fredenhagen theorem, implies that it is a type III$_{1}$ von Neumann algebra. At the moment this will be our comment, leaving more precise argument for future work. 

Therefore, what we have found is a result that implies ( or at the very least consistent with) the fact that the state $\ket{\Omega_{PI}}$ corresponds to a bulk state with two regions exterior to black hole horizons connected by a long wormhole. The fact that we have black holes horizons here is suggested, as is usual in AdS/CFT, from the fact that the entropy of the each boundary theories in the state $\ket{\Omega_{PI}}$ is $O(N^{a})$ for some positive $a$ that depends on the actual boundary theories, while the partition function diverges. The algebra of operators associated with the two exterior regions are $\mathcal{A}_{PI}^{1}$ and $\mathcal{A}_{PI}^{2}$ while the interior region of the long wormhole corresponds to $\mathcal{A}_{In}$\footnote{By this we mean, we pick the $t=0$ slice at the boundary and consider a bulk Cauchy slice anchored at this boundary slice with zero extrinsic curvature. Then $\mathcal{A}_{In}$ is associated with the causal diamond spacetime of the subregion of this bulk Cauchy slice in the interior of the black hole.}.

The one sided versions of these states first discussed in \cite{Balasubramanian:2025zey} can also be described in a manner parallel to the above discussion. 
\begin{equation}
    \ket{\Psi_{N}} = \frac{1}{\sqrt{Z_{N}}}e^{-\beta H_{N}} O_{N}\ket{\Psi},
\end{equation}
where $\ket{\Psi}$ is a CFT state with a semiclassical holographic dual and $Z$ is the normalization. The expectation values of simple operators will not have a standard large $N$ limit due to a chaotic/oscillatory nature of the operator chosen $O_{N}$. However the expectation values will have a well defined ultralimits. These will define quantum mechanical states on the algebra of simple operators labeled physically distinguished free ultrafilters. For large $\beta$ the von Neumann completion of the algebra of simple operators will be a type $I$ von Neumann algebra. The state computed by the gravitational path integral is given by a convex combination/direct integral of the states with a positive measure of the states we labeled by the ultrafilters. The resulting state will also be a type I von Neumann algebra, with a non trivial commutant, that is bigger than the center of the algebra (as long as the result states are not disjoint). The big commutant will also be type I and will correspond to the baby universe algebra.

When the inverse temperature $\beta$ is small, a phase transition like the Hawking-Page transition happens for the above state and the $O(N^{a})$ entropy and the diverging $Z$ indicate the appearance of a black hole horizon. Here the algebra of simple operators completed with respect to the states along the physically distinguishable free ultrafilters will be type III$_{1}$ von Neumann algebras. Thus the commutant will also a type III von Neumann algebra. However, if the resulting state along the free ultrafilters are not disjoint, the interior region will be bigger that the interior region of the thermal state $\frac{1}{\sqrt{Z}}e^{-\beta H_{N}} \ket{\Psi}$.

\section{Some comments on relation to ultraproducts}\label{disc}

We have introduced in this paper the CFT$_{p}$ boundary theories in the large $N$ limit which include the usual large $N$ of holographic CFTs as a subsector. As we have discussed above when we say the large $N$ holographic CFT, we do not just mean the GFFs that arise in the large $N$ of operators around the vacuum or any other classical background in particular, but a theory that includes all of such sectors that correspond to a given background geometry and perturbative quantum (gravitational) fields acting on top of it. The CFT$_{p}$ theory extends this to include operators with oscillating or chaotic expectation values which may be heavy or $O(1)$. The new operators that we have included are operators that give the ultralimits of these chaotic operators along the free ultrafilter $p$. Thus CFT$_{p}$ can thought as the set of the ultralimits of these operators, in some sense. It would be interesting to see how CFT$_{p}$ fit similar objects in the standard mathematical literature such as the set of the ultralimits of states in a Hilbert space (called Hilbert space ultraproducts) and the ultralimits of the operators acting on these states (generally called von Neumann algebra ultraproducts). These mathematical notions are introduced in  appendix \ref{ultra}, specifically on the second subsection of the appendix.  

It is usually stated in the literature that the sequence of black hole microstates, for example energy eigen states at high enough energy ($O(N^{2})$ for $\mathcal{N}=4$ super Yang Mills for instance), do not have a large $N$ limit. This is primarily because the black hole Hilbert space does not have a large $N$ limit therefore there is no vector or state in general this sequence of states converge to. This statement is made however under the assumption that the Hilbert space should be a separable Hilbert space\footnote{Separable Hilbert spaces as Hilbert spaces generated by a countably many basis vectors.}. There are of course several reasons why separable Hilbert spaces are preferred \cite{Witten:2021jzq,Streater:1989vi} however we do not expect the large $N$ holographic CFT we described in the previous paragraph to be described by a separable Hilbert space (see \cite{Liu:2026fnd} for a recent related discussion), even though each of the sectors on a given background geometry correspond to a separable Hilbert space. In particular, the above sequence of energy eigen states will have a large $N$ limit in the Hilbert space ultraproduct. Each of the states in the sequence are states $(\ket{E}_{n})_{n\in\mathbb{N}}$ in the corresponding Hilbert space, $\mathcal{H}_{n}$, of for instance the $\mathcal{N}=4$ super Yang Mills theory with gauge group $SU(n+1)$, i.e, $n$ is the rank of the gauge group. The limiting state which we call $\ket{E_{p}}$ is a state in the Hilbert space ultraproduct $\mathcal{H}_{p}$, and it is an equivalence class of sequences of states like $(\ket{E_{n}}+\ket{\psi_{n}})_{n}$ so that 
\begin{equation}
    \lim_{n\rightarrow p} |\ket{\psi_{n}}| = \lim_{n\rightarrow p} \sqrt{\langle\psi_{n}|\psi_{n}\rangle}=0
\end{equation}
Of course, the states in the large $N$ holographic CFT are also included in this Hilbert space. More precisely, each of the Hilbert spaces corresponding to the QFT on the fixed background discussed in the standard large $N$ holographic CFT are isometrically embedded in $\mathcal{H}_{p}$. The reason is that inner products in $\mathcal{H}_{p}$ are given by
\begin{equation}
    \langle\psi_{p}|\phi_{p}\rangle = \lim_{n\rightarrow p}\langle\psi_{n}|\phi_{n}\rangle
\end{equation}
and if $\lim_{n\rightarrow \infty}\langle\psi_{n}|\phi_{n}\rangle$ is well defined then $\lim_{n\rightarrow p}\langle\psi_{n}|\phi_{n}\rangle$ always agrees with it as long as $p$ is a free ultrafilter. The general expectation for the exponentially small overlap for semi-classically distinct backgrounds $\Psi$ and $\Phi$,
\begin{equation}
  \langle\Phi|\Psi\rangle = O(e^{-fN^{2}})  
\end{equation}
where $f$ is $O(1)$ is also translates to the inner product of the corresponding state given by an exponentially small infinitesimal number,
\begin{equation}
    \langle\psi_{p}|\phi_{p}\rangle = [e^{-fN^{2}}].
\end{equation}
Similarly, the different separable sectors of the CFT$_{p}$ theory are also isometrically embedded in $\mathcal{H}_{p}$. $\mathcal{H}_{p}$ is a highly non separable Hilbert spaces which is consistent with the fact that roughly speaking CFT$_{p}$ and black hole microstates in the large $N$ limit are embedded in it.   

However we expect that $\mathcal{H}_{p}$ includes more states than that. To see this consider the von Neumann algebra given by Groh-Raynaud(GR) ultraproduct, which is given by the strong closure (within $\mathcal{B(H}_{p})$) of the representation of operators that act point-wise on the sequence of Hilbert spaces $(\mathcal{H}_{n})_{n}$ like \eqref{operator act}. This also includes non convergent sequences of operators (not just expectation values) with no relation whatsoever between consecutive terms in the sequence of the operators for instance $a_{i}= e^{it\text{Tr}F^{2}}$ while $a_{i+1}= e^{it\text{Tr}\Phi^{2}}$. It is not clear what role such sequences have in the large $N$ limit of AdS/CFT however Groh-Raynaud(GR) ultraproduct includes them. In addition states created by acting with these operators are also in $\mathcal{H}_{p}$. As a concluding remark we note that Ocneanu ultraproduct, which is a small corner of the GR ultraproduct may play a role in the wormhole story or in general in AdS/CFT.  

\appendix

\section{The Stone–$\check{\text{C}}$ech compactification, ultrafilters and ultraproducts}\label{ultra}

Here we discuss some of the properties of the Stone–$\check{\text{C}}$ech compactification, $\beta\mathbb{N}$, of the set of natural numbers from the perspective ultrafilters, and ultraproducts which were mentioned in section \ref{disc}. Interested reader may find more detailed discussions on the topic on these references, which we closely follow here, \cite{hindman1998algebra,ando2014ultraproducts} among others. We will not proving most of the theorems, the goal is to precisely state the theorems and the definition of the objects used in the main body. 

We have introduced the definition of ultrafilters in the main body of the paper and we revise it here.
\begin{definition}
    A proper filter $p$ on $\mathbb{N}$ is a non empty subset of the power set of $\mathbb{N}$, $\mathcal{P}(\mathbb{N})$, such that,
    \begin{enumerate}
        \item if $A$ and $B$ are in $p$, then $A \cap B$ is in $p$,
        \item if $A$ is in $p$ and $A\subseteq B \subseteq \mathbb{N}$, then $B\in p$,
        \item $\emptyset \notin p$
    \end{enumerate}
\end{definition}
A set that satisfies $1$ and $2$ but contains the empty set, like $\mathcal{P}(\mathbb{N})$, is called an improper filter. We never discuss improper filters beyond this point and when we say filters in the following discussion, we actually mean proper filters.  
\begin{definition}
    An ultrafilter $p$ on $\mathbb{N}$ is a filter that is not properly contained in any other filter on $\mathbb{N}$. 
\end{definition}

A set of sets like $p$ is said to have the finite intersection property if and only if any finite intersection of its elementary sets is non empty, i.e, $\cap \mathcal{X} \neq \emptyset$ for any finite $\mathcal{X} \subseteq p$. 

\begin{theorem}
    A set of subsets of $\mathbb{N}$, $p$, is an ultrafilter if and only if it satisfies the finite intersection property and for each $A \in \mathcal{P}(\mathbb{N})\backslash p$, there is $B \in p$ such that $A\cap B=\emptyset$. In addition, $p$ is the maximal subset of $\mathcal{P}(\mathbb{N})$ with the finite intersection property.
\end{theorem}

Following the above theorem it is clear that for $x\in \mathbb{N}$, the set $\{A\in \mathcal{P}(\mathbb{N})|x\in A\}$ is an ultrafilter. Such an ultrafilter is called a principal ultrafilter.

\begin{theorem}
    An ultrafilter $e(x)$ is a principal ultrafilter if and only if $\cap\; e(x) \neq \emptyset$, in particular, there is some $x\in \mathbb{N}$ such that $\cap\; e(x)=\{x\}$.
\end{theorem}
As discussed in section \ref{sec 3}, within the Zermelo-Fraenkel set theory, only such kinds of ultrafilter can be constructed. However, with the axiom of choice one can construct many more ultrafilters called non principal/free ultrafilters. In particular we have the following theorem,
\begin{theorem}
    Let $\mathcal{X}$ be any subset of $\mathcal{P}(\mathbb{N})$ with the finite intersection property. Then, there is an ultrafilter $p$ such that $\mathcal{X}\subseteq p$.
\end{theorem}
In proving the above theorem one uses Zorn's lemma\footnote{The theorem and proof can be found in \cite{hindman1998algebra} as theorem 3.8.}, which is equivalent to the axiom of choice within the Zermelo-Fraenkel set theory.

\begin{definition}
    Let $p$ is an ultrafilter on $\mathbb{N}$, then the norm of $p$, $|p|$, is defined as;
    \begin{equation}
        |p|= \text{min }\{|A|, \text{ such that } A\in p\}.
    \end{equation}
\end{definition}

We can see that $|p|$ is either $1$, for a principal ultrafilter, or infinite for a free ultrafilter.

In particular we have the following theorem,
\begin{theorem}
    If $\mathcal{X}$ is a subset of $\mathcal{P}(\mathbb{N})$, and the intersection of any finite number of elements of $\mathcal{X}$ is infinite; then $\mathcal{X}$ is contained in an ultrafilter with all of its members infinite, in other words, there is a free ultrafilter, $p$, such that $\mathcal{X}\subseteq p$.   
\end{theorem}

\subsection{$\beta\mathbb{N}$ as the Stone–$\check{\text{C}}$ech compactification of $\mathbb{N}$}

First we define the symbol $\beta\mathbb{N}$ as follows,
\begin{definition}
\begin{align}
    \beta\mathbb{N} &= \{p| p \text{ is an ultrafilter on }\mathbb{N}\},\\
    \text{For } A& \subseteq \mathbb{N}, \beta A = \{p| p\in \beta\mathbb{N} \text{ and }A \in p\}.
\end{align}
    
\end{definition}

We note the following important properties 
\begin{lemma}
For any $A,B \subseteq \mathbb{N}$, 
    \begin{enumerate}
        \item $\beta(A\cap B) = \beta A \cap \beta B$,
        \item $\beta(A\cup B) = \beta A \cup \beta B$,
        \item $\beta A = \emptyset$ if and only if $A=\emptyset$,
        \item $\beta A = \beta\mathbb{N}$ if and only if $A=\mathbb{N}$.
    \end{enumerate}
\end{lemma}

Thus, the set $\{\beta A| A \subseteq\mathbb{N}\}$ can be taken to be the base for a topology on $\beta\mathbb{N}$. With this topology we have the following,
\begin{theorem}
.
    \begin{enumerate}
        \item $\beta\mathbb{N}$ is a compact Hausdorff space.
        \item The sets of the form $\beta A$, for $A\subseteq \mathbb{N}$, are the clopen subsets of $\beta\mathbb{N}$.
        \item The set $\{e(x) \in \beta\mathbb{N}| \text{ for } x \in \mathbb{N}\}$ of principal ultrafilters are the isolated points in $\beta\mathbb{N}$ and are dense in $\beta\mathbb{N}$.
    \end{enumerate}
\end{theorem}
The space $\beta\mathbb{N}$ is called Hausdorff if for distinct elements, $p$ and $q$, of $\beta\mathbb{N}$, there are two disjoint open sets of $\beta\mathbb{N}$ containing each $p$ and $q$. By clopen sets we mean subsets of $\beta\mathbb{N}$ that are both closed and open. In addition, an isolated point in $\beta\mathbb{N}$ is a point for which a non empty open set exists such that the open set includes only that point. In the context of above theorem $6.3$, for the point $e(x)\in \beta\mathbb{N}$, the open set $\beta x$ (by which we mean $\beta A$ where $A=\{x\}$) only includes $e(x)$. Conversely, assume $\beta A$ is a non empty open subset of $\beta\mathbb{N}$ that only includes the isolated point $\{p\}$. Then since $\beta A$ is non empty, for $x \in A$, $e(x)\in \beta A$. Therefore $p =e(x)$. The set of principal ultrafilters are also dense in $\beta\mathbb{N}$ since, for any non empty open subset $\beta A$, there is at least one $e(x)$ such that $\{e(x)\} \cap \beta A \neq \emptyset$. 

An important point is that the set of natural numbers $\mathbb{N}$ can be embedded in $\beta\mathbb{N}$ in a particular way such that $\beta\mathbb{N}$ can be taken as the Stone–$\check{\text{C}}$ech compactification of $\mathbb{N}$. By an embedding of $\mathbb{N}$ into $\beta\mathbb{N}$ we mean there is a function $\varphi$ from $\mathbb{N}$ to $\beta\mathbb{N}$ such that it is homeomorphic (i.e, bijective, continuous and with an inverse that is also continuous) from $\mathbb{N}$ onto $\varphi(\mathbb{N})$. 

On the other hand, for a topological space like $\mathbb{N}$, the Stone–$\check{\text{C}}$ech compactification is defined as follows,

\begin{definition}
    Let $X$ be a completely regular topological space. Then the Stone–$\check{\text{C}}$ech compactification of $X$ is a pair $(\varphi,Z)$ such that $\varphi$ is an embedding of $X$ onto a compact space $Z$ while $\varphi(X)$ is dense in $Z$. In addition, for any compact space $Y$ and any continuous function $f: X \rightarrow Y$, there is a continuous function $g: Z\rightarrow Y$ such that $f = g\; \circ\;\varphi $.  
\end{definition}

In addition the Stone–$\check{\text{C}}$ech compactification provides the maximal or largest compactification of any completely regular (and Hausdorff) space. Now we are in a position to state the central theorem of the subsection.

\begin{theorem}
    The space $\beta\mathbb{N}$ is the Stone–$\check{\text{C}}$ech compactification of $\mathbb{N}$ with the embedding of $\mathbb{N}$ onto $\beta\mathbb{N}$ given by $x \mapsto e(x)$.
\end{theorem}

Next we move on to discussions on ultralimits, we note that they can be seen as the standard limit on the space $\beta\mathbb{N}$. But, we start with revising their definition.

\begin{definition}
    Let $p\in\beta\mathbb{N}$, and $(a_{n})_{n\in\mathbb{N}}$ a sequence in a topological space $X$, with some $y\in X$. Then, $p-\lim a_{n} = y$ if and only if for any neighborhood $U$ of $y$, $\{n | a_{n}\in U\} \in p$. And $p-\lim$ denotes an ultralimit along an ultrafilter $p$. 
\end{definition}

We also recall the following definition of limits of functions on a topological space. 

\begin{definition}
    Let $X$ and $Y$ be topological spaces and $f: A \rightarrow Y$ where $A\subseteq X$, then we say $\lim_{x\rightarrow a}f(a)=y$, for $x \in \bar{A}$ and $y\in Y$, if and only if for any neighborhood $V$ of $y$, there is a neighborhood $U$ of $X$ such that $f(A\cap U)\in V$.  
\end{definition}

Then we have the following important theorem,
\begin{theorem}\label{limm}
    Let $Y$ be a topological space, while $p\in\beta\mathbb{N}$ and $y \in Y$. If $A\in p$ and a function $f: A \rightarrow Y$, $p-\lim_{a\in A} f(a) = y$ if and only if $\lim_{a\rightarrow p}f(a)=y$. In addition, if $Y$ is compact, $p-\lim f(a)$ always exists. And when it exists, it is unique. 
\end{theorem}

Here, obviously, we can take $A$ to be the full $\mathbb{N}$ and we have identified $a\in \mathbb{N}$ with $e(a)\in \beta\mathbb{N}$. 

\subsection{Ultraproducts}

In the above theorem \ref{limm}, when the topological space $Y$ is not compact, the ultralimit, $p-\lim a_{n}$ or $\lim_{n\rightarrow p}a_{n}$ does not always exist in $Y$. However it exists in a much bigger space, which can be considered as an extension of the space $Y$. This extension is called the ultraproduct of $Y$ and it is, roughly speaking, the set $\{\lim_{n\rightarrow p}a_{n}\}$ for any sequence $(a_{n})_{n\in \mathbb{N}}$ in $Y$, convergent or not and bounded or not. For the case where $Y$ is the set of real numbers, the ultraproduct is called hyperreals and was discussed in the main body of the paper. The ultraproduct of the real numbers along some free ultrafilter $p$, $\mathbb{R}^{p}$ is defined as the infinite direct product $\prod_{\mathbb{N}}\mathbb{R}$ up to equivalence relation,
\begin{equation}
   \mathbb{R}^{p} =  \prod_{\mathbb{N}}\mathbb{R}/ \sim
\end{equation}
where, for any two elements of $\prod_{\mathbb{N}}\mathbb{R}$, $(x_{i})_{i\in \mathbb{N}}$ and $(y_{i})_{i\in \mathbb{N}}$, $\sim$ is defined by
\begin{equation}
    (x_{i}) \sim (y_{i}) \text{ if and only if } \{n| x_{n} = y_{n}\} \in p.
\end{equation}
elementary operations in $\mathbb{R}$ can be extended to $\mathbb{R}^{p}$ and $\mathbb{R}$ itself can be embedded onto $\mathbb{R}^{p}$ by the constant sequence $x \mapsto (x,x, \cdots)$, as we mentioned previously.  

We have assumed above that the infinite product is of the same space $Y$. However, this is not necessarily the case. In fact an astonishing advantage of ultralimits is the possibility to take limits of sequences with terms of the sequence belonging to distinct spaces. There is no sense in which the standard limit of such a sequence exists, however along a given ultrafilter, the ultralimit of such a sequence exists in the ultraproduct space.  The spaces could be fields or groups, however we are more concerned with Hilbert spaces and von Neumann algebras which would correspond to the Hilbert spaces of the boundary CFT, say $\mathcal{N}=4$ super Yang Mills theory in four dimensions with $SU(N)$ gauge group, and the algebra of observables of the CFT acting on the Hilbert space respectively. In the following $n=N-1$ is the rank of the gauge group. 

\begin{definition}\label{ultrapro}
    Let $(\mathcal{H}_{n})_{n\in\mathbb{N}}$ be a sequence of Hilbert spaces, $p$ be a free ultrafilter and let the space $l^{\infty}(\mathbb{N},\mathcal{H}_{n})$ be the space sequences of vectors states like $(\psi_{n})_{n} \in \Pi_{n=1}^{\infty} \mathcal{H}_{n}$, such that sup$_{n}|\psi_{n}|<\infty$, where $|\psi_{n}|$ is the norm of the state vector. The Hilbert space ultraproduct is $\mathcal{H}_{p}:=l^{\infty}(\mathbb{N},\mathcal{H}_{n})/ \mathcal{I}_{p}$, where $\mathcal{I}_{p}$ is a closed subspace of sequences $(\psi_{n})_{n}$ with $\lim_{n\rightarrow p}|\psi_{n}|=0$ .
\end{definition}

The Hilbert space ultraproduct $\mathcal{H}_{p}$ is itself a Hilbert space with the inner product between elements $\psi_{p} = [\psi_{n}]_{n}$ and $\phi_{p}= [\phi_{n}]_{n}$\footnote{The square bracket is to represent the equivalence class.} given by
\begin{equation}\label{operator act}
   \langle\psi_{p}|\phi_{p}\rangle = \lim_{n\rightarrow p}\langle\psi_{n}|\phi_{n}\rangle
\end{equation}

Unfortunately, the ultraproduct of $(\mathcal{B(H}_{n}))_{n}$, the sequence of the algebra of all bounded observables acting on each $\mathcal{H}_{n}$, that one would naively define similarly as above, will not be a von Neumann algebra. What we mean by similarly as above is that we consider a sequence of operators with terms in each $\mathcal{B(H}_{n})$ rather than states vectors in each $\mathcal{H}_{n}$, and replace the norm of the state vector by the norm of the operators in definition \ref{ultrapro}. The resulting space $\mathcal{B(H})_{p}$ is not a von Neumann algebra even though it is a Banach space.

A slightly more straight forward procedure to construct a von Neumann algebra from $\mathcal{B(H})_{p}$ is what is called Groh-Raynaud(GR) ultraproduct. Here we consider a representation of the operators in $\mathcal{B(H})_{p}$ as bounded operators acting pointwise on the Hilbert space ultraproduct, $\mathcal{H}_{p}$. That is we define 
\begin{equation}
    \pi_{p} (a_{p}) \psi_{p} = \pi_{p}([a_{n}]_{n})[\psi_{n}]_{n} :=[a_{n}\psi_{n}]_{n} = (a\psi)_{p}
\end{equation}
where $a_{p} \in \mathcal{B(H})_{p}$ and $\psi_{p}\in\mathcal{H}_{p}$. 
\begin{lemma}
    The representation $\pi_p{}$ defined as above is injective and $\pi_{p}(\mathcal{B(H})_{p})$ is strongly dense in $\mathcal{B(H}_{p})$. 
\end{lemma}

Following the above lemma we see that the strong operator closure of $\pi_{p}(\mathcal{B(H})_{p})$ within $\mathcal{B(H})_{p}$ is a von Neumann algebra and is in fact $\mathcal{B(H})_{p}$ itself. Furthermore, for a sequence of von Neumann algebras $(M_{n})_{n}$ with $M_{n}\subset \mathcal{B(H}_{n})$, the ultraproduct given by the strong operator closure of $\pi_{p}(M_{p})$ within $\mathcal{B(H})_{p}$ is also a von Neumann algebra.  

\begin{definition}
    Given a sequence of von Neumann algebras $(M_{n})_{n}$, one considers the sequence of Hilbert spaces provided by the standard representation of $M_{n}$, i.e, $\mathcal{H}_{n} = L^{2}(M_{n})$. The Groh-Raynaud ultraproduct is the ultraproduct constructed by the closure of $\pi_{p}(M_{p})$, now constructed with respect to the sequence of the standard representations, as above.
\end{definition}

One can look into a small corner of the GR ultraproduct \cite{ando2014ultraproducts} by introducing the Ocneanu ultraproduct. The main distinction of the Ocneanu ultraproduct, from the ultraproduct we constructed above, is that it is defined for a sequence of von Neumann algebras with a normal faithful state. Most of the von Neumann algebras appearing in physics are of this type. Then the ideal $\mathcal{I}_{p}$ in construction of the Ocneanu ultraproduct is defined with respect to the sequence of this particular state, while instead of $l^{\infty}(\mathbb{N}, M_{n})$ we consider a subspace of which this $\mathcal{I}_{p}$ is an ideal.   

\begin{definition}
    Let $(M_{n},\omega_{n})_{n\in \mathbb{N}}$ is a sequence of von Neumann algebras with a faithful normal state $\omega_{n}$ on $M_{n}$ for each $n$, and let $p$ be a free ultrafilter on $\mathbb{N}$, then define
    \begin{equation}
        \mathcal{I}_{p} = \{(a_{n})_{n} \in l^{\infty}(\mathbb{N},M_{n}) \text{, such that } \lim_{n\rightarrow p}\sqrt{\omega_{n}(a^{\dagger}_{n}a_{n} + a_{n}a_{n}^{\dagger})}=0\}.
    \end{equation}
In addition, define
\begin{equation}
    \mathcal{M}_{p} =  \{(a_{n})_{n} \in l^{\infty}(\mathbb{N},M_{n}) \text{, such that } (a_{n})_{n} \mathcal{I}_{p}\subset \mathcal{I}_{p} \text{ and } \mathcal{I}_{p}(a_{n})_{n}\subset \mathcal{I}_{p}\}.
\end{equation}

The Ocneanu ultraproduct $(M_{n},\omega_{n})_{p}$ is thus defined as

\begin{equation}
    (M_{n},\omega_{n})^{p} = \mathcal{M}_{p}/\mathcal{I}_{p}
\end{equation}
\end{definition}

Obviously both $\mathcal{M}_{p}$ and $\mathcal{I}_{p}$ depend on the sequence $(M_{n},\omega_{n})_{n\in \mathbb{N}}$ even though we have not explicitly written it above. The `weird' definition of $\mathcal{I}_{p}$ involving $\omega_{n}(a^{\dagger}_{n}a_{n} + a_{n}a_{n}^{\dagger})$ is so that $\mathcal{I}_{p}$ is both left and right ideal.   

A proof by Ocneanu \cite{ocneanu2006actions} implies that

\begin{theorem}
    $(M_{n},\omega_{n})^{p}$ defined above is a $W^{*}$ algebra and a state $\omega_{p}$ on  $(M_{n},\omega_{n})^{p}$ defined as,
    \begin{equation}
        \omega_{p}(a^{p}) := \lim_{n\rightarrow p}\omega_{n}(a_{n}) \text{, for } a^{p}=[a_{n}]_{n} \in  (M_{n},\omega_{n})^{p},
    \end{equation}
is a faithful normal state. 
    
\end{theorem}

A $W^{*}$ algebra can be understood as a C$^{*}$ algebra with a distinguished folium of states. In particular, the GNS representation of a $W^{*}$ algebra with respect to an associated faithful normal state gives us a von Neumann algebra.  

\acknowledgments

I want to thank Amos Yarom for many discussions on related topics and also Kyriakos Papadodimas for useful discussions. The final parts of this work was performed at Aspen Center for Physics, which is supported by National Science Foundation grant PHY-2210452; and a grant from the Simons Foundation (1161654, Troyer).




\bibliographystyle{JHEP}
\bibliography{biblio}

@article{Penington:2019npb,
    author = "Penington, Geoffrey",
    title = "{Entanglement Wedge Reconstruction and the Information Paradox}",
    eprint = "1905.08255",
    archivePrefix = "arXiv",
    primaryClass = "hep-th",
    doi = "10.1007/JHEP09(2020)002",
    journal = "JHEP",
    volume = "09",
    pages = "002",
    year = "2020"
}

@article{Almheiri:2019hni,
    author = "Almheiri, Ahmed and Mahajan, Raghu and Maldacena, Juan and Zhao, Ying",
    title = "{The Page curve of Hawking radiation from semiclassical geometry}",
    eprint = "1908.10996",
    archivePrefix = "arXiv",
    primaryClass = "hep-th",
    doi = "10.1007/JHEP03(2020)149",
    journal = "JHEP",
    volume = "03",
    pages = "149",
    year = "2020"
}

@article{Almheiri:2019psf,
    author = "Almheiri, Ahmed and Engelhardt, Netta and Marolf, Donald and Maxfield, Henry",
    title = "{The entropy of bulk quantum fields and the entanglement wedge of an evaporating black hole}",
    eprint = "1905.08762",
    archivePrefix = "arXiv",
    primaryClass = "hep-th",
    doi = "10.1007/JHEP12(2019)063",
    journal = "JHEP",
    volume = "12",
    pages = "063",
    year = "2019"
}

@article{Kudler-Flam:2026nzz,
    author = "Kudler-Flam, Jonah and Witten, Edward",
    title = "{Wormholes and Averaging over N}",
    eprint = "2605.15180",
    archivePrefix = "arXiv",
    primaryClass = "hep-th",
    month = "5",
    year = "2026"
}

@article{Liu:2026fnd,
    author = "Liu, Hong",
    title = "{Ramp, Plateau, and Wormholes without Averaging, and Hyper-non-perturbative Structures in Gravity}",
    eprint = "2608.02743",
    archivePrefix = "arXiv",
    primaryClass = "hep-th",
    reportNumber = "MIT-CTP/6074",
    month = "8",
    year = "2026"
}

@article{Marolf:2020xie,
    author = "Marolf, Donald and Maxfield, Henry",
    title = "{Transcending the ensemble: baby universes, spacetime wormholes, and the order and disorder of black hole information}",
    eprint = "2002.08950",
    archivePrefix = "arXiv",
    primaryClass = "hep-th",
    doi = "10.1007/JHEP08(2020)044",
    journal = "JHEP",
    volume = "08",
    pages = "044",
    year = "2020"
}

@article{Antonini:2023hdh,
    author = "Antonini, Stefano and Sasieta, Martin and Swingle, Brian",
    title = "{Cosmology from random entanglement}",
    eprint = "2307.14416",
    archivePrefix = "arXiv",
    primaryClass = "hep-th",
    doi = "10.1007/JHEP11(2023)188",
    journal = "JHEP",
    volume = "11",
    pages = "188",
    year = "2023"
}

@article{Minahan:2010js,
    author = "Minahan, Joseph A.",
    title = "{Review of AdS/CFT Integrability, Chapter I.1: Spin Chains in N=4 Super Yang-Mills}",
    eprint = "1012.3983",
    archivePrefix = "arXiv",
    primaryClass = "hep-th",
    reportNumber = "UUITP-38-10",
    doi = "10.1007/s11005-011-0522-9",
    journal = "Lett. Math. Phys.",
    volume = "99",
    pages = "33--58",
    year = "2012"
}

@article{Sieg:2010jt,
    author = "Sieg, C.",
    title = "{Review of AdS/CFT Integrability, Chapter I.2: The spectrum from perturbative gauge theory}",
    eprint = "1012.3984",
    archivePrefix = "arXiv",
    primaryClass = "hep-th",
    reportNumber = "HU-MATH-2010-23, HU-EP-10-88",
    doi = "10.1007/s11005-011-0508-7",
    journal = "Lett. Math. Phys.",
    volume = "99",
    pages = "59--84",
    year = "2012"
}

@article{Rej:2010ju,
    author = "Rej, Adam",
    title = "{Review of AdS/CFT Integrability, Chapter I.3: Long-range spin chains}",
    eprint = "1012.3985",
    archivePrefix = "arXiv",
    primaryClass = "hep-th",
    reportNumber = "IMPERIAL-TP-AR-2010-2",
    doi = "10.1007/s11005-011-0509-6",
    journal = "Lett. Math. Phys.",
    volume = "99",
    pages = "85--102",
    year = "2012"
}

@article{Tseytlin:2010jv,
    author = "Tseytlin, A. A.",
    title = "{Review of AdS/CFT Integrability, Chapter II.1: Classical AdS5xS5 string solutions}",
    eprint = "1012.3986",
    archivePrefix = "arXiv",
    primaryClass = "hep-th",
    reportNumber = "IMPERIAL-TP-AT-2010-05",
    doi = "10.1007/s11005-011-0466-0",
    journal = "Lett. Math. Phys.",
    volume = "99",
    pages = "103--125",
    year = "2012"
}

@article{McLoughlin:2010jw,
    author = "McLoughlin, Tristan",
    title = "{Review of AdS/CFT Integrability, Chapter II.2: Quantum Strings in AdS5xS5}",
    eprint = "1012.3987",
    archivePrefix = "arXiv",
    primaryClass = "hep-th",
    reportNumber = "AEI-2010-177",
    doi = "10.1007/s11005-011-0510-0",
    journal = "Lett. Math. Phys.",
    volume = "99",
    pages = "127--148",
    year = "2012"
}

@article{Magro:2010jx,
    author = "Magro, Marc",
    title = "{Review of AdS/CFT Integrability, Chapter II.3: Sigma Model, Gauge Fixing}",
    eprint = "1012.3988",
    archivePrefix = "arXiv",
    primaryClass = "hep-th",
    reportNumber = "AEI-2010-132",
    doi = "10.1007/s11005-011-0481-1",
    journal = "Lett. Math. Phys.",
    volume = "99",
    pages = "149--167",
    year = "2012"
}

@article{Schafer-Nameki:2010qho,
    author = "Schafer-Nameki, Sakura",
    title = "{Review of AdS/CFT Integrability, Chapter II.4: The Spectral Curve}",
    eprint = "1012.3989",
    archivePrefix = "arXiv",
    primaryClass = "hep-th",
    reportNumber = "KCL-MTH-10-18",
    doi = "10.1007/s11005-011-0525-6",
    journal = "Lett. Math. Phys.",
    volume = "99",
    pages = "169--190",
    year = "2012"
}

@article{Schlenker:2022dyo,
    author = "Schlenker, Jean-Marc and Witten, Edward",
    title = "{No ensemble averaging below the black hole threshold}",
    eprint = "2202.01372",
    archivePrefix = "arXiv",
    primaryClass = "hep-th",
    doi = "10.1007/JHEP07(2022)143",
    journal = "JHEP",
    volume = "07",
    pages = "143",
    year = "2022"
}

@article{Saad:2019lba,
    author = "Saad, Phil and Shenker, Stephen H. and Stanford, Douglas",
    title = "{JT gravity as a matrix integral}",
    eprint = "1903.11115",
    archivePrefix = "arXiv",
    primaryClass = "hep-th",
    month = "3",
    year = "2019"
}

@article{Stanford:2019vob,
    author = "Stanford, Douglas and Witten, Edward",
    title = "{JT gravity and the ensembles of random matrix theory}",
    eprint = "1907.03363",
    archivePrefix = "arXiv",
    primaryClass = "hep-th",
    doi = "10.4310/ATMP.2020.v24.n6.a4",
    journal = "Adv. Theor. Math. Phys.",
    volume = "24",
    number = "6",
    pages = "1475--1680",
    year = "2020"
}

@article{Coleman:1988cy,
    author = "Coleman, Sidney R.",
    title = "{Black holes as red herrings: Topological fluctuations and the loss of quantum coherence}",
    reportNumber = "HUTP-88/A008",
    doi = "10.1016/0550-3213(88)90110-1",
    journal = "Nucl. Phys. B",
    volume = "307",
    pages = "867--882",
    year = "1988"
}

@article{Giddings:1988wv,
    author = "Giddings, Steven B. and Strominger, Andrew",
    title = "{Baby Universes, Third Quantization and the Cosmological Constant}",
    reportNumber = "HUTP-88/A036",
    doi = "10.1016/0550-3213(89)90353-2",
    journal = "Nucl. Phys. B",
    volume = "321",
    pages = "481--508",
    year = "1989"
}

@article{Klebanov:1988eh,
    author = "Klebanov, Igor R. and Susskind, Leonard and Banks, Tom",
    title = "{Wormholes and the Cosmological Constant}",
    reportNumber = "SLAC-PUB-4705",
    doi = "10.1016/0550-3213(89)90538-5",
    journal = "Nucl. Phys. B",
    volume = "317",
    pages = "665--692",
    year = "1989"
}

@article{Arkani-Hamed:2007cpn,
    author = "Arkani-Hamed, Nima and Orgera, Jacopo and Polchinski, Joseph",
    title = "{Euclidean wormholes in string theory}",
    eprint = "0705.2768",
    archivePrefix = "arXiv",
    primaryClass = "hep-th",
    doi = "10.1088/1126-6708/2007/12/018",
    journal = "JHEP",
    volume = "12",
    pages = "018",
    year = "2007"
}

@article{Giddings:1987cg,
    author = "Giddings, Steven B. and Strominger, Andrew",
    title = "{Axion Induced Topology Change in Quantum Gravity and String Theory}",
    reportNumber = "HUTP-87-A067",
    doi = "10.1016/0550-3213(88)90446-4",
    journal = "Nucl. Phys. B",
    volume = "306",
    pages = "890--907",
    year = "1988"
}

@article{Maldacena:2004rf,
    author = "Maldacena, Juan Martin and Maoz, Liat",
    title = "{Wormholes in AdS}",
    eprint = "hep-th/0401024",
    archivePrefix = "arXiv",
    reportNumber = "ITFA-2003-57",
    doi = "10.1088/1126-6708/2004/02/053",
    journal = "JHEP",
    volume = "02",
    pages = "053",
    year = "2004"
}

@article{Bergman:2007ss,
    author = "Bergman, Aaron and Distler, Jacques",
    title = "{Wormholes in Maximal Supergravity}",
    eprint = "0707.3168",
    archivePrefix = "arXiv",
    primaryClass = "hep-th",
    month = "7",
    year = "2007"
}

@article{Bergshoeff:2004pg,
    author = "Bergshoeff, E. and Collinucci, Andres and Gran, U. and Roest, D. and Vandoren, S.",
    editor = "Kiritsis, E.",
    title = "{Non-extremal instantons and wormholes in string theory}",
    eprint = "hep-th/0412183",
    archivePrefix = "arXiv",
    reportNumber = "KCL-MTH-04-16, ITP-UU-04-51, SPIN-04-33, UG-04-04",
    doi = "10.1002/prop.200410227",
    journal = "Fortsch. Phys.",
    volume = "53",
    pages = "990--996",
    year = "2005"
}

@article{Loges:2023ypl,
    author = "Loges, Gregory J. and Shiu, Gary and Van Riet, Thomas",
    title = "{A 10d construction of Euclidean axion wormholes in flat and AdS space}",
    eprint = "2302.03688",
    archivePrefix = "arXiv",
    primaryClass = "hep-th",
    reportNumber = "KEK-TH-2495",
    doi = "10.1007/JHEP06(2023)079",
    journal = "JHEP",
    volume = "06",
    pages = "079",
    year = "2023"
}

@article{Hertog:2017owm,
    author = "Hertog, Thomas and Trigiante, Mario and Van Riet, Thomas",
    title = "{Axion Wormholes in AdS Compactifications}",
    eprint = "1702.04622",
    archivePrefix = "arXiv",
    primaryClass = "hep-th",
    doi = "10.1007/JHEP06(2017)067",
    journal = "JHEP",
    volume = "06",
    pages = "067",
    year = "2017"
}

@article{Astesiano:2023iql,
    author = "Astesiano, Davide and Gautason, Fridrik Freyr",
    title = "{Supersymmetric Wormholes in String Theory}",
    eprint = "2309.02481",
    archivePrefix = "arXiv",
    primaryClass = "hep-th",
    doi = "10.1103/PhysRevLett.132.161601",
    journal = "Phys. Rev. Lett.",
    volume = "132",
    number = "16",
    pages = "161601",
    year = "2024"
}

@article{Krasnov:2006jb,
    author = "Krasnov, Kirill and Schlenker, Jean-Marc",
    title = "{On the renormalized volume of hyperbolic 3-manifolds}",
    eprint = "math/0607081",
    archivePrefix = "arXiv",
    doi = "10.1007/s00220-008-0423-7",
    journal = "Commun. Math. Phys.",
    volume = "279",
    pages = "637--668",
    year = "2008"
}

@article{Banados:1992wn,
    author = "Banados, Maximo and Teitelboim, Claudio and Zanelli, Jorge",
    title = "{The Black hole in three-dimensional space-time}",
    eprint = "hep-th/9204099",
    archivePrefix = "arXiv",
    reportNumber = "PRINT-92-0151 (CHILE), IASSNS-HEP-92-29",
    doi = "10.1103/PhysRevLett.69.1849",
    journal = "Phys. Rev. Lett.",
    volume = "69",
    pages = "1849--1851",
    year = "1992"
}

@article{Bahiru:2023zlc,
    author = "Bahiru, Eyoab and Belin, Alexandre and Papadodimas, Kyriakos and Sarosi, Gabor and Vardian, Niloofar",
    title = "{Holography and localization of information in quantum gravity}",
    eprint = "2301.08753",
    archivePrefix = "arXiv",
    primaryClass = "hep-th",
    reportNumber = "CERN-TH-2023-003",
    doi = "10.1007/JHEP05(2024)261",
    journal = "JHEP",
    volume = "05",
    pages = "261",
    year = "2024"
}

@book{keisler2012elementary,
  title={Elementary calculus: An infinitesimal approach},
  author={Keisler, H Jerome},
  year={2012},
  publisher={Courier Corporation}
}

@article{Hawking:1978jz,
    author = "Hawking, S. W.",
    title = "{Quantum Gravity and Path Integrals}",
    doi = "10.1103/PhysRevD.18.1747",
    journal = "Phys. Rev. D",
    volume = "18",
    pages = "1747--1753",
    year = "1978"
}

@article{Dorigoni:2024dhy,
    author = "Dorigoni, Daniele and Treilis, Rudolfs",
    title = "{Large-N integrated correlators in $ \mathcal{N} $ = 4 SYM: when resurgence meets modularity}",
    eprint = "2405.10204",
    archivePrefix = "arXiv",
    primaryClass = "hep-th",
    doi = "10.1007/JHEP07(2024)235",
    journal = "JHEP",
    volume = "07",
    pages = "235",
    year = "2024"
}

@article{Dorigoni:2014hea,
    author = "Dorigoni, Daniele",
    title = "{An Introduction to Resurgence, Trans-Series and Alien Calculus}",
    eprint = "1411.3585",
    archivePrefix = "arXiv",
    primaryClass = "hep-th",
    reportNumber = "DAMTP-2014-44",
    doi = "10.1016/j.aop.2019.167914",
    journal = "Annals Phys.",
    volume = "409",
    pages = "167914",
    year = "2019"
}

@article{Aniceto:2018bis,
    author = "Aniceto, In{\^e}s and Basar, Gokce and Schiappa, Ricardo",
    title = "{A Primer on Resurgent Transseries and Their Asymptotics}",
    eprint = "1802.10441",
    archivePrefix = "arXiv",
    primaryClass = "hep-th",
    reportNumber = "NSF-ITP-17-153",
    doi = "10.1016/j.physrep.2019.02.003",
    journal = "Phys. Rept.",
    volume = "809",
    pages = "1--135",
    year = "2019"
}

@article{Dyson:1952tj,
    author = "Dyson, F. J.",
    title = "{Divergence of perturbation theory in quantum electrodynamics}",
    doi = "10.1103/PhysRev.85.631",
    journal = "Phys. Rev.",
    volume = "85",
    pages = "631--632",
    year = "1952"
}

@article{Lipatov:1976ny,
    author = "Lipatov, L. N.",
    title = "{Divergence of the perturbation-theory series and the quasi-classical theory}",
    reportNumber = "LENINGRAD-76-255",
    journal = "Sov. Phys. JETP",
    volume = "45",
    pages = "216--223",
    year = "1977"
}

@article{caliceti2007useful,
  title={From useful algorithms for slowly convergent series to physical predictions based on divergent perturbative expansions},
  author={Caliceti, Emanuela and Meyer-Hermann, Michael and Ribeca, Paolo and Surzhykov, Andrey and Jentschura, Ulrich D},
  journal={Physics reports},
  volume={446},
  number={1-3},
  pages={1--96},
  year={2007},
  publisher={Elsevier}
}

@article{Kudler-Flam:2025cki,
    author = "Kudler-Flam, Jonah and Witten, Edward",
    title = "{Emergent mixed states for baby universes and black holes}",
    eprint = "2510.06376",
    archivePrefix = "arXiv",
    primaryClass = "hep-th",
    doi = "10.1007/JHEP05(2026)090",
    journal = "JHEP",
    volume = "05",
    pages = "090",
    year = "2026"
}

@book{bratteli2012operator,
  title={Operator algebras and quantum statistical mechanics: Volume 1: C*-and W*-Algebras. Symmetry Groups. Decomposition of States},
  author={Bratteli, Ola and Robinson, Derek William},
  year={2012},
  publisher={Springer Science \& Business Media}
}

@article{Goel:2018ubv,
    author = "Goel, Akash and Lam, Ho Tat and Turiaci, Gustavo J. and Verlinde, Herman",
    title = "{Expanding the Black Hole Interior: Partially Entangled Thermal States in SYK}",
    eprint = "1807.03916",
    archivePrefix = "arXiv",
    primaryClass = "hep-th",
    doi = "10.1007/JHEP02(2019)156",
    journal = "JHEP",
    volume = "02",
    pages = "156",
    year = "2019"
}

@article{Horowitz:1998ha,
    author = "Horowitz, Gary T. and Myers, Robert C.",
    title = "{The AdS / CFT correspondence and a new positive energy conjecture for general relativity}",
    eprint = "hep-th/9808079",
    archivePrefix = "arXiv",
    reportNumber = "NSF-ITP-98-076, MCGILL-98-13",
    doi = "10.1103/PhysRevD.59.026005",
    journal = "Phys. Rev. D",
    volume = "59",
    pages = "026005",
    year = "1998"
}

@article{Belin:2018fxe,
    author = "Belin, Alexandre and Lewkowycz, Aitor and S{\'a}rosi, G{\'a}bor",
    title = "{The boundary dual of the bulk symplectic form}",
    eprint = "1806.10144",
    archivePrefix = "arXiv",
    primaryClass = "hep-th",
    doi = "10.1016/j.physletb.2018.10.071",
    journal = "Phys. Lett. B",
    volume = "789",
    pages = "71--75",
    year = "2019"
}

@article{Leutheusser:2021frk,
    author = "Leutheusser, Samuel Aaron Wehlau and Liu, Hong",
    title = "{Emergent Times in Holographic Duality}",
    eprint = "2112.12156",
    archivePrefix = "arXiv",
    primaryClass = "hep-th",
    reportNumber = "MIT-CTP/5382",
    doi = "10.1103/PhysRevD.108.086020",
    journal = "Phys. Rev. D",
    volume = "108",
    number = "8",
    pages = "086020",
    year = "2023"
}

@article{Leutheusser:2021qhd,
    author = "Leutheusser, Samuel and Liu, Hong",
    title = "{Causal connectability between quantum systems and the black hole interior in holographic duality}",
    eprint = "2110.05497",
    archivePrefix = "arXiv",
    primaryClass = "hep-th",
    reportNumber = "MIT-CTP/5335",
    doi = "10.1103/PhysRevD.108.086019",
    journal = "Phys. Rev. D",
    volume = "108",
    number = "8",
    pages = "086019",
    year = "2023"
}

@article{Witten:2021unn,
    author = "Witten, Edward",
    title = "{Gravity and the crossed product}",
    eprint = "2112.12828",
    archivePrefix = "arXiv",
    primaryClass = "hep-th",
    doi = "10.1007/JHEP10(2022)008",
    journal = "JHEP",
    volume = "10",
    pages = "008",
    year = "2022"
}

@article{Furuya:2023fei,
    author = "Furuya, Keiichiro and Lashkari, Nima and Moosa, Mudassir and Ouseph, Shoy",
    title = "{Information loss, mixing and emergent type III$_{1}$ factors}",
    eprint = "2305.16028",
    archivePrefix = "arXiv",
    primaryClass = "hep-th",
    doi = "10.1007/JHEP08(2023)111",
    journal = "JHEP",
    volume = "08",
    pages = "111",
    year = "2023"
}

@inbook{Witten:2021jzq,
    author = "Witten, Edward",
    title = "{Why does quantum field theory in curved spacetime make sense? And what happens to the algebra of observables in the thermodynamic limit?}",
    eprint = "2112.11614",
    archivePrefix = "arXiv",
    primaryClass = "hep-th",
    doi = "10.1007/978-3-031-17523-7_11",
    year = "2022"
}

@article{Chandrasekaran:2022eqq,
    author = "Chandrasekaran, Venkatesa and Penington, Geoff and Witten, Edward",
    title = "{Large N algebras and generalized entropy}",
    eprint = "2209.10454",
    archivePrefix = "arXiv",
    primaryClass = "hep-th",
    doi = "10.1007/JHEP04(2023)009",
    journal = "JHEP",
    volume = "04",
    pages = "009",
    year = "2023"
}

@article{Balasubramanian:2025zey,
    author = "Balasubramanian, Vijay and Yildirim, Tom",
    title = "{The nonperturbative Hilbert space of quantum gravity with one boundary}",
    eprint = "2506.04319",
    archivePrefix = "arXiv",
    primaryClass = "hep-th",
    doi = "10.1007/JHEP03(2026)040",
    journal = "JHEP",
    volume = "03",
    pages = "040",
    year = "2026"
}

@book{Streater:1989vi,
    author = "Streater, R. F. and Wightman, A. S.",
    title = "{PCT, spin and statistics, and all that}",
    isbn = "978-0-691-07062-9",
    year = "1989"
}

@book{hindman1998algebra,
  title={Algebra in the Stone-{\v{C}}ech compactification: theory and applications},
  author={Hindman, Neil and Strauss, Dona},
  volume={27},
  year={1998},
  publisher={Walter de Gruyter}
}

@article{ando2014ultraproducts,
  title={Ultraproducts of von Neumann algebras},
  author={Ando, Hiroshi and Haagerup, Uffe},
  journal={Journal of Functional Analysis},
  volume={266},
  number={12},
  pages={6842--6913},
  year={2014},
  publisher={Elsevier}
}

@book{ocneanu2006actions,
  title={Actions of discrete amenable groups on von Neumann algebras},
  author={Ocneanu, Adrian},
  year={2006},
  publisher={Springer}
}

@article{Maldacena:2001kr,
    author = "Maldacena, Juan Martin",
    title = "{Eternal black holes in anti-de Sitter}",
    eprint = "hep-th/0106112",
    archivePrefix = "arXiv",
    reportNumber = "NSF-ITP-01-59",
    doi = "10.1088/1126-6708/2003/04/021",
    journal = "JHEP",
    volume = "04",
    pages = "021",
    year = "2003"
}

@article{Drukker:2011zy,
    author = "Drukker, Nadav and Marino, Marcos and Putrov, Pavel",
    title = "{Nonperturbative aspects of ABJM theory}",
    eprint = "1103.4844",
    archivePrefix = "arXiv",
    primaryClass = "hep-th",
    reportNumber = "IMPERIAL-TP-2011-ND-01",
    doi = "10.1007/JHEP11(2011)141",
    journal = "JHEP",
    volume = "11",
    pages = "141",
    year = "2011"
}

@article{Hatsuda:2015gca,
    author = "Hatsuda, Yasuyuki and Moriyama, Sanefumi and Okuyama, Kazumi",
    title = "{Exact instanton expansion of the ABJM partition function}",
    eprint = "1507.01678",
    archivePrefix = "arXiv",
    primaryClass = "hep-th",
    reportNumber = "DESY-15-101, OCU-PHYS-428",
    doi = "10.1093/ptep/ptv145",
    journal = "PTEP",
    volume = "2015",
    number = "11",
    pages = "11B104",
    year = "2015"
}

@article{Grassi:2014vwa,
    author = "Grassi, Alba and Marino, Marcos",
    title = "{M-theoretic matrix models}",
    eprint = "1403.4276",
    archivePrefix = "arXiv",
    primaryClass = "hep-th",
    doi = "10.1007/JHEP02(2015)115",
    journal = "JHEP",
    volume = "02",
    pages = "115",
    year = "2015"
}
\end{document}